\documentclass[english]{article}
\usepackage[T1]{fontenc}
\usepackage[utf8]{inputenc}
\usepackage{textcomp}
\usepackage{amsmath}
\usepackage{cases}
\usepackage{amsthm}
\usepackage{amssymb}
\usepackage{graphicx}
\usepackage{eurosym}
\usepackage{color}
\usepackage[shortlabels]{enumitem}
\usepackage[toc,page]{appendix}%
\makeatletter
  \theoremstyle{remark}
  \newtheorem{rem}{\protect\remarkname}
  \theoremstyle{plain}
  \newtheorem{lem}{\protect\lemmaname}
  \theoremstyle{plain}
  \newtheorem{prop}{\protect\propositionname}
  \theoremstyle{plain}
  \newtheorem*{theo*}{\protect\theoremname}
  \theoremstyle{definition}
  \newtheorem{defn}{\protect\definitionname}
  \theoremstyle{plain}
  
\theoremstyle{plain}
\newtheorem{thm}{\protect\theoremname}

\usepackage {a4wide}
\usepackage{mathrsfs}
\usepackage{bbm}
\makeatother

\usepackage{babel}
  \providecommand{\definitionname}{Definition}
  \providecommand{\propositionname}{Proposition}
  \providecommand{\theoremname}{Theorem}
  \providecommand{\remarkname}{Remark}
\providecommand{\corollaryname}{Corollary}
\providecommand{\lemmaname}{Lemma}
\newcommand*\samethanks[1][\value{footnote}]{\footnotemark[#1]}

\newcommand{\norm}[1]{\left\lVert#1\right\rVert}

\begin{document}

\title{Portfolio choice, portfolio liquidation, and portfolio transition under drift uncertainty\thanks{This research has been conducted with the support of the Research Initiative ``Modélisation des marchés actions et dérivés'' financed by HSBC France under the aegis of the Europlace Institute of Finance. The authors would like to thank Rama Cont (Imperial College), Nicolas Grandchamp des Raux (HSBC France), Charles-Albert Lehalle (CFM and Imperial College), Jean-Michel Lasry (Institut Louis Bachelier), Huyên Pham (Université Paris-Diderot), and Christopher Ulph (HSBC London) for the conversations they had on the subject.}}
\date{}
\author{Alexis Bismuth\thanks{Université Paris 1 Panthéon-Sorbonne. Centre d'Economie de la Sorbonne. 106, Boulevard de l'Hôpital, 75013 Paris.}\;\thanks{Den-Service de thermo-hydraulique et de mécanique des fluides - Laboratoire de Génie Logiciel pour la Simulation (DEN/STMF/LGLS), CEA, Université Paris-Saclay, F-91191, Gif-sur-Yvette, France.}, Olivier Guéant\samethanks[2], Jiang Pu\thanks{Institut Europlace de Finance. 28 Place de la Bourse, 75002 Paris.}}
\maketitle


\begin{abstract}
This paper presents several models addressing optimal portfolio choice, optimal portfolio liquidation,
and optimal portfolio transition issues, in which the expected returns
of risky assets are unknown. Our approach is based on a coupling between
Bayesian learning and dynamic programming techniques that leads to partial differential equations. It enables to
recover the well-known results of Karatzas and Zhao in a framework \emph{à la} Merton, but also to deal with cases where martingale methods are no longer
available. In particular, we address optimal portfolio choice, portfolio liquidation, and portfolio transition problems in a
framework \emph{à  la} Almgren-Chriss, and we build therefore a model
in which the agent takes into account in his decision
process both the liquidity of assets and the uncertainty with respect
to their expected return.

\vspace{5mm}

 \vspace{5mm}
\end{abstract}

\section{Introduction}

The modern theory of portfolio selection started in 1952 with the seminal
paper \cite{markowitz1952portfolio} of Markowitz.\footnote{Markowitz was awarded the Nobel Prize in 1990 for his work. For a
brief history of portfolio theory, see \cite{markowitz1999early}.} In his paper, Markowitz considered the problem of an agent who
wishes to build a portfolio with the maximum possible level of expected
return, given a limit level of variance. He then coined the concept
of efficient portfolio and described how to find such portfolios.
Markowitz paved the way for studying theoretically the optimal portfolio
choice of risk-averse agents. A few years after Markowitz's
paper, Tobin published indeed his famous research work on the
liquidity preferences of agents and the separation theorem (see \cite{tobin1958liquidity}), which is based
on the ideas developed by Markowitz. A few years later, in the sixties,
Treynor, Sharpe, Lintner, and Mossin introduced independently the
Capital Asset Pricing Model (CAPM) which is also built on top of the ideas of Markowitz. The ubiquitous notions of $\alpha$ and $\beta$ owe a lot
therefore to Markowitz modern portfolio theory.\\

Although initially written within a mean-variance optimization framework,
the so-called Markowitz problem can also be written within the Von
Neumann-Morgenstern expected utility framework. This was for instance
done by Samuelson and Merton (see \cite{merton1969lifetime,merton1971optimum,samuelson1969lifetime}),
who, in addition, generalized Markowitz problem by extending the
initial one-period framework to a multi-period one. Samuelson did it
in discrete time, whereas Merton did it in continuous time. It is
noteworthy that they both embedded the intertemporal portfolio choice
problem into a more general optimal investment/consumption problem.\footnote{This problem in continuous time is now referred to as Merton's problem.}\\

In \cite{merton1969lifetime}, Merton used partial differential equation (PDE) techniques for characterizing the optimal consumption process of an agent and its
optimal portfolio choices. In particular, Merton managed to find closed-form
solutions in the constant absolute risk aversion (CARA) case (i.e., for exponential
utility functions), and in the constant relative risk aversion (CRRA) case
(i.e., for power and log utility functions). Merton's problem has
then been extended to incorporate several features such as transaction
costs (proportional and fixed) or credit constraints. Major
advances to solve Merton's problem in full generality have been
made in the eighties by Karatzas \emph{et al.} using (dual) martingale
methods. In \cite{karatzas1987optimal}, Karatzas, Lehoczky, and Shreve
used a martingale method to solve Merton's problem for almost any
smooth utility function and showed
how to partially disentangle the consumption maximization problem
and the terminal wealth maximization problem. Constrained problems
and extensions to incomplete markets were then considered \textendash{}
see for instance the paper \cite{cvitanic1992convex} by Cvitani\'{c}
and Karatzas.\\

In the literature on portfolio selection or in the slightly more general
literature on Merton's problem, input parameters (for instance the
expected returns of risky assets) are considered known constants,
or stochastic processes with known initial values and dynamics. In
practice however, one cannot state for sure that price returns will
follow a given distribution. Uncertainty on model parameters is the
\emph{raison d'être} of the celebrated Black-Litterman model (see \cite{black1992global}), which is built on top of Markowitz model and the CAPM. Nevertheless,
like Markowitz model, Black-Litterman model is a static one. In particular,
the agent of Black-Litterman model does not use empirical returns to dynamically
learn the distribution of asset returns.\\

Generalizations of optimal allocation models (or models dealing with Merton's problem) involving filtering and learning techniques in a partial information framework have been proposed in the literature. The problems that are addressed are of three types depending on the assumptions regarding the drift: unknown constant drift (e.g. \cite{rogers2001}, \cite{danilova}, \cite{karatzas1998bayesian}), unobserved drift with Ornstein-Uhlenbeck dynamics (e.g. \cite{brendle2006portfolio}, \cite{fouque}, \cite{li}, \cite{rishel1999optimal}), and unobserved drift modelled by a hidden Markov chain (e.g. \cite{cj}, \cite{vroum}, \cite{rieder2005portfolio}, \cite{sh2004}). In the different models, filtering (or learning) enables to estimate the unknown parameters from the dynamics of the prices, and sometimes also from additional information such as analyst views or expert opinions (see \cite{davislleo} and \cite{fouque2}) or inside information (see \cite{danilova} and \cite{monoyios2009}).\\

Most models (see \cite{bdl}, \cite{rogers2001}, \cite{danilova}, \cite{karatzas1998bayesian}, \cite{lak95}, \cite{lak98}, \cite{monoyios2009}, \cite{ps2008}) use martingale (dual) methods to solve optimal allocation problems under partial information. For instance, in a framework similar to ours, Karatzas and Zhao \cite{karatzas1998bayesian} considered a model where the asset returns are Gaussian with unknown mean and they used martingale methods under the filtration of observables to compute, for almost any utility function, the optimal portfolio allocation (there is no consumption in their model).\\

Some models, like ours, use instead Hamilton-Jacobi-Bellman (HJB) equations and therefore PDE techniques. Rishel \cite{rishel1999optimal} proposed a model with one risky asset where the drift has an Ornstein-Uhlenbeck dynamics and solved the HJB equation associated with CRRA utility functions. Interestingly, it is one of the rare references to tackle the question of explosion when Bayesian filtering and optimization are carried out simultaneously. Brendle \cite{brendle2006portfolio} generalized the results of \cite{rishel1999optimal} to a multi-asset framework and also considered the case of CARA utility functions. Fouque \emph{et al.} \cite{fouque} solved a related problem with correlation between the noise process of the price and that of the drift and used perturbation analysis to obtain approximations. Li \emph{et al.} \cite{li} also studied a similar problem with a mean-variance objective function. Rieder and Bäuerle \cite{rieder2005portfolio} proposed a model with one risky asset where the drift is modelled by a hidden Markov chain and solved it with PDEs in the case of CRRA utility function.\\

Outside of the optimal portfolio choice literature, several authors proposed financial models in which both online learning and stochastic optimal control coexist. For instance, Laruelle \emph{et al.} proposed in \cite{laruelle} a model in which an agent optimizes its execution strategy with limit orders and simultaneously learns the parameters of the Poisson process modelling the execution of limit orders. Interesting ideas in the same field of algorithmic trading can also be found in the work of Fernandez-Tapia (see \cite{joaquin}). An interesting paper is also that of Ekstr\"om and Vaicenavicius \cite{ekstrom} who tackled the problem of the optimal time at which to sell an asset with unknown drift. Recently, Casgrain and Jaimungal \cite{cj} also used similar ideas for designing algorithmic trading strategies.\\

In this paper, we consider several problems of portfolio choice, portfolio liquidation, and portfolio transition in
continuous time in which the (constant) expected returns of the risky
assets are unknown but estimated online. In the first sections, we consider a multidimensional portfolio choice problem similar to the one
tackled by Karatzas and Zhao in \cite{karatzas1998bayesian} with a rather general Bayesian prior for the drifts (our family of priors includes compactly supported and Gaussian distributions).\footnote{It is noteworthy that this approach can be carried out in the frequentist case as well.} For this problem, with general Bayesian prior, we derive HJB equations and show that, in the CARA and CRRA cases, these equations can be transformed into linear parabolic PDEs. The interest of the paper lies here in the fact that our framework is multidimensional and general in terms of possible priors. Moreover, unlike other papers, we provide a verification result and this is important in view of the explosion occurring for some couples of priors and utility functions. We then specify our results in the case of a Gaussian prior for the drifts and recover formulas already present in the literature (see \cite{karatzas1998bayesian} or limit cases of \cite{brendle2006portfolio}). The Gaussian prior case is discussed in depth, (i) because the associated PDEs can be simplified into simple ODEs (at least for CARA and CRRA utility functions) that can be solved in closed form by using classical tricks, and (ii) because Gaussian priors provide examples of explosion: the problem may not be well posed in the CRRA case when the relative risk aversion parameter is too small.\\

The PDE approach is interesting in itself  and we believe that it enables to avoid the laborious computations needed to simplify the general expressions of Karatzas and Zhao. However, our message is of course not limited to that one. The PDE approach
can indeed be used in situations where the (dual) martingale approach
cannot be used. In the last section of this paper, we use our approach to solve the optimal
allocation problem in a trading framework \emph{à  la} Almgren-Chriss. The Almgren-Chriss framework was initially
built for solving optimal execution problems (see \cite{almgren1999value,almgren2001optimal})
but it is also very useful outside of the cash-equity world. For instance,
Almgren and Li \cite{almgren2016option}, and Guéant and Pu \cite{gueant2015option}
used it for the pricing and hedging of vanilla options when liquidity
matters.\footnote{Guéant \emph{et al.} also used the Almgren-Chriss framework to tackle
the pricing, hedging, and execution issues of Accelerated Share
Repurchase contracts \textendash{} see \cite{gueant2014optimal,gueant2015accelerated}.} The model we propose is one of the first models that uses the Almgren-Chriss
framework for addressing an asset management problem, and definitely the
first paper in this area in which the Almgren-Chriss framework is
used in combination with Bayesian learning techniques.\footnote{Almgren and Lorenz used Bayesian techniques in optimal execution (see
\cite{almgren2006bayesian}), but they considered myopic agents with
respect to learning.} We also show how our framework can be slightly modified for addressing optimal portfolio liquidation and transition issues.\\

This paper aims at proving that online learning -- in our case on-the-fly Bayesian estimations -- combined with stochastic
optimal control can be very efficient to tackle a lot of financial
problems. It is essential to understand that online learning is
a forward process whereas dynamic programming classically relies on
backward induction. By using these two classical tools
simultaneously, we do not only benefit from the power of online and Bayesian
learning to continuously learn the value of unknown parameters,
but we also develop a framework in which agents learn and make decisions
knowing that they will go on learning in the future in the same manner
as they have learnt in the past. The same ideas are for instance at
play in the literature on Bayesian multi-armed bandits where the unknown
parameters are the parameters of the prior distributions of the different rewards.\\

In Section 2, we provide the main results related to our Bayesian framework. We first compute the Bayesian estimator of the drifts entering the dynamics of prices (more precisely the conditional mean given the prices trajectory and the prior). We then derive the dynamics of that Bayesian estimator. These results are classical and can be found in \cite{bain} or \cite{lipster}, but they are recalled for the sake of completeness. In Section 3, we consider the portfolio allocation problem of an agent in a context with one risk-free asset and $d$ risky assets, and we show how the associated HJB equations can be transformed into linear parabolic PDEs in the case of a CARA utility function and of a CRRA utility function. As opposed to most of the papers in the literature, we also provide verification theorems. This is of particular importance because the Bayesian framework leads to blowups for some of the optimal control problems. In Section 4, we solve the same portfolio allocation problem as in Section 3 but in the specific case of a Gaussian prior. We show that a more natural set of state variables can be used to solve the same problem. We also provide an example of blowup in the Gaussian case. In Section 4, thanks to closed-form solutions, we also analyze the role of learning on the dynamics of the allocation process of the agent. In Section 5, we introduce liquidity costs through a modelling framework \emph{\`a la} Almgren-Chriss and we use our combination of Bayesian learning and stochastic optimal control techniques for solving various portfolio choice, portfolio liquidation, and portfolio transition problems.\\

\section{Bayesian learning}

\subsection{Notations and first properties}

We consider an agent facing a portfolio allocation problem with one risk-free asset and $d$~risky assets.\\

Let $\left(\Omega,\left(\mathcal{F}_{t}^{W}\right)_{t\in\mathbb{R}_{+}},\mathbb{P}\right)$
be a filtered probability space, with $\left(\mathcal{F}_{t}^{W}\right)_{t\in\mathbb{R}_{+}}$
satisfying the usual conditions. Let $\left(W_{t}\right)_{t\in\mathbb{R}_{+}}$
be a $d$-dimensional Brownian motion adapted to $\left(\mathcal{F}_{t}^{W}\right)_{t\in\mathbb{R}_{+}}$, with correlation structure given by
$ d\left\langle W^{i},W^{j}\right\rangle_t =\rho^{ij}dt$ for all $i,j$ in $\left\{1,\hdots,d\right\}$.\\

The risk-free interest rate is denoted by $r$.
We index by $i \in \left\{1,\hdots,d\right\}$ the $d$ risky assets. For $i\in\left\{1,\hdots,d\right\} $, the price of the $i^\text{th}$ asset $S^i$ has the classical log-normal dynamics
\begin{eqnarray}
 dS_{t}^{i}  &=&  \mu^{i}S_{t}^{i}dt+\sigma^{i}S_{t}^{i}dW_{t}^{i},
\end{eqnarray}
where the volatility vector $\sigma = (\sigma^{1}, \ldots, \sigma^d)'$ satisfies $\forall i \in \left\{1,\hdots,d\right\}, \sigma^{i} > 0$, and where the drift vector $\mu=(\mu^{1}, \ldots, \mu^d)'$ is unknown.\\

We assume that the prior distribution of $\mu$, denoted by $m_\mu$, is sub-Gaussian.\footnote{This assumption can be slightly relaxed, but we consider this simple one to simplify the statement of our results.} In particular, it satisfies the following property:
\begin{align}
\exists \eta>0,\quad \mathbb{E}[e^{\eta \norm{\mu}^2}] = \int_{z\in\mathbb{R}^d} e^{\eta \norm{z}^2}m_\mu (dz)
<+\infty.
\label{eq:priorcondition}
\end{align}

Throughout, we shall respectively denote by $\rho = (\rho^{ij})_{1\le i,j \le d}$ and $\Sigma = (\rho^{ij}\sigma^{i}\sigma^{j})_{1\le i,j \le d}$ the correlation and covariance matrices associated with the dynamics of prices.\\

We also denote by $(Y_t)_{t\in\mathbb{R}_+}$ the process defined by
\begin{align}
 \forall i\in \left\{1,\hdots,d\right\},\forall t\in\mathbb{R}_+,\quad
& Y^i_t=\log S^i_t.
\end{align}

\begin{rem}
Both $\mu$ and $(W_{t})_{t\in\mathbb{R}_+}$ are unobserved by the agent,
but for each index $i\in\left\{1,\hdots,d\right\}$, $\mu^i t+\sigma^i W_{t}^i$ is observed at time $t\in\mathbb{R}_+$ because
\begin{eqnarray}
\mu^i t+\sigma^i
W_{t}^i&=&
Y^i_t-Y^i_0
+\dfrac{1}{2}
{\sigma^i}^{2}t.
\end{eqnarray}
\end{rem}

The evolution of the prices reveals information to the agent about the true value of the drift vector $\mu$. In what follows we denote by $\mathcal{F}^{S} = \left(\mathcal{F}_{t}^{S}\right)_{t\in\mathbb{R}_{+}}$ the filtration generated by $\left(S_{t}\right)_{t\in\mathbb{R}_{+}}$ or equivalently by $\left(Y_{t}\right)_{t\in\mathbb{R}_{+}}$.\\

\begin{rem}
$(W_t)_{t\in\mathbb{R}_+}$ is not an $\mathcal{F}^{S}$-Brownian motion, because it is not $\mathcal{F}^{S}$-adapted.
\end{rem}

We introduce the process $(\beta_t)_{t\in\mathbb{R}_+}$ defined by
\begin{eqnarray}
    \forall t \in \mathbb{R}_+,\quad \beta_t&=&\mathbb{E}\left[\left.
\mu
\right|
 \mathcal{F}_t^S
\right].
\end{eqnarray}

\begin{rem}
$(\beta_t)_{t\in\mathbb{R}_+}$ is well defined because of the assumption (\ref{eq:priorcondition}) on the prior $m_\mu$.
\end{rem}

From an investor's point of view, $(\beta_t)_{t\in\mathbb{R}_+}$ is of main concern. It encapsulates the information gathered so far about the returns one can expect from the assets.

The first result stated in Theorem 1 is a formula for $\beta_t$.

\begin{thm}
\label{propbayes}
Let us define
\begin{equation}
F:\; (t,y)\in\mathbb{R}_+\times\mathbb{R}^d
\mapsto
\int_{\mathbb{R}^d}
\exp\left(
(z-r\vec{1})'
\Sigma^{-1}
\left[
        y-Y_0 +\!
            \left(\!
            -r\vec{1}
            +\frac12
            \sigma \odot \sigma\!
        \right)t
-\frac{t}{2}
(z-r\vec{1})
\right]
\right)
    m_\mu(dz),
    \label{eq:F}
\end{equation}
where $\odot$ denotes the element-wise multiplication of vectors.\\

$F$ is a well-defined finite-valued $C^\infty(\mathbb{R}_+\times\mathbb{R}^d)$ function.\\

We have
\begin{eqnarray}
\forall t\in\mathbb{R}_+,\quad \beta_t&=&
\Sigma
G
(t,Y_t)+r\vec{1} ,\label{eq:eq42}\end{eqnarray}
where
\begin{eqnarray}
G&=&\frac{\nabla_y F}{F},
\end{eqnarray} and where we denote by $\vec{1}$ the vector $\left(
1, \hdots, 1
\right)'\in\mathbb{R}^{d}$.
\end{thm}

Before we prove Theorem 1, let us introduce the probability measure $\mathbb{Q}$ defined by
\begin{eqnarray}
\frac{d\mathbb{Q}}{d\mathbb{P}}
&=&
\exp\left(
-\alpha(\mu)' \rho^{-1}W_T-\frac12
\alpha(\mu)' \rho^{-1} \alpha(\mu)
T
\right),
\end{eqnarray}
where $\alpha : z = (z^1, \ldots, z^d)' \in \mathbb{R}^d \mapsto \left(
\frac{z^1-r}{\sigma^1},\ldots,\frac{z^d-r}{\sigma^d}
\right)'$ and $T$ is an arbitrary constant in $\mathbb{R}^*_+$.

Girsanov's theorem implies that the process $\left(W_t^\mathbb{Q}\right)_{t\in[0,T]}$ defined by
\begin{eqnarray}
\forall i \in\left\{1,\hdots,d\right\},\forall t\in [0,T],\quad
\left(W_t^{\mathbb{Q}}\right)^i&=&W^i_t+\frac{\mu^i-r}{\sigma^i}t,
\end{eqnarray}
is a $d$-dimensional Brownian motion with correlation structure given by $\rho$ under $\mathbb{Q}$ and adapted to the filtration $\left(\mathcal{F}^S_t\right)_{t\in[0,T]}$. Moreover
\begin{eqnarray}
\forall i\in \left\{1,\hdots,d\right\},\quad
    \frac{dS^i_t}{S^i_t}&=&rdt+\sigma^i \left(dW_t^{\mathbb{Q}}\right)^i\quad\!\!
\text{and}\!\!\quad dY_t^i=
\left(
r-\frac{{\sigma^{i}}^2}{2}
\right)dt+\sigma^i\left(dW_t^\mathbb{Q}\right)^i.
\end{eqnarray}

The following proposition will be used in the proof of Theorem 1.
\begin{prop}
\label{prop:indepmu}
Under the probability measure $\mathbb{Q}$, $\mu$ is independent of $W_t^\mathbb{Q}$ for all $t\in[0,T]$.
\end{prop}

\begin{proof}
Since, for all $t\in[0,T]$, $\mu$ is independent of $W_t$ under the probability measure $\mathbb{P}$, we have, for $(t,a,b)\!\in\![0,T]\times\mathbb{R}^d\times\mathbb{R}^d$,
\begin{eqnarray*}
&&\mathbb{E}^{\mathbb{Q}}\left[
\exp\left(
i a'\mu+ i b'W_t^\mathbb{Q}
\right)
\right]
\\
&=&
\mathbb{E}
\left[
\exp\left(
i a'\mu+ i b'
\left(
W_t+\alpha(\mu)t
\right)
-\alpha(\mu)' \rho^{-1}W_T-\frac12
\alpha(\mu)' \rho^{-1} \alpha(\mu)T
\right)
\right]
\\
&=&
\mathbb{E}
\bigg[
\exp
\left(
i a'\mu
+i b'\alpha(\mu)t
-\frac12
\alpha(\mu)' \rho^{-1} \alpha(\mu)
T
\right)
\\
&&
\mathbb{E}
\left[
\left.
\exp\left(
ib'W_t
-\alpha(\mu)' \rho^{-1}W_T
\right)
\right|
\mu
\right]
\bigg]\\
&=&
\mathbb{E}
\bigg[
\exp
\left(
i a'\mu
+i b'\alpha(\mu)t
-\frac12
\alpha(\mu)' \rho^{-1} \alpha(\mu)
T
\right)
\\
&&
\mathbb{E}
\left[
\left.
\exp\left(
ib'W_t
-\alpha(\mu)' \rho^{-1}
\left(
W_T-W_t
\right)
-\alpha(\mu)' \rho^{-1}W_t
\right)
\right|
\mu
\right]
\bigg]\\
&=&
\mathbb{E}
\bigg[
\exp
\left(
i a'\mu
+i b'\alpha(\mu)t
-\frac12
\alpha(\mu)' \rho^{-1} \alpha(\mu)
T
\right)
\\
&&
\exp\left(\frac 12 \alpha(\mu)' \rho^{-1} \alpha(\mu) (T-t)\right) \exp\left( \frac 12 \left(ib -  \rho^{-1} \alpha(\mu)\right)'\rho \left(ib -  \rho^{-1} \alpha(\mu)\right) t \right)\bigg]\\
&=&
\mathbb{E}
\left[
\exp
\left(
i a'\mu
\right)
\right]
\exp
\left(-
\frac{t}{2}b'\rho b
\right).
\end{eqnarray*}

Now, let us notice that
\begin{eqnarray*}
\mathbb{E}^\mathbb{Q}
\left[
\exp
\left(
ia'\mu
\right)
\right]
=
\mathbb{E}
\left[
\exp
\left(
ia'\mu
\right)
\frac{d\mathbb{Q}}{d\mathbb{P}}
\right]
=
\mathbb{E}
\left[
\exp
\left(
ia'\mu
\right)
\underbrace{
\mathbb{E}
\left[
\left.
\frac{d\mathbb{Q}}{d\mathbb{P}}
\right|
\mu
\right]}_{=1}
\right]
=
\mathbb{E}
\left[
\exp
\left(
ia'\mu
\right)
\right]
\end{eqnarray*}
and $\exp\left(-\frac{t}{2}b'\rho b\right)$ is the Fourier transform of $W_t^\mathbb{Q}$ under the probability measure $\mathbb{Q}$.
\\

Therefore,
\begin{eqnarray*}
\mathbb{E}^\mathbb{Q}
\left[
\exp
\left(
ia'\mu
+ib' W_t^\mathbb{Q}
\right)
\right]
&=&\mathbb{E}^\mathbb{Q}
\left[
\exp
\left(
ia'\mu
\right)
\right]
\mathbb{E}^\mathbb{Q}
\left[
\exp
\left(
ib' W_t^\mathbb{Q}
\right)
\right],
\end{eqnarray*}
hence the result.
\end{proof}

We are now ready to prove Theorem 1.

\begin{proof}[Proof of Theorem 1.]

Let us first show that $F$ is a well-defined finite-valued $C^\infty(\mathbb{R}_+\times\mathbb{R}^d)$ function.

We have
\begin{align*}
\forall (t,y,z)\in \mathbb{R}_+\times\mathbb{R}^d\times\mathbb{R}^d,\quad&
\exp\left(
(z-r\vec{1})'
\Sigma^{-1}
\left[
        y-Y_0
            +\left(
            -r\vec{1}
            +\frac12
            \sigma \odot \sigma
        \right)t
-\frac{t}{2}
(z-r\vec{1})
\right]
\right)
\\
\leq&
\exp\left(
(z-r\vec{1})'
\Sigma^{-1}
\left[
        y-Y_0+
            \left(
            -r\vec{1}
            +\frac12
            \sigma \odot \sigma
        \right)t
\right]
\right).
\end{align*}
Therefore, to show that $F$ takes finite values, we just need to prove that for $a\in\mathbb{R}^d$,
\begin{align*}
\int_{\mathbb{R}^d} \exp(a'z)m_\mu(dz)=\mathbb{E}\left[\exp(a'\mu)\right]<+\infty.
\end{align*}

Thanks to condition (\ref{eq:priorcondition}) on the prior, there exists $\eta > 0$ such that $\mathbb{E}
\left[
\exp\left(\eta\norm{\mu}^2\right)
\right] < +\infty$. Therefore,
\begin{align*}
\forall a\in \mathbb{R}^d,
\quad
\mathbb{E}
\left[
\exp\left( \mu'a\right)
\right]
=&\mathbb{E}
\left[
\exp\left(a'\mu-\eta\norm{\mu}^2\right)\exp\left(\eta
\norm{\mu}^2\right)
\right]
\\ \leq&
\exp\left(\sup_{z \in \mathbb{R}^d} a'z-\eta\norm{z}^2 \right)
\mathbb{E}
\left[
\exp\left(\eta\norm{\mu}^2\right)
\right]\\
\leq&
\exp\left(\frac{\norm{a}^2}{4\eta}\right)
\mathbb{E}
\left[
\exp\left(\eta\norm{\mu}^2\right)
\right]\\
<&+\infty.
\end{align*}

Consequently, $F$ is well defined and takes finite values.\\

For proving that $F$ is in fact a $C^\infty(\mathbb{R}_+\times\mathbb{R}^d)$ function, we see by formal derivation that it is sufficient to prove that, for all $n\in \mathbb{N}$,
\begin{align*}
a \in \mathbb{R}^d \mapsto \int_{\mathbb{R}^d} \norm{z}^n\exp(a'z)m_\mu(dz)=\mathbb{E}\left[\norm{\mu}^n\exp(a'\mu)\right]
\end{align*}
is bounded over all compact sets of $\mathbb{R}^d$.\\

We have
\begin{align*}
\forall a\in \mathbb{R}^d, \forall n \in \mathbb{N},
\quad
\mathbb{E}
\left[\norm{\mu}^n
\exp\left( a'\mu\right)
\right]
=&\mathbb{E}
\left[
\norm{\mu}^n\exp\left(a'\mu-\eta\norm{\mu}^2\right)\exp\left(\eta
\norm{\mu}^2\right)
\right]\\ \leq&
\sup_{z \in \mathbb{R}^d} \left( \norm{z}^n \exp\left(a'z-\eta\norm{z}^2 \right)\right)
\mathbb{E}
\left[
\exp\left(\eta\norm{\mu}^2\right)
\right]\\
\leq&
\sup_{z \in \mathbb{R}^d} \norm{z}^n \left(\exp\left(\norm{z}\norm{a}-\eta\norm{z}^2 \right)\right)
\mathbb{E}
\left[
\exp\left(\eta\norm{\mu}^2\right)
\right]\\
<&+\infty,
\end{align*}
hence the result.\\

We are now ready to prove the formula for $\beta_t$.\\

By Bayes' theorem we have, for all $t$ in $[0,T]$,
\begin{eqnarray*}
\beta_t&=&\frac{\mathbb{E}^{\mathbb{Q}}
    \left[\left.
    \mu \frac{d\mathbb{P}}{d\mathbb{Q}}
    \right| \mathcal{F}^S_t
    \right]}
    {\mathbb{E}^{\mathbb{Q}}
    \left[\left.
    \frac{d\mathbb{P}}{d\mathbb{Q}}
    \right| \mathcal{F}^S_t
    \right]}.\nonumber
\end{eqnarray*}
Since
\begin{eqnarray*}
\frac{d\mathbb{P}}{d\mathbb{Q}}
&=&\exp\left(
\alpha(\mu)' \rho^{-1}
W_T^\mathbb{Q}
-\frac{T}{2}
\alpha(\mu)' \rho^{-1} \alpha(\mu)
\right),\nonumber
\end{eqnarray*}
we have
\begin{eqnarray*}
\beta_t&=&\frac{\mathbb{E}^{\mathbb{Q}}
    \left[
    \mu
\exp\left(
\alpha(\mu)' \rho^{-1}
W_T^\mathbb{Q}
-\frac{T}{2}
\alpha(\mu)' \rho^{-1} \alpha(\mu)
\right)
    | \mathcal{F}^S_t
    \right]}
    {\mathbb{E}^{\mathbb{Q}}
    \left[
    \exp\left(
\alpha(\mu)' \rho^{-1}
W_T^\mathbb{Q}
-\frac{T}{2}
\alpha(\mu)' \rho^{-1} \alpha(\mu)
\right)
    | \mathcal{F}^S_t
    \right]}.
\end{eqnarray*}

Proposition \ref{prop:indepmu} now yields
\begin{eqnarray*}
    \beta_t&=&
    \frac{\mathbb{E}^{\mathbb{Q}}
    \left[
    \mu
    \exp\left(
\alpha(\mu)' \rho^{-1}
W_t^\mathbb{Q}
-\frac{t}{2}
\alpha(\mu)' \rho^{-1} \alpha(\mu)
\right)
    | \mathcal{F}^S_t
    \right]}
    {\mathbb{E}^{\mathbb{Q}}
    \left[
\exp\left(
\alpha(\mu)' \rho^{-1}
W_t^\mathbb{Q}
-\frac{t}{2}
\alpha(\mu)' \rho^{-1} \alpha(\mu)
\right)
    | \mathcal{F}^S_t
    \right]}
    \nonumber\\
    &=& \frac{
    \displaystyle \int\limits_{\mathbb{R}^d} z
\exp\left(
\alpha(z)' \rho^{-1}
W_t^\mathbb{Q}
-\frac{t}{2}
\alpha(z)' \rho^{-1} \alpha(z)
\right)
    m_\mu(dz)
    }
    {
    \displaystyle \int\limits_{\mathbb{R}^d}
\exp\left(
\alpha(z)' \rho^{-1}
W_t^\mathbb{Q}
-\frac{t}{2}
\alpha(z)' \rho^{-1} \alpha(z)
\right)
    m_\mu(dz)
    }
\\
    &=&\frac{
    \displaystyle \int\limits_{\mathbb{R}^d} z
\exp\left(
(z-r\vec{1})'\Sigma^{-1}
    \left(
    Y_t
    - Y_0+
        \left(
            -r\vec{1}
            +\frac12
            \sigma \odot \sigma
        \right)t
    \right)
-\frac{t}{2}
\alpha(z)' \rho^{-1} \alpha(z)
\right)
    m_\mu(dz)
    }
    {
    \displaystyle \int\limits_{\mathbb{R}^d}
\exp\left(
(z-r\vec{1})'\Sigma^{-1}
    \left(
        Y_t-Y_0+
        \left(
            -r\vec{1}
            +\frac12
            \sigma \odot \sigma
        \right)t
    \right)
-\frac{t}{2}
\alpha(z)' \rho^{-1} \alpha(z)
\right)
    m_\mu(dz)
    }
    \nonumber.
 \end{eqnarray*}

Consequently
\begin{eqnarray*}
&&\Sigma^{-1}
\!\left(
\beta_t-r\vec{1}
\right)\\
&=&
\frac{
    \displaystyle \int\limits_{\mathbb{R}^d} \Sigma^{-1}(z-r\vec{1})
\exp\left(
(z-r\vec{1})'\Sigma^{-1}
    \left(
    Y_t
    - Y_0+
        \left(
            -r\vec{1}
            +\frac12
            \sigma \odot \sigma
        \right)t
    \right)
-\frac{t}{2}
\alpha(z)' \rho^{-1} \alpha(z)
\right)
    m_\mu(dz)
    }
    {
    \displaystyle \int\limits_{\mathbb{R}^d}
\exp\left(
(z-r\vec{1})'\Sigma^{-1}
    \left(
        Y_t-Y_0+
        \left(
            -r\vec{1}
            +\frac12
            \sigma \odot \sigma
        \right)t
    \right)
-\frac{t}{2}
\alpha(z)' \rho^{-1} \alpha(z)
\right)
    m_\mu(dz)
    }\\
&=&\frac{\nabla_y F}{F}(t,Y_t)\\
&=& G(t,Y_t).
\end{eqnarray*}

Therefore, and because $T$ is arbitrary, we have
\begin{eqnarray*}
\forall t\in\mathbb{R}_+,\quad
\beta_t &=& \Sigma G(t,Y_t)+r\vec{1}.
\end{eqnarray*}
\end{proof}

Throughout this article, we assume that the prior $m_\mu$ is such that $G$ has the following Lipschitz property with respect to $y$:
\begin{equation}
\label{eq:lip}
\forall T>0, \exists K_T > 0, \forall t \in [0,T], \forall y \in \mathbb{R}^d, \norm{D_y G(t,y)} \le K_T.
\end{equation}

As we shall see below, this assumption is verified if $m_\mu$ has a compact support. It is also verified for $m_\mu$ Gaussian (see Proposition \ref{diffprop} for instance).\footnote{Because we are dealing with asset returns, the class of compactly supported distributions is sufficient, from a financial point of view, to deal with almost all relevant cases. Gaussian distributions are not in that class but Gaussian priors are approximations of real-life beliefs that are used mainly for their convenience in computations.} However, it is not true in general for all sub-Gaussian priors.

\subsection{Dynamics of $(\beta_t)_{t\in\mathbb{R}_+}$}

Let us define the process $\left(\widehat W_t\right)_{t\in\mathbb{R}_+}$ by
\begin{eqnarray}
\forall i\in\left\{1,\hdots,d\right\},\forall t \in\mathbb{R}_+,\quad
    \widehat W_t^i& =& W_t^i+\int_0^t \frac{\mu^i-\beta^i_s}{\sigma^i} ds.
\end{eqnarray}
\begin{rem}
The process $(\widehat W_t)_{t \in \mathbb{R}_+}$ is called the innovation process in filtering theory. As shown below for the sake of completeness, it is classically known to be a Brownian motion (see for instance \cite{bain} on continuous Kalman filtering).
\end{rem}

\begin{prop}
$\left(\widehat{W}_{t}\right)_{t\in\mathbb{R}_+}$ is a $d$-dimensional Brownian motion
adapted to $\left(\mathcal{F}_{t}^{S}\right)_{t\in\mathbb{R}_+}$, with the same correlation structure as $\left({W}_{t}\right)_{t\in\mathbb{R}_+}$, i.e.,
$$
\forall i,j\in \left\{1,\hdots,d\right\},
\quad
d\langle\widehat W^{i}, \widehat W^{j} \rangle_t
=
d\langle W^{i}, W^{j} \rangle_t = \rho^{ij} dt.
$$
\end{prop}
\begin{proof}
To prove this result, we use Lévy's characterization of a Brownian motion.\\

Let $t\in \mathbb{R}_+$. By definition, we have
\begin{eqnarray*}
\forall i\in\left\{1,\hdots,d\right\},\quad
\widehat{W}_{t}^i&=&\dfrac{1}{\sigma^i}\left(\log\left(\dfrac{S_{t}^i}{S_{0}^i}\right)+\dfrac{1}{2}{\sigma^{i}}^2 t\right)-\int_{0}^{t}\dfrac{\beta^i_{s}}{\sigma^i}ds,
\end{eqnarray*}
hence the $\mathcal{F}_{t}^{S}$-measurability of $\widehat{W}_{t}$.\\

Let $s,t\in \mathbb{R}_+$, with $s<t$. For $i\in
\{ 1,\hdots ,d\}$,
\begin{eqnarray*}
\mathbb{E}\left[\left.\widehat{W}^i_{t}-\widehat{W}^i_{s}\right|\mathcal{F}_{s}^{S}\right] & = & \mathbb{E}\left[\left.W^i_{t}-W_{s}^i\right|\mathcal{F}_{s}^{S}\right]+\mathbb{E}\left[\left.
\int_{s}^{t}
\frac{1}{\sigma^i}
(\mu^i-\beta^i_{u})du\right|\mathcal{F}_{s}^{S}\right].
\end{eqnarray*}
For the first term, the increment $W^i_{t}-W^i_{s}$ is independent of $\mathcal{F}_{s}^{W}$ and independent of $\mu$. Therefore, it is independent of $\mathcal{F}_{s}^{S}$ and we have
\begin{eqnarray*}
\mathbb{E}\left[\left.W^i_{t}-W^i_{s}\right|\mathcal{F}_{s}^{S}\right] = \mathbb{E}[W^i_{t}-W^i_{s}] = 0.
\end{eqnarray*}
Regarding the second term, we have
\begin{eqnarray*}
\mathbb{E}\left[\left.\int_{s}^{t} \frac{1}{\sigma^i}
(\mu^i-\beta^i_{u})du\right|\mathcal{F}_{s}^{S}\right] & = & \int_{s}^{t}\mathbb{E}\left[\left.
\frac{1}{\sigma^i}
(\mu^i-\beta^i_{u})
\right|\mathcal{F}_{s}^{S}\right]du\\
 & = & \int_{s}^{t}\mathbb{E}\left[\left.\mathbb{E}\left[\left.
 \frac{1}{\sigma^i}(\mu^i-\beta^i_{u})
 \right|\mathcal{F}_{u}^{S}\right]\right|\mathcal{F}_{s}^{S}\right]du\\
 & = & 0,
\end{eqnarray*}
by definition of $\beta^i_{u}$.\\

We obtain that $\left(\widehat{W}_{t}\right)_{t\in\mathbb{R}_+}$ is an $\mathcal{F}^{S}$-martingale.\\

Since $(\widehat{W}_t)_{t\in\mathbb{R}_+}$ has continuous paths and $d\langle \widehat{W}^i,\widehat{W}^j\rangle_{t} =\rho^{ij}dt$, we conclude that $(\widehat{W}_t)_{t\in\mathbb{R}_+}$ is a $d$-dimensional $\mathcal{F}^{S}$-Brownian motion with correlation structure given by $\rho$.
\end{proof}

We are now ready to state the dynamics of $\left(\beta_t\right)_{t \in\mathbb{R}_+}$.

\begin{thm}
\label{propbayesdyn}
$\left(\beta_t\right)_{t \in\mathbb{R}_+}$ has the following dynamics:
\begin{equation}
\label{eq:dbeta}
d\beta_t = \Sigma D_y G(t,Y_t) \left(\sigma \odot d\widehat W_t\right).
\end{equation}
\end{thm}
\begin{proof}
By It\=o's formula and Theorem 1, we have
\begin{eqnarray*}
&&\Sigma^{-1}d\beta_t\\
&=&
\partial_t G(t,Y_t)dt
+\sum\limits_{i=1}^d
\partial_{y_i}G(t,Y_t) \;
dY_t^i
+\frac{1}{2}
\sum\limits_{i,j=1}^d\rho^{ij}
\sigma^i\sigma^j
\partial^2_{ y_i  y_j}G(t,Y_t)dt \nonumber\\
\end{eqnarray*}
\begin{eqnarray*}
&=&
\partial_t G(t,Y_t)dt
+
\sum\limits_{i=1}^d
\partial_{y_i} G(t,Y_t)
\left(
\beta_t^i dt+\sigma^i d\widehat W_t^i -\frac{{\sigma^{i}}^2}{2}dt
\right)
+
\frac{1}{2}
\sum\limits_{i,j=1}^d
\rho^{ij}\sigma^i\sigma^j
\partial^2_{ y_i y_j}G(t,Y_t)dt
\nonumber
\\
&=&\!\!\!
\left(
\partial_tG(t,Y_t)
+\sum\limits_{i=1}^d
\partial_{y_i} G(t,Y_t)
\left(
r+(\Sigma G)^i(t,Y_t)
-\frac{{\sigma^{i}}^2}{2}
\right)
+\frac{1}{2}
\sum\limits_{i,j=1}^d
\rho^{ij}\sigma^i\sigma^j
\partial^2_{y_i y_j}G(t,Y_t)
\right)\!
dt \nonumber
\\
&&+
\sum\limits_{i=1}^d
\sigma^i
\partial_{y_i} G(t,Y_t)
d\widehat W_t^i. \nonumber
\end{eqnarray*}

Because $\left(\beta_t\right)_{t\in\mathbb{R}_+}$ is a martingale under $(\mathbb{P},\mathcal{F}^S)$, we have
\begin{equation*}
 d\beta_t
=
\sum\limits_{i=1}^d
\sigma^i
\Sigma
\partial_{y_i} G(t,Y_t)
d\widehat W_t^i = \Sigma D_y G(t,Y_t) \left(\sigma \odot d\widehat W_t\right).
\end{equation*}
\end{proof}

The results obtained above (Theorems \ref{propbayes} and \ref{propbayesdyn}) will be useful in the next section on optimal portfolio choice. The process $(\beta_t)_{t \in \mathbb{R}_+}$ indeed represents the best estimate of the drift in the dynamics of the prices.

\subsection{A few remarks on the compact support case}

The results presented in the next sections of this paper are valid for sub-Gaussian prior distributions $m_\mu$ satisfying~\eqref{eq:lip}. A special class of such prior distributions is that of distributions with compact support.\\

We have indeed the following proposition:\\

\begin{prop}
\label{compact}
If $m_\mu$ has a compact support, then $G$ and all its derivatives are bounded over $\mathbb{R}_+\times\mathbb{R}^d$.
\end{prop}

\begin{proof}
Let us consider $i \in \{1, \ldots, d\}$. By definition, the $i^\textrm{th}$ coordinate of $G$ is $G^i = \frac{\partial_{y^i} F}{F}$. Therefore, by immediate induction,
$$\forall n \in \mathbb{N}, \forall j_1, \ldots, j_n \in \{1, \ldots, d\},  \forall n' \in \mathbb{N},\quad \partial^{n'}_{t\cdots t}\partial^n_{y^{j_1}\cdots y^{j_n}} G^i$$ is the sum and product of terms of the form $$\frac{\partial^{m'}_{t\cdots t}\partial^{m}_{y^{k_1}\cdots y^{k_m}} F}{F},\quad  \textrm{for }  m,m' \in \mathbb{N}, k_1, \ldots, k_m \in \{1, \ldots, d\}.$$
Now, for $(t,y) \in \mathbb{R}_+\times\mathbb{R}^d$, and for $m,m' \in \mathbb{N}, k_1, \ldots, k_m \in \{1, \ldots, d\}$,
\begin{eqnarray*}
&&\frac{\partial^{m'}_{t\cdots t}\partial^{m}_{y^{k_1}\cdots y^{k_m}} F(t,y)}{F(t,y)}\\
&=&\frac{
    \displaystyle \int\limits_{\mathbb{R}^d} \left((z-r\vec{1})'\Sigma^{-1}\left(\!
            -r\vec{1}
            +\frac12
            \sigma \odot \sigma\!
        \right)-\frac{1}{2}
\alpha(z)' \rho^{-1} \alpha(z)\right)^{m'}\! \prod_{p=1}^m (z-r\vec{1})'\Sigma^{-1}e_{k_p} f(t,y,z)
    m_\mu(dz)
    }
    {
    \displaystyle \int\limits_{\mathbb{R}^d} f(t,y,z)
    m_\mu(dz)
    },
\end{eqnarray*}
where $$f(t,y,z) = \exp\left(
(z-r\vec{1})'\Sigma^{-1}
    \left(
    y
    - Y_0+
        \left(
            -r\vec{1}
            +\frac12
            \sigma \odot \sigma
        \right)t
    \right)
-\frac{t}{2}
\alpha(z)' \rho^{-1} \alpha(z)
\right),$$
and where $(e_k)_{1\le k \le d}$ is the canonical basis of $\mathbb{R}^d$.\\

Therefore
\begin{eqnarray*}
&&\left|\frac{\partial^{m'}_{t\cdots t}\partial^{m}_{y^{k_1}\cdots y^{k_m}} F(t,y)}{F(t,y)}\right|\\
& \le& \sup_{z \in \textrm{support}(m_\mu)} \left| \left((z-r\vec{1})'\Sigma^{-1}\left(
            -r\vec{1}
            +\frac12
            \sigma \odot \sigma
        \right)-\frac{1}{2}
\alpha(z)' \rho^{-1} \alpha(z)\right)^{m'} \prod_{p=1}^m (z-r\vec{1})'\Sigma^{-1}e_{k_p}\right|\\
& <& +\infty,
\end{eqnarray*}
hence the result.
\end{proof}

In addition to showing that the Lipschitz hypothesis $(\ref{eq:lip})$ is true when $m_\mu$ has a compact support, Proposition \ref{compact} will be useful in Section 3 to provide a large class of priors for which there is no blowup phenomenon in the equations characterizing the optimal portfolio choice of an agent.\\

\section{Optimal portfolio choice}

In this section we proceed with the study of optimal portfolio choice. For that purpose, let us set an investment horizon $T\in \mathbb{R}_+^*$.\\

Let us also introduce the notion of ``linear growth'' for a process in our $d$-dimensional context. This notion plays an important part in the verification theorems.

\begin{defn}
Let us consider $t \in [0,T]$. An $\mathbb{R}^{d}$-valued, measurable, and $\mathcal{F}^{S}$-adapted process $\left(\mathcal{\zeta}_{s}\right)_{s\in[t,T]}$
is said to satisfy the linear growth condition with respect to $\xi = (\xi_s)_{s \in [t,T]}$ if,
\begin{eqnarray*}
\exists C_T > 0, \forall s\in [t,T], \qquad \norm{\mathcal{\zeta}_{s}} & \leq & C_{T}\left(1+\sup_{\tau\in\left[t,s\right]}\norm{\xi_{\tau}}\right).
\end{eqnarray*}
\end{defn}

The first subsection is devoted to the CARA case, and the second one focuses on the CRRA case.

\subsection{CARA case}

We consider the portfolio choice of the agent in the CARA case. We denote by $\gamma > 0$ his absolute risk aversion parameter.\\

We define, for $t \in [0,T]$ the set
\begin{eqnarray*}
\mathcal{A}_{t} & = & \left\{ \left(M_{s}\right)_{s\in\left[t,T\right]}, \mathbb{R}^{d}\textrm{-valued\;} \mathcal{F}^{S}\textrm{-adapted process}\right.\\
&& \left.\textrm{satisfying the linear growth condition with respect to } (Y_s)_{s \in [t,T]}\right\}.
\end{eqnarray*}

We denote by $\left(M_{t}\right)_{t\in[0,T]} \in \mathcal{A} = \mathcal{A}_0$ the $\mathbb{R}^d$-valued process modelling the strategy of the agent. More precisely, $\forall i \in\left\{1,\hdots,d\right\}$, $M^i_t$ represents the amount invested in the
$i^\text{th}$ asset at time $t$. The resulting value of the agent's portfolio is modelled by a process $(V_t)_{t\in[0,T]}$ with $V_0 > 0$. The dynamics of $\left(V_t\right)_{t \in [0,T]}$ is given by the following stochastic differential equation (SDE):
\begin{eqnarray}
\label{dV1}
dV_{t} & = & \left(M_{t}'\left(\mu-r\vec{1}\right)+rV_{t}\right)dt+M_{t}'\left( \sigma \odot dW_{t}\right).
\end{eqnarray}
With the notations introduced in Section 2, we have
\begin{eqnarray}
\nonumber dV_t & = & \left(M_{t}'\left(\beta_{t}-r\vec{1}\right)+rV_{t}\right)dt+M_{t}'\left( \sigma \odot d\widehat{W}_{t}\right) \\
\label{dV2}  & = & \left(M_{t}'
 \Sigma
  G(t,Y_t)
  +rV_{t}\right)dt+M_{t}'\left( \sigma \odot d\widehat{W}_{t}\right),\nonumber
\end{eqnarray}
and
\begin{eqnarray*}
  dY_t &=&
  \left(
  r\vec{1}+\Sigma G\left(t,Y_t\right) - \frac12
  \sigma \odot \sigma
  \right) dt
+\sigma \odot d\widehat W_t.
\end{eqnarray*}

Given $M\in\mathcal{A}_{t}$ and $s\geq t$, we define therefore
\begin{flalign}
Y_s^{t,y}&=y+\int_t^s
\left(
  r\vec{1}+\Sigma G(\tau,Y^{t,y}_{\tau}) - \frac12
  \sigma \odot \sigma
  \right) d\tau
+\sigma \odot
(\widehat W_s-\widehat W_t),
&\\
V_{s}^{t,V,y,M} & =  V+\int_{t}^{s}\left(M_{\tau}'\Sigma
G(\tau,Y^{t,y}_{\tau})
+rV_{\tau}^{t,V,y,M}\right)d\tau+\int_{t}^{s}M_{\tau}'(\sigma \odot d\widehat{W}_{\tau}).
\end{flalign}

For an arbitrary initial state $(V_{0},y_0)$, the agent maximizes, over $M$ in the set of admissible strategies $\mathcal{A}$, the expected utility of his portfolio value at time $T$, i.e.,
\[
\mathbb{E}\left[-\exp\left(-\gamma V_{T}^{0,V_{0},y_0,M}\right)\right].
\]

The value function $v$ associated with this problem is then defined
by
\begin{eqnarray}
\label{eq:value}
v:\left(t,V,y\right) \in[0,T]\times\mathbb{R}\times\mathbb{R}^d & \mapsto &   \sup_{(M_s)_{s \in [t,T]}\in\mathcal{A}_{t}}\mathbb{E}\left[-\exp\left(-\gamma V_{T}^{t,V,y,M}\right)\right].
\end{eqnarray}

The HJB equation associated with this problem is
\begin{align}
        &\partial_t u
        +rV\partial_V u
        +
	\left(
           \nabla_y u
	\right)'
	\left(
	r\vec{1}
	+\Sigma
	G
	- \frac{1}{2}
	\sigma\odot \sigma \right)
        +\frac12
	\mathrm{Tr}
	\left(\Sigma
	 \nabla^2_{yy}u
	\right)\nonumber\\
 	&+\sup\limits_{M\in\mathbb{R}^d}
	\left\{
        \partial_V u M' \Sigma G
        +\frac12 M'\Sigma M
	\partial_{VV}^2u
	+M'\Sigma\partial_V\nabla_yu
        \right\}=0,
\label{eq:HJB_cara_u}
\end{align}
with terminal condition
\begin{align}
\forall (V,y)\in\mathbb{R}\times\mathbb{R}^d,\quad        u(T,V,y)&=-\exp(-\gamma V)
    \label{eq:terminal_CARA_u}
. \end{align}

To solve the HJB equation, we use the following ansatz:
\begin{eqnarray}
u\left(t,V,y\right) & = & -\exp\left[-\gamma\left(e^{r\left(T-t\right)}V+\phi\left(t,y\right)\right)\right].\label{eq:ansatz_CARA}
\end{eqnarray}

\begin{prop}
\label{prop:u-phi_CARA}Suppose there exists $\phi\in C^{1,2}\left(\left[0,T\right]\times\mathbb{R}^d\right)$
satisfying
\begin{align}
\partial_t\phi
+\left(\nabla_y\phi\right)'
	\left(
	r\vec{1}
	- \frac{1}{2}
	\sigma\odot \sigma
	\right)
+\frac12
	\mathrm{Tr}
	\left(\Sigma
	\nabla^2_ {yy} \phi
	\right)
+
\frac{1}{2\gamma}
G'\Sigma G  =  0,\label{eq:HJB_cara_phi}
\end{align}

with terminal condition
\begin{eqnarray}
\forall y\in\mathbb{R}^d,\quad
\phi\left(T,y\right) & = & 0.\label{eq:terminal_CARA_phi}
\end{eqnarray}
Then $u$ defined by (\ref{eq:ansatz_CARA}) is solution of the HJB
equation (\ref{eq:HJB_cara_u}) with terminal condition (\ref{eq:terminal_CARA_u}).

Moreover, the supremum in (\ref{eq:HJB_cara_u}) is achieved at:
\begin{eqnarray}
M^{\star} (t,y)& = & e^{-r(T-t)}\left(
    \frac{G(t,y)}{\gamma}-\nabla_y\phi (t,y)
    \right).\label{eq:optimizer_CARA1}
\end{eqnarray}
\end{prop}

\begin{proof}
Let us consider $\phi \in C^{1,2}([0,T]\times\mathbb{R}^d)$ solution of \eqref{eq:HJB_cara_phi} with terminal condition \eqref{eq:terminal_CARA_phi}. For $u$ defined by \eqref{eq:ansatz_CARA} and by considering $\widetilde{M}=Me^{r(T-t)}$, we have
\begin{eqnarray*}
 &&\partial_t u +
 rV \partial_V u +
        	\left(
	\nabla_y u
	\right)'
	\left(
	r\vec{1}
	+\Sigma
	G
	- \frac{1}{2}
	\sigma\odot \sigma\
	\right)
        +\frac12
	\mathrm{Tr}
	\left(\Sigma
	 \nabla^2_{yy}u
	\right)\nonumber\\
 	&&
        +\sup\limits_{M\in\mathbb{R}^d} \left\{
        \partial_V u M' \Sigma G
        +\frac12 M'\Sigma M
	\partial_{VV}^2u
	+M'\Sigma\partial_V\nabla_yu
        \right\}
\nonumber\\
&=&
-\gamma u \left(-rV
e^{r(T-t)}+\partial_t \phi
\right)
-\gamma u e^{r(T-t)}rV
-\gamma u
	\left(
	\nabla_y\phi
	\right)'
	\left(
	r\vec{1}
	+\Sigma
	G
	- \frac{1}{2}
	\sigma\odot \sigma\
	\right)
\\
&&-\frac{\gamma u}{2}
	\left(
	-\gamma
	\mathrm{Tr}
	\left(
	\Sigma
	\nabla_y \phi
	( \nabla_y \phi)'
	\right)
	+
	\mathrm{Tr}
	\left(
	\Sigma
	\nabla^2_ {yy} \phi
	\right)
	\right)
-\gamma u
\sup\limits_{\widetilde{M} \in\mathbb{R}^d}\left\{
\widetilde{M}
'\Sigma G
-
\frac{\gamma}{2}\widetilde{M}'\Sigma
\widetilde{M}
-\gamma  \widetilde{M}'\Sigma
\nabla_y\phi
\right\}.
\end{eqnarray*}

The supremum in the above expression is reached at
\begin{align*}
\widetilde{M}^\star=
\frac{G}{\gamma}
- \nabla_y \phi ,
\end{align*}
corresponding to
\begin{eqnarray}
M^{\star}(t,y)=
e^{-r(T-t)}
\left(
\frac{G(t,y)}{\gamma}
- \nabla_y \phi(t,y)
\right).
\end{eqnarray}
Plugging this expression in the partial differential equation, we get:
\begin{eqnarray*}
 &&\partial_t u +
 rV \partial_V u +
        \left(
	\nabla_y u
	\right)'
	\left(
	r\vec{1}
	+\Sigma
	G
	- \frac{1}{2}
	\sigma\odot \sigma\
	\right)
        +\frac12
	\mathrm{Tr}
	\left(\Sigma
	 \nabla^2_{yy}u
	\right)\nonumber\\
 	&&
        +\sup\limits_{M\in\mathbb{R}^d}
	\left\{
        \partial_V u M' \Sigma G
        +\frac12 M'\Sigma M
	\partial_{VV}^2u
	+M'\Sigma\partial_V\nabla_yu
        \right\}
\nonumber\\
&=&
-\gamma u
\bigg[
-rV e^{r(T-t)}\!+\partial_t\phi
+e^{r(T-t)}rV
+\left(\nabla_y\phi
\right)'
	\left(
	r\vec{1}\!
	+\Sigma
	G
	- \frac{1}{2}
	\sigma\odot \sigma\!
	\right)\! \\
&&
	-\frac{\gamma}{2}
	\left(\nabla_y\phi\right)'
	\Sigma
	\nabla_y\phi
	+
	\frac{1}{2}
	\mathrm{Tr}
	\left(\Sigma
	\nabla^2_ {yy} \phi
	\right)\!
+\frac{\gamma}{2}
\left[
\frac{G}{\gamma}-  \nabla_y\phi
\right]'
\Sigma
\left[
\frac{G}{\gamma}-  \nabla_y\phi
\right]
\bigg]
 \\
&=&
-\gamma u
\left[
\partial_t\phi
+\left(\nabla_y\phi\right)'
	\left(
	r\vec{1}
	+\Sigma
	G
	- \frac{1}{2}
	\sigma\odot \sigma
	-\Sigma G
	\right)
+\frac12
	\mathrm{Tr}
	\left(\Sigma
	\nabla^2_{yy} \phi
	\right)
+
\frac{1}{2\gamma}
G'\Sigma G
\right]\\
&=&
0.
\end{eqnarray*}

As it is straightforward to verify that $u$ satisfies the terminal condition
(\ref{eq:terminal_CARA_u}), the result is proved.
\end{proof}

From the previous proposition, we see that solving the HJB equation \eqref{eq:HJB_cara_u} with terminal condition \eqref{eq:terminal_CARA_u} boils down to solving \eqref{eq:HJB_cara_phi} with terminal condition \eqref{eq:terminal_CARA_phi}. Because \eqref{eq:HJB_cara_phi} is a simple parabolic PDE, we can easily build a strong solution.

\begin{prop}
Let us define
\begin{equation}
\label{eq:FK-cara}
\phi : (t,y)\in[0,T]\times\mathbb{R}^d \mapsto \mathbb{E}^\mathbb{Q}\left[
    \int_t^T \frac{1}{2\gamma}G(s, Y^{t,y}_s)'\Sigma G(s, Y^{t,y}_s)ds\right],
 \end{equation}
where $\forall (t,y)\in[0,T]\times\mathbb{R}^d$, $\forall s \in [t,T]$,
\begin{align*}
Y^{t,y}_s= y +
	\left(
	r\vec{1}
	- \frac{1}{2}
	\sigma\odot \sigma
	\right)(s-t)+\sigma\odot \left(W_s^\mathbb{Q} - W_t^\mathbb{Q}\right).
\end{align*}
Then $\phi$ is a $C^{1,2}([0,T]\times\mathbb{R}^d)$ function, solution of \eqref{eq:HJB_cara_phi} with terminal condition \eqref{eq:terminal_CARA_phi}.\\
Furthermore,
\begin{equation}
\label{eq:gradphi}
\exists A_T > 0, \forall t \in [0,T], \forall y \in \mathbb{R}^d, \forall i \in \{1, \ldots, d\}, \norm{\nabla_y \phi(t,y)} \le A_T (1 + \norm{y}).
\end{equation}
\end{prop}

\begin{proof}
Because of the assumption \eqref{eq:lip} on $G$, the first part of the proposition is a consequence of classical results for parabolic PDEs and of the classical Feynman-Kac representation (see for instance \cite{af,ks}).\\

For the second part, we notice first that
$$\forall (t,y) \in [0,T]\times\mathbb{R}^d, \nabla_y \phi(t,y) = \mathbb{E}^\mathbb{Q}\left[\int_t^T \frac{1}{\gamma}D_YG(s, Y^{t,y}_s)\Sigma G(s, Y^{t,y}_s)ds\right].$$
Therefore, by \eqref{eq:lip}, there exists a constant $C \ge 0$ such that
\begin{equation}
\label{eq:g1}\forall (t,y) \in [0,T]\times\mathbb{R}^d, \norm{\nabla_y \phi(t,y)} \le C \sup_{s \in [t,T]} \mathbb{E}^{\mathbb{Q}}\left[\norm{G(s, Y^{t,y}_s)} \right].
\end{equation}
By \eqref{eq:lip} again, there exists a constant $C' \ge 0$ such that
\begin{eqnarray}
\nonumber \norm{G(s,Y^{t,y}_s)} &\le& \norm{G(s,Y^{t,y}_s) - G(t,y)} + \norm{G(t,y)}\\
&\le & \norm{G(s,Y^{t,y}_s) - G(t,y)} + C' (1+\norm{y})\label{eq:g2}
\end{eqnarray}
Now, by Theorem \ref{propbayesdyn}, $\forall s \in [t,T]$,
$$G(s,Y^{t,y}_s) - G(t,y) = \int_t^s D_YG(\tau, Y^{t,y}_\tau) (\sigma \odot d\widehat{W}_\tau).$$
Therefore,
$$\mathbb{E}^{\mathbb{Q}}\left[\norm{G(s, Y^{t,y}_s)- G(t,y)} \right] = \mathbb{E}\left[\norm{\int_t^s D_YG(\tau, Y^{t,y}_\tau) (\sigma \odot d\widehat{W}_\tau)} \frac{d\mathbb{Q}}{d\mathbb{P}} \right].$$
Now, for $p \ge 1$, we have
\begin{eqnarray*}
\mathbb{E}\left[\left(\frac{d\mathbb{Q}}{d\mathbb{P}}\right)^p \right] &=& \mathbb{E}\left[\exp\left(-p\alpha(\mu)' \rho^{-1}W_T-\frac p2\alpha(\mu)' \rho^{-1} \alpha(\mu)T\right)\right]\\
&=&\mathbb{E}\left[\exp\left(\frac {p(p-1)}2\alpha(\mu)' \rho^{-1} \alpha(\mu)T\right)\right].
\end{eqnarray*}
Because $m_\mu$ is sub-Gaussian, there exists $p > 1$ such that $\frac{d\mathbb{Q}}{d\mathbb{P}} \in L^p(\Omega,\mathbb{P})$. Because of the Lipschitz assumption on $G$, we have for any $q>1$, and in particular for $q$ such that $\frac 1p + \frac 1q = 1$, that
$$\sup_{(t,y) \in [0,T] \times \mathbb{R}^d} \sup_{s \in [t,T]} \mathbb{E}\left[\norm{\int_t^s D_YG(\tau, Y^{t,y}_\tau) (\sigma \odot d\widehat{W}_\tau)}^q \right] < +\infty.$$
Therefore,
$$\sup_{(t,y) \in [0,T] \times \mathbb{R}^d} \sup_{s \in [t,T]} \mathbb{E}\left[\norm{\int_t^s D_YG(\tau, Y^{t,y}_\tau) (\sigma \odot d\widehat{W}_\tau)} \frac{d\mathbb{Q}}{d\mathbb{P}} \right] < + \infty.$$
We can conclude that $\mathbb{E}^{\mathbb{Q}}[\norm{G(s,Y^{t,y}_s) - G(t,y)}]$ is bounded uniformly, and therefore using Eqs.~\eqref{eq:g1} and \eqref{eq:g2} that $\norm{\nabla_y \phi}$ is indeed at most linear in $y$ uniformly in $t \in [0,T]$.
\end{proof}

Using the above results, we know that there exists a $C^{1,2,2}([0,T]\times\mathbb{R}\times\mathbb{R}^d)$ function $u$ solution of the HJB equation (\ref{eq:HJB_cara_u}) with terminal condition (\ref{eq:terminal_CARA_u}). By using a verification argument, we can show that $u$ is in fact the value function $v$ defined in Eq. (\ref{eq:value}) and then solve the problem faced by the agent. This is the purpose of the following theorem.

\begin{thm}
\label{verif_cara_gl}
Let us consider the $C^{1,2}([0,T]\times\mathbb R^d)$ function $\phi$ defined by \eqref{eq:FK-cara}. Let us then consider the associated function $u$ defined by (\ref{eq:ansatz_CARA}).\\

For all $\left(t,V,y\right)\in\left[0,T\right]\times\mathbb{R}\times\mathbb{R}^{d}$
and $M = (M_s)_{s \in [t,T]}\in\mathcal{A}_t$, we have
\begin{eqnarray}
\mathbb{E}\left[-\exp\left(-\gamma V_{T}^{t,V,y,M}\right)\right] & \leq & u\left(t,V,y\right).\label{eq:verif_ineq_cara_gl}
\end{eqnarray}
Moreover, equality in (\ref{eq:verif_ineq_cara_gl}) is obtained by taking the
optimal control $(M^{\star}_s)_{s \in [t,T]}\in\mathcal{A}_t$ given by \eqref{eq:optimizer_CARA1}, i.e.,
$$\forall s \in [t,T], M^{\star}_s = e^{-r(T-s)}
\left(
\frac{G(s,Y^{t,y}_s)}{\gamma}
- \nabla_y \phi(s,Y^{t,y}_s)
\right).$$
In particular $u=v$.
\end{thm}

\begin{proof}

From the Lipschitz property of $G$ stated in Eq. (\ref{eq:lip}) and the property of $\phi$ stated in Eq. (\ref{eq:gradphi}), we see that $(M_s^\star)_{s \in [t,T]}$ is indeed admissible (i.e., $(M_s^\star)_{s \in [t,T]} \in \mathcal{A}_t$).\\

Let us then consider $\left(t,V,y\right)\in [0,T]\times\mathbb{R}\times\mathbb{R}^{d}$ and $M = (M_s)_{s \in [t,T]}\in\mathcal{A}_t$.\\

By It\=o's formula, we have for all $s\in[t,T]$
\begin{eqnarray*}
&&du\left(s,V^{t,V,y,M}_s,Y^{t,y}_s\right)\\
\!\!&=&\!\!
\partial_t u\left(s,V^{t,V,y,M}_s,Y^{t,y}_s\right) ds+
\partial_V u\left(s,V^{t,V,y,M}_s,Y^{t,y}_s\right) dV^{t,V,y,M}_s+
\nabla_y u\left(s,V^{t,V,y,M}_s,Y^{t,y}_s\right)' dY^{t,y}_s
\\&&+\frac12\partial^2_{VV} u\left(s,V^{t,V,y,M}_s,Y^{t,y}_s\right) M_s'\Sigma M_s ds
+\frac12
\mathrm{Tr}
\left(\Sigma
\nabla^2_{yy} u\left(s,V^{t,V,y,M}_s,Y^{t,y}_s\right)
\right)ds
\\&&+M_s' \Sigma \partial_{V} \nabla_y u\left(s,V^{t,V,y,M}_s,Y^{t,y}_s\right) ds
\\
\!\!&=&\!\!
\mathcal{L}^Mu\left(s,V^{t,V,y,M}_s,Y^{t,y}_s\right)ds\\
&&
+\left(\partial_V u\left(s,V^{t,V,y,M}_s,Y^{t,y}_s\right)M_s
+\nabla_y u\left(s,V^{t,V,y,M}_s,Y^{t,y}_s\right)\right)'
\left(\sigma\odot d\widehat W_s\right),
\end{eqnarray*}
where
\begin{eqnarray*}
&&\mathcal{L}^Mu(s,V^{t,V,y,M}_s,Y^{t,y}_s)\\
&=& \partial_t u\left(s,V^{t,V,y,M}_s,Y^{t,y}_s\right) +
\partial_V u\left(s,V^{t,V,y,M}_s,Y^{t,y}_s\right) (rV_{s}^{t,V,y,M} + M_{s}'\Sigma
G(s,Y^{t,y}_s)
)\\
&&+
\nabla_y u\left(s,V^{t,V,y,M}_s,Y^{t,y}_s\right)' \left(r\vec{1}+\Sigma G(s,Y^{t,y}_{s}) - \frac12
  \sigma \odot \sigma\right)
\\&&+\frac12\partial^2_{VV} u\left(s,V^{t,V,y,M}_s,Y^{t,y}_s\right) M_s'\Sigma M_s
+\frac12
\mathrm{Tr}
\left(
\Sigma \nabla^2_{yy} u\left(s,V^{t,V,y,M}_s,Y^{t,y}_s\right)
\right)\\
&&+M_s' \Sigma \partial_{V} \nabla_y u\left(s,V^{t,V,y,M}_s,Y^{t,y}_s\right)'
\end{eqnarray*}
Note that we have
$$
\partial_V u\left(s,V^{t,V,y,M}_s,Y^{t,y}_s\right)M_s
+\nabla_y u\left(s,V^{t,V,y,M}_s,Y^{t,y}_s\right)
$$$$=
-\gamma u\left(s,V^{t,V,y,M}_s,Y^{t,y}_s\right)
\left( e^{r(T-s)} M_s
+\nabla_y\phi(s,Y^{t,y}_s)\right).
$$
Let us subsequently define, for all $s\in [t,T]$,
\begin{align*}
\kappa^M_s&=
-\gamma
\left(
 e^{r(T-s)} M_s
+\nabla_y\phi(s,Y^{t,y}_s)
\right),
\end{align*}
and
\begin{align*}
\xi^M_{t,s}&=\exp
\left(
\int_t^s{\kappa^M_{\tau}}'
\left(\sigma \odot d\widehat W_\tau\right)
-\frac12
\int_t^s {\kappa^M_{\tau}}'\Sigma\kappa^M_{\tau}d\tau
\right).
\end{align*}

We have
$$d\xi^M_{t,s} = \xi^M_{t,s} {\kappa^M_s}' \left(\sigma \odot d\widehat W_s\right)$$
and
$$d\left(\xi^M_{t,s}\right)^{-1} = - \left(\xi^M_{t,s}\right)^{-1} \kappa^M_s \left(\sigma \odot d\widehat W_s\right) + \left(\xi^M_{t,s}\right)^{-1} {\kappa^M_{s}}'\Sigma\kappa^M_{s}ds.$$
Therefore
\begin{eqnarray*}
&&d\left(
u\left(s,V^{t,V,y,M}_s,Y^{t,y}_s\right)\left(\xi^M_{t,s}\right)^{-1}
\right)\\
&=&
u\left(s,V^{t,V,y,M}_s,Y^{t,y}_s\right)\left(- \left(\xi^M_{t,s}\right)^{-1} {\kappa^M_s}' \left(\sigma \odot d\widehat W_s\right) + \left(\xi^M_{t,s}\right)^{-1} {\kappa^M_{s}}'\Sigma\kappa^M_{s}ds\right)\\
&&+\left(\xi^M_{t,s}\right)^{-1}\left(\mathcal{L}^Mu\left(s,V^{t,V,y,M}_s,Y^{t,y}_s\right)ds
+ u\left(s,V^{t,V,y,M}_s,Y^{t,y}_s\right) {\kappa^M_s}' \left(\sigma \odot d\widehat W_s\right)\right)\\
&&- \left(\xi^M_{t,s}\right)^{-1} u\left(s,V^{t,V,y,M}_s,Y^{t,y}_s\right) {\kappa^M_s}' \Sigma \kappa^M_s ds \\
&=& \left(\xi^M_{t,s}\right)^{-1}
\mathcal{L}^Mu\left(s,V^{t,V,y,M}_s,Y^{t,y}_s\right)ds.
\end{eqnarray*}
By definition of $u$, $\mathcal{L}^Mu\left(s,V^{t,V,y,M}_s,Y^{t,y}_s\right)\leq 0$ and  $\mathcal{L}^Mu\left(s,V^{t,V,y,M}_s,Y^{t,y}_s\right)=0$ if $M_s=M^{\star}_s$. As a consequence,
$\left(
u\left(s,V^{t,V,y,M}_s,Y^{t,y}_s\right)\left(\xi^M_{t,s}\right)^{-1}
\right)_{s\in[t,T]}$ is nonincreasing, and therefore
$$u(T,V^{t,V,y,M}_T,Y^{t,y}_T) \le u(t,V,y)\xi^M_{t,T},$$
with equality when $(M_s)_{s \in [t,T]} = (M_s^\star)_{s \in [t,T]}$.\\

Subsequently,
$$\mathbb{E}\left[-\exp\left(-\gamma V_{T}^{t,V,y,M}\right)\right] = \mathbb{E}\left[u(T,V^{t,V,y,M}_T,Y^{t,y}_T)\right]\leq u(t,V,y)\mathbb{E}\left[\xi^M_{t,T}\right],
$$
with equality when $(M_s^\star)_{s \in [t,T]} = (M_s)_{s \in [t,T]}$.\\

To conclude the proof let us show that $\mathbb{E}\left[\xi^M_{t,T}\right] = 1$. To do so, we will use the fact that $\xi^M_{t,t}=1$ and prove that $\left(\xi^M_{t,s}\right)_{s\in[t,T]}$ is a martingale under $\left(\mathbb{P},\left(\mathcal{F}_s^{S}\right)_{s \in [t,T]}\right)$.\\

Because $M \in \mathcal{A}_t$, and because of Eq. (\ref{eq:gradphi}), we know that there exists a constant $C$ such that
$$\sup_{s \in [t,T]} \norm{\kappa^M_s}^2 \le C\left(1+ \sup_{s \in [t,T]} \norm{Y^{t,y}_s}^2\right).$$
By definition of $(Y^{t,y}_s)_{s \in [t,T]}$, there exists therefore a constant $C'$ such that
$$\sup_{s \in [t,T]} \norm{\kappa^M_s}^2 \le C'\left(1+ \norm{\mu}^2 + \sup_{s \in [t,T]} \norm{W_s-W_t}^2\right).$$

Now, by using the above inequality along with Eq. (\ref{eq:priorcondition}) and classical properties of the Brownian motion, we easily prove that
\begin{equation}
\label{eq:sub}
\exists \epsilon > 0, \forall s \in [t,T], \quad \mathbb{E}\left[ \exp\left(\frac 12 \int_s^{(s+\epsilon)\wedge T} {\kappa^M_s}'\Sigma \kappa^M_s ds\right)\right] < +\infty.
\end{equation}

From Novikov's condition (or more exactly one of its corollary -- see for instance Corollary 5.14 in \cite{ks}), we see that $\left(\xi^M_{t,s}\right)_{s\in[t,T]}$ is a martingale under $\left(\mathbb{P},\left(\mathcal{F}_s^{S}\right)_{s \in [t,T]}\right)$, hence the result
$$\mathbb{E}\left[-\exp\left(-\gamma V_{T}^{t,V,y,M}\right)\right] \leq u(t,V,y),
$$
with equality when $(M_s^\star)_{s \in [t,T]} = (M_s)_{s \in [t,T]}$. Therefore,
$$u(t,V,y) = v(t,V,y) =\!\!\! \sup_{(M_s)_{s \in [t,T]} \in \mathcal{A}_t}\!\!\! \mathbb{E}\left[-\exp\left(-\gamma V_{T}^{t,V,y,M}\right)\right] = \mathbb{E}\left[-\exp\left(-\gamma V_{T}^{t,V,y,M^\star}\right)\right].$$
\end{proof}

The optimal portfolio choice of an agent with a CARA utility function is therefore fully characterized. Let us now turn to the case of an agent with a CRRA utility function.

\subsection{CRRA case}

We consider the portfolio choice of the agent in the CRRA case. We denote by $\gamma > 0$ the relative risk aversion parameter.\\

We denote by $U^{\gamma}$ the utility function of the agent, i.e.,
$$U^{\gamma} : V \in \mathbb{R}_+^*  \mapsto \begin{cases}
 \dfrac{V^{1-\gamma}}{1-\gamma} & \text{if }\gamma \neq 1\\
 \log\left(V\right) & \text{if }\gamma=1.\\
\end{cases}
$$

If $\gamma < 1$, we define for $t \in [0,T]$ the set
\begin{eqnarray*}
\mathcal{A}^\gamma_{t} & = & \left\{ \left(\theta_{s}\right)_{s\in\left[t,T\right]}, \mathbb{R}^{d}\textrm{-valued\;} \mathcal{F}^{S}\textrm{-adapted process}, \mathbb{E}\left[\int_t^T \theta^2_{s} ds\right] < +\infty \right\}.\\
\end{eqnarray*}

If $\gamma \ge 1$, we define for $t \in [0,T]$ the set
\begin{eqnarray*}
\mathcal{A}^\gamma_{t} & = & \left\{ \left(\theta_{s}\right)_{s\in\left[t,T\right]}, \mathbb{R}^{d}\textrm{-valued\;} \mathcal{F}^{S}\textrm{-adapted process}\right.\\
&& \left.\textrm{satisfying the linear growth condition with respect to } (Y_s)_{s \in [t,T]}\right\}.
\end{eqnarray*}

We denote by $\left(\theta_{t}\right)_{t\in[0,T]} \in \mathcal{A}^\gamma = \mathcal{A}^\gamma_0$ the $\mathbb{R}^d$-valued process modelling the strategy of the agent. More precisely, $\forall i \in\left\{1,\hdots,d\right\}$, $\theta^i_t$ represents the part of the wealth invested in the
$i^\text{th}$ risky asset at time $t$. The resulting value of the agent's portfolio is modelled by a process $(V_t)_{t\in[0,T]}$ with $V_0 > 0$. The dynamics of $\left(V_t\right)_{t \in [0,T]}$ is given by the following stochastic differential equation (SDE):
\begin{eqnarray}
\label{dV1_crra}
dV_{t} & = & \left(\theta_{t}'\left(\mu-r\vec{1}\right)+r\right)V_{t}dt+V_t\theta_{t}'\left( \sigma \odot dW_{t}\right).
\end{eqnarray}
With the notations introduced in Section 2, we have
\begin{eqnarray}
\nonumber dV_t & = & \left(\theta_{t}'\left(\beta_t-r\vec{1}\right)+r\right)V_{t}dt+V_t\theta_{t}'\left( \sigma \odot d\widehat{W}_{t}\right) \\
\label{dV2_crra}  & = & \left(\theta_{t}'
 \Sigma
  G(t,Y_t)
  +r\right)V_{t}dt+V_t\theta_{t}'\left( \sigma \odot d\widehat{W}_{t}\right),\nonumber
\end{eqnarray}
and
\begin{eqnarray*}
  dY_t &=&
  \left(
  r\vec{1}+\Sigma G\left(t,Y_t\right) - \frac12
  \sigma \odot \sigma
  \right) dt
+\sigma \odot d\widehat W_t.
\end{eqnarray*}

Given $\theta\in\mathcal{A}^\gamma_{t}$ and $s\geq t$, we define
\begin{flalign}
Y_s^{t,y}&=y+\int_t^s
\left(
  r\vec{1}+\Sigma G(\tau,Y^{t,y}_{\tau}) - \frac12
  \sigma \odot \sigma
  \right) d\tau
+\sigma \odot
(\widehat W_s-\widehat W_t),
&\\
V_{s}^{t,V,y,\theta} & =  V+\int_{t}^{s}\left(\theta_{\tau}'\Sigma
G(\tau,Y^{t,y}_{\tau})
+r\right)V_{\tau}^{t,V,y,\theta}d\tau+\int_{t}^{s}V_{\tau}^{t,V,y,\theta} \theta_{\tau}'(\sigma \odot d\widehat{W}_{\tau}).
\end{flalign}

For an arbitrary initial state $(V_{0},y_0)$, the agent maximizes, over $\theta$ in the set of admissible strategies $\mathcal{A}^\gamma$, the expected utility of his portfolio value at time $T$, i.e.,
\[
\mathbb{E}\left[U^\gamma\left(V_{T}^{0,V_{0},y_0,\theta}\right)\right].
\]

The value function $v$ associated with this problem is then defined
by
\begin{eqnarray}
\label{eq:value_crra}
v:\left(t,V,y\right) \in[0,T]\times\mathbb{R}_+^*\times\mathbb{R}^d & \mapsto &   \sup_{(\theta_s)_{s \in [t,T]}\in\mathcal{A}^\gamma_{t}}\mathbb{E}\left[U^\gamma\left(V_{T}^{t,V,y,\theta}\right)\right].
\end{eqnarray}

The HJB equation associated with this problem is given by
\begin{eqnarray}
\partial_t u +
        (\nabla_y u)'
	\left(r\vec{1}+\Sigma G - \frac{1}{2}\sigma \odot\sigma
	\right)
        +\frac12
	\mathrm{Tr}\left(\Sigma\nabla^2_{yy}u
	\right)
        +
rV\partial_V u&&
\nonumber\\
+\sup\limits_{\theta\in \mathbb{R}^d} \left\{
V\partial_V u
\theta'
\Sigma G
+\frac{V^2}{2}
\theta'\Sigma\theta
\partial_{VV}^2u
+V\theta'\Sigma\partial_V\nabla_yu
        \right\}&=&0\label{eq:HJB_crra_u}
, \end{eqnarray}
with terminal condition
\begin{eqnarray}
\forall V\in\mathbb{R}^*_+,\forall y\in \mathbb{R}^d,\quad
u(T,V,y)&=&U^{\gamma}(V).
\label{eq:terminal_CRRA_u}
\end{eqnarray}

To solve the HJB equation and then solve the optimal portfolio choice problem, we need to consider separately the cases $\gamma=1$ and $\gamma \neq 1$.

\subsubsection{The $\gamma \neq 1$ case}

To solve the HJB equation when $\gamma \neq 1$, we use the following ansatz:
\begin{eqnarray}
u\left(t,V,y\right) & = &
U^\gamma\left(
e^{r(T-t)}V
\right)\phi(t,y)^\gamma.
\label{eq:ansatz_CRRA}
\end{eqnarray}

\begin{prop}
\label{prop:u_phi_crra}Suppose there exists a positive function $\phi\in C^{1,2}\left(\left[0,T\right]\times\mathbb{R}^d\right)$
satisfying
\begin{eqnarray}
\partial_t \phi
+
        (\nabla_y \phi)'
	\!
	\left(r\vec{1}+\frac{1}{\gamma}\Sigma G - \frac{1}{2}\sigma \odot\sigma
	\!
	\right)
        \!+
\frac{1}{2}
	\mathrm{Tr}\left(
\Sigma
\nabla^2_{yy}\phi
	\right)
+\phi\frac{(1-\gamma)}{2\gamma^2}
G'\Sigma G
   &=& 0,\label{eq:HJB_crra_phi}
\end{eqnarray}
with terminal condition\textup{
\begin{eqnarray}
\forall y\in\mathbb{R}^d,\quad \phi\left(T,y\right) & = & 1.\label{eq:terminal_crra_phi}
\end{eqnarray}
}Then $u$ defined by (\ref{eq:ansatz_CRRA}) is solution of the HJB
equation (\ref{eq:HJB_crra_u}) with terminal condition (\ref{eq:terminal_CRRA_u}).\\

Moreover, the supremum in (\ref{eq:HJB_crra_u}) is achieved at
\begin{eqnarray}
\theta^{\star}(t,y) & = &
\frac{G(t,y)}{\gamma}  +
 \frac{\nabla_y\phi(t,y)}{\phi(t,y)}.\label{eq:optimizer_crra1}
\end{eqnarray}
\end{prop}

\begin{proof}
Let us consider $\phi\in C^{1,2}\left(\left[0,T\right]\times\mathbb{R}^d\right)$, positive solution of \eqref{eq:HJB_crra_phi} with terminal condition \eqref{eq:terminal_crra_phi}. For $u$ defined by \eqref{eq:ansatz_CRRA}, we have:
\begin{eqnarray*}
&&\partial_t u +
        (\nabla_y u)'
	\left(r\vec{1}+\Sigma G - \frac{1}{2}\sigma \odot\sigma
	\right)
        +\frac12
	\mathrm{Tr}\left(\Sigma\nabla^2_{yy}u
	\right)
        +r V\partial_V u
\\
&&+\sup\limits_{\theta\in \mathbb{R}^d} \left\{
V\partial_V u
\theta'
\Sigma G
+\frac{V^2}{2}
\theta'\Sigma\theta
\partial_{VV}^2u
+V\theta'\Sigma\partial_V\nabla_yu
        \right\}
\nonumber\\
&=&
\frac{\gamma u}{\phi}\partial_t \phi
+
        \frac{\gamma u(\nabla_y \phi)' }{\phi}
	\left(r\vec{1}+\Sigma G - \frac{1}{2}\sigma \odot\sigma
	\right)
+\frac{\gamma u}{2 \phi}
	\mathrm{Tr}\left(
\Sigma
\nabla^2_{yy}\phi
\right)
+
\frac{\gamma (\gamma -1)u}{2\phi^2}
\mathrm{Tr}
\left(
\Sigma
\nabla_y\phi(\nabla_y\phi)'
\right)
\\
&&
+
(1-\gamma)u
\sup\limits_{\theta \in \mathbb{R}^d} \left\{
\theta'
\Sigma G
-\frac{\gamma}{2}
\theta'\Sigma\theta
+\gamma\theta'\Sigma
\frac{\nabla_y\phi }{ \phi}
        \right\}.
\end{eqnarray*}

The supremum in the above expression is reached at
\begin{align*}
\theta^\star(t,y)=
\frac{G(t,y)}{\gamma}  +
 \frac{\nabla_y\phi(t,y)}{\phi(t,y)}.
\end{align*}

Plugging this expression in the partial differential equation, we get:
\begin{eqnarray*}
&&\partial_t u +
        (\nabla_y u)'
	\left(r\vec{1}+\Sigma G - \frac{1}{2}\sigma \odot\sigma
	\right)
        +\frac12
	\mathrm{Tr}\left(\Sigma\nabla^2_{yy}u
	\right)
        +r V\partial_V u
\\
&&+\sup\limits_{\theta\in \mathbb{R}^d}
\left\{
V\partial_V u
\theta'
\Sigma G
+\frac{V^2}{2}
\theta'\Sigma\theta
\partial_{VV}^2u
+V\theta'\Sigma\partial_V\nabla_yu
        \right\}
\\
&=&
\frac{\gamma u}{\phi}
\Bigg[
\partial_t \phi
+
        (\nabla_y \phi)'
	\left(r\vec{1}+\Sigma G - \frac{1}{2}\sigma \odot\sigma
	\right)
+\frac{1}{2}
	\mathrm{Tr}\left(
\Sigma
\nabla^2_{yy}\phi
\right)
+
\frac{ (\gamma -1)}{2\phi}
(\nabla_y\phi)'
\Sigma
\nabla_y\phi
\\
&&
+(1-\gamma)\frac{\phi}{2}
\left(
\frac{G}{\gamma}+\frac{\nabla_y \phi}{\phi}
\right)'
\Sigma
\left(
\frac{G}{\gamma}+\frac{\nabla_y \phi}{\phi}
\right)
\Bigg]
\\
&=&
\frac{\gamma u}{\phi}
\Bigg[
\partial_t \phi
+
        (\nabla_y \phi)'
	\!
	\left(r\vec{1}+\frac{1}{\gamma}\Sigma G - \frac{1}{2}\sigma \odot\sigma
	\!
	\right)
        \!+
\frac{1}{2}
	\mathrm{Tr}\left(
\Sigma
\nabla^2_{yy}\phi
	\right)
+\frac{(1-\gamma)\phi}{2\gamma^2}
G'\Sigma G
\Bigg]
\\
&=&0.
\end{eqnarray*}

As it is straightforward to verify that $u$ satisfies the terminal condition
(\ref{eq:terminal_CRRA_u}), the result is proved.
\end{proof}

For solving our problem, we would like to prove that there exists a (positive) $C^{1,2}\left(\left[0,T\right]\times\mathbb{R}^d\right)$ function $\phi$ solution of \eqref{eq:HJB_crra_phi} with terminal condition \eqref{eq:terminal_crra_phi} such that $\frac{\nabla_y\phi}{\phi}$ is at most linear in $y$. However, unlike what happened in the CARA case, there is no guarantee, in general, that such a function exists. We will even show in Section 4 that there are blowup cases for some Gaussian priors in the case $\gamma < 1$.\\

Even though there is no general result, we can state for instance a result in the case of a prior distribution $m_\mu$ with compact support.

\begin{prop}
Let us suppose that the prior distribution $m_\mu$ has compact support.\\

Let us define
\begin{equation}
\label{eq:FK-crra}
\phi : (t,y)\in[0,T]\times\mathbb{R}^d \mapsto \mathbb{E}\left[\exp\left(
	\frac{(1-\gamma)}{2\gamma^2}
    \int_t^T
G(s,Z^{t,y}_s)'\Sigma G (s,Z^{t,y}_s)
ds
    \right)
    \right],
 \end{equation}
where $\forall (t,y)\in[0,T]\times\mathbb{R}^d$, we introduce for $s \in [t,T]$,
\begin{align*}
dZ_s^{t,y}= \left(r\vec{1}+\frac{1}{\gamma}\Sigma G (t,Z_s^{t,y})- \frac{1}{2}\sigma \odot\sigma\right)
ds +\sigma\odot d W_s, \quad Z^{t,y}_t = y.
\end{align*}
Furthermore, in that case
\begin{equation}
\label{compact_bound}
\exists A_T > 0, \forall t \in [0,T], \forall y \in \mathbb{R}^d, \forall i \in \{1, \ldots, d\}, \norm{\frac{\nabla_y \phi(t,y)}{\phi(t,y)} } \le A_T.
\end{equation}
\end{prop}

\begin{proof}
By using Theorem 1 and Proposition \ref{compact}, we easily see that formal differentiations are authorized. Therefore $\phi$ is a $C^{1,2}\left(\left[0,T\right]\times\mathbb{R}^d\right)$ function solution of \eqref{eq:HJB_crra_phi} with terminal condition \eqref{eq:terminal_crra_phi}.\\

For the second point, we write, for $(t,y) \in [0,T]\times\mathbb{R}^d$,
\begin{eqnarray*}
&&\nabla_y \phi(t,y)=\\
&&\mathbb{E}\left[\frac{(1-\gamma)}{\gamma^2} \int_t^T\!\!
D_yZ^{t,y}_s D_ZG(s,Z^{t,y}_s)\Sigma G (s,Z^{t,y}_s)
ds \exp\left(
	\frac{(1-\gamma)}{2\gamma^2}
    \int_t^T\!\!
G(s,Z^{t,y}_s)'\Sigma G (s,Z^{t,y}_s)
ds
    \right)
    \right].
\end{eqnarray*}
We have
$$d D_y Z^{t,y}_s = \frac{1}{\gamma}\Sigma D_Z G(t,Z_s^{t,y}) D_y Z^{t,y}_s ds, \quad D_yZ^{t,y}_t = I_d.$$
Because of the Lipschitz property of $G$ and Gr\"onwall inequality, $\sup_{s \in [t,T]} \norm{D_y Z^{t,y}_s}$ is uniformly bounded on $[0,T]\times\mathbb{R}^d$. By Proposition \ref{compact}, we then deduce that there exists $C \ge 0$ such that
$$\forall (t,y) \in [0,T]\times\mathbb{R}^d, \sup_{s \in [t,T]} \norm{D_yZ^{t,y}_s D_ZG(s,Z^{t,y}_s)\Sigma G (s,Z^{t,y}_s)} \le C.$$
Therefore,
\begin{eqnarray*}\norm{\nabla_y \phi(t,y)} &\le& \mathbb{E}\left[\frac{|1-\gamma|}{\gamma^2} C(T-t)\exp\left(
	\frac{(1-\gamma)}{2\gamma^2}
    \int_t^T
G(s,Z^{t,y}_s)'\Sigma G (s,Z^{t,y}_s)
ds
    \right)
    \right]\\
    &\le& \frac{|1-\gamma|}{\gamma^2} C T \phi(t,y).
\end{eqnarray*}
Hence the result.
\end{proof}

We now write a verification theorem and provide a result for solving the problem faced by the agent under additional hypotheses.

\begin{thm}
\label{verif_crra_gl}
Let us suppose that there exists a positive function $\phi \in C^{1,2}([0,T]\times\mathbb R^d)$ solution of (\ref{eq:HJB_crra_phi}) with terminal condition (\ref{eq:terminal_crra_phi}). Let us also suppose that
\begin{equation}
\label{eq:gradlogphi}
\exists A_T > 0, \forall t \in [0,T], \forall y \in \mathbb{R}^d, \forall i \in \{1, \ldots, d\}, \norm{\frac{\nabla_y \phi(t,y)}{\phi(t,y)} } \le A_T (1 + \norm{y}).
\end{equation}
Let us then consider the function $u$ defined by (\ref{eq:ansatz_CRRA}).\\

For all $\left(t,V,y\right)\in\left[0,T\right]\times\mathbb{R}_+^*\times\mathbb{R}^{d}$
and $\theta = (\theta_s)_{s \in [t,T]}\in\mathcal{A}^\gamma_t$, we have
\begin{eqnarray}
\label{eq:CRRAverif}
\mathbb{E}
\left[
U^\gamma\left(V_T^{t,V,y,\theta}
\right)
\right]
&\leq&
u(t,V,y)
\end{eqnarray}
Moreover, equality in (\ref{eq:CRRAverif}) is obtained by taking the
optimal control $(\theta^{\star}_s)_{s \in [t,T]}\in\mathcal{A}^\gamma_t$ given by \eqref{eq:optimizer_crra1}, i.e.,
\begin{equation}\label{eq:optcrraverif}\forall s \in [t,T], \theta^{\star}_s = \frac{G(s,Y^{t,y}_s)}{\gamma}
+ \frac{\nabla_y \phi(s,Y^{t,y}_s)}{\phi(s,Y^{t,y}_s)}.
\end{equation}
In particular $u=v$.
\end{thm}

\begin{proof}

The proof is similar to that of the CARA case, therefore we do not detail all the computations.\\

From the Lipschitz property of $G$ stated in Eq. (\ref{eq:lip}) and assumption (\ref{eq:gradlogphi}) on $\phi$, we see that $(\theta_s^\star)_{s \in [t,T]}$ is indeed admissible (i.e., $(\theta_s^\star)_{s \in [t,T]} \in \mathcal{A}^\gamma_t$).\\

Let us then consider $\left(t,V,y\right)\in [0,T]\times\mathbb{R}_+^*\times\mathbb{R}^{d}$ and $\theta = (\theta_s)_{s \in [t,T]}\in\mathcal{A}^\gamma_t$.\\

By It\=o's formula, we have for all $s\in[t,T]$
\begin{eqnarray*}
&&d\left(u\left(s,V_s^{t,V,y,\theta},Y_s^{t,y}\right)\right)
\\ & = &
\mathcal{L}^{\theta}u\left(s,V_s^{t,V,y,\theta},Y_s^{t,y}\right)ds\\
&&+
\left(
\partial_V u\left(s,V_s^{t,V,y,\theta},Y_s^{t,y}\right) \theta_s V_s^{t,V,y,\theta}
+
\nabla_y u\left(s,V_s^{t,V,y,\theta},Y_s^{t,y}\right)
\right)'
\left(
\sigma\odot d\widehat W_s
\right)
,\label{eq:verif_ito_crra}
\end{eqnarray*}
where
\begin{eqnarray*}
&&\mathcal{L}^{\theta} u
\left(s,V_s^{t,V,y,\theta},Y_s^{t,y}\right)\\
& = &
\partial_{t} u\left(s,V_s^{t,V,y,\theta},Y_s^{t,y}\right)
+\partial_V  u\left(s,V_s^{t,V,y,\theta},Y_s^{t,y}\right)
\left(
\theta_s'\Sigma G(s,Y_s^{t,y})+r
\right)V_s^{t,V,y,\theta}
\\&&+\nabla_y u\left(s,V_s^{t,V,y,\theta},Y_s^{t,y}\right) '
\left(
r\vec{1}+\Sigma G(s,Y_s^{t,y})-\dfrac{1}{2}\sigma\odot\sigma
\right)\\
&&+\dfrac{1}{2}V_s^{t,V,y,\theta}\partial_V\nabla_y u\left(s,V_s^{t,V,y,\theta},Y_s^{t,y}\right)' \Sigma \theta_s
+\dfrac{1}{2}\partial^2_{VV}u\left(s,V_s^{t,V,y,\theta},Y_s^{t,y}\right)
\left(V_s^{t,V,y,\theta}\right)^2\theta_s'\Sigma\theta_s
\\&&+\dfrac{1}{2}\text{Tr}\left(\Sigma\nabla^{2}_{yy}u\left(s,V_s^{t,V,y,\theta},Y_s^{t,y}\right)\right).
\end{eqnarray*}

Note that we have
\begin{eqnarray*}
&&\partial_V u\left(s,V_s^{t,V,y,\theta},Y_s^{t,y}\right)\theta_sV_s^{t,V,y,\theta}
+
\nabla_y u\left(s,V_s^{t,V,y,\theta},Y_s^{t,y}\right) \\
&=&
u\left(s,V_s^{t,V,y,\theta},Y_s^{t,y}\right)
\left(
(1-\gamma)\theta_s
+\gamma\dfrac{\nabla_y\phi\left(s,Y_s^{t,y}\right) }{\phi\left(s,Y_s^{t,y}\right)}
\right).
\end{eqnarray*}
Let us subsequently define, for all $s\in [t,T]$,
\begin{align*}
\kappa^\theta_s&=
(1-\gamma)\theta_s
+\gamma\dfrac{\nabla_y\phi\left(s,Y_s^{t,y}\right) }{\phi\left(s,Y_s^{t,y}\right)},
\end{align*}
and
\begin{align*}
\xi^\theta_{t,s}&=\exp
\left(
\int_t^s{\kappa^\theta_{\tau}}'
\left(\sigma \odot d\widehat W_\tau\right)
-\frac12
\int_t^s {\kappa^\theta_{\tau}}'\Sigma\kappa^\theta_{\tau}d\tau
\right).
\end{align*}

We have
\begin{eqnarray*}
d\left(
u\left(s,V_s^{t,V,y,\theta},Y_s^{t,y}\right)
\left(\xi_{t,s}^{\theta}\right)^{-1}\right) & = & \left(\xi_{t,s}^{\theta}\right)^{-1}\mathcal{L}^{\theta}u\left(s,V_s^{t,V,y,\theta},Y_s^{t,y}\right) ds.
\end{eqnarray*}

By definition of $u$, $\mathcal{L}^{\theta}u\left(s,V_s^{t,V,y,\theta},Y_s^{t,y}\right) \leq 0$ and $\mathcal{L}^\theta u\left(s,V_s^{t,V,y,\theta},Y_s^{t,y}\right)=0$ if $\theta_s=\theta_s^\star$. As a consequence, $\left(u\left(s,V_s^{t,V,y,\theta},Y_s^{t,y}\right)
\left(\xi_{t,s}^{\theta}\right)^{-1}
\right)_{s\in[t,T]}$ is nonincreasing, and therefore
\begin{eqnarray}
u\left(T,V_T^{t,V,y,\theta},Y_T^{t,y}\right) & \leq & u\left(t,V,y\right)\xi_{t,T}^{\theta},\label{eq:tp2}
\end{eqnarray}
with equality when $\left(\theta_s\right)_{s\in[t,T]}=\left(\theta_s^\star\right)_{s\in[t,T]}$.\\

Subsequently,
$$\mathbb{E}\left[U^\gamma\left(V_{T}^{t,V,y,\theta}\right)\right] = \mathbb{E}\left[u(T,V^{t,V,y,\theta}_T,Y^{t,y}_T)\right]\leq u(t,V,y)\mathbb{E}\left[\xi_{t,T}^\theta\right],
$$
with equality when $(\theta_s^\star)_{s \in [t,T]} = (\theta_s)_{s \in [t,T]}$.\\

Using the same method as in the proof of Theorem \ref{verif_cara_gl}, we see that $\mathbb{E}[\xi_{t,T}^{\theta^\star}] = 1$. Therefore,

$$u(t,V,y) = \mathbb{E}\left[U^\gamma\left(V_{T}^{t,V,y,\theta^\star}\right)\right].$$

We have just shown the second part of the theorem. For the first part, we consider the cases  $\gamma\geq1$ and $\gamma<1$ separately because the set of admissible strategies is larger in the second case.

\begin{enumerate}[(a)]
\item If $\gamma\geq1$, $\left(\theta_s\right)_{s\in[t,T]}\in\mathcal{A}_{t}^{\gamma}$ verifies the linear growth condition. Therefore, using the assumption on  $\dfrac{\nabla_y\phi}{\phi}$ and the same argument as in Theorem \ref{verif_cara_gl}, we see that $\left(\xi_{t,s}^{\theta}\right)_{s\in[t,T]}$ is a martingale
with $\mathbb{E}\left[\xi_{t,s}^{\theta}\right]=1$
for all $s\in[t,T]$.\\

We obtain
\begin{eqnarray*}
\mathbb{E}\left[U^{\gamma}\left(V_{T}^{t,V,y,\theta}\right)\right] & \leq & u\left(t,V,y\right).
\end{eqnarray*}

\item If $\gamma<1$, then we define the stopping time
\begin{eqnarray*}
\tau_{n} & = & T\wedge\inf\left\{ s\in\left[t,T\right],\quad\norm{\kappa_{s}^{\theta}}\geq n\right\}.
\end{eqnarray*}
We use this stopping time in order to localize Eq. (\ref{eq:tp2})
\begin{eqnarray*}
u\left(\tau_{n},V_{\tau_{n}}^{t,V,y,\theta},Y_{\tau_{n}}^{t,y}\right) & \leq & \xi_{t,\tau_{n}}^{\theta}u\left(t,V,y\right).
\end{eqnarray*}

By taking the expectation, we have, for all $n\in\mathbb{N}$,
\begin{eqnarray*}
\mathbb{E}\left[u\left(\tau_{n},V_{\tau_{n}}^{t,V,y,\theta},Y_{\tau_{n}}^{t,y}\right)\right] & \leq & u\left(t,V,y\right).
\end{eqnarray*}

As $u$ is nonnegative when $\gamma<1$, we can apply Fatou's lemma
\begin{eqnarray*}
\mathbb{E}\left[\liminf_{n\to+\infty}\ u\left(\tau_{n},V_{\tau_{n}}^{t,V,y,\theta},Y_{\tau_{n}}^{t,y}\right)\right] & \leq & \liminf_{n\to+\infty}\ \mathbb{E}\left[u\left(\tau_{n},V_{\tau_{n}}^{t,V,y,\theta},Y_{\tau_{n}}^{t,y}\right)\right]\\
 & \leq & u\left(t,V,y\right).
\end{eqnarray*}

Because $\left(\theta_s\right)_{s\in[t,T]}\in\mathcal{A}_{t}^{\gamma}$, $\tau_{n}\to_{n \to +\infty} T$ almost surely. Therefore
\begin{eqnarray*}
\mathbb{E}\left[U^{\gamma}\left(V_{s}^{t,V,y,\theta}\right)\right] & \leq & u\left(t,V,y\right).
\end{eqnarray*}
\end{enumerate}

In both cases, we conclude that

$$u(t,V,y) = v(t,V,y) =\!\!\! \sup_{(\theta_s)_{s \in [t,T]} \in \mathcal{A}^\gamma_t}\!\!\! \mathbb{E}\left[U^\gamma\left( V_{T}^{t,V,y,\theta}\right)\right] = \mathbb{E}\left[U^\gamma\left( V_{T}^{t,V,y,\theta^\star}\right)\right].$$
\end{proof}

The above verification theorem can be used for instance when $m_\mu$ has a compact support because of \eqref{compact_bound}. In the next section, we address the case of Gaussian priors and we shall see that there is a blowup phenomenon associated with the solution of the partial differential equation (\ref{eq:HJB_crra_phi}) with terminal condition (\ref{eq:terminal_crra_phi}) when $\gamma$ is too small.\\

Before we turn to the Gaussian case, let us consider the specific case $\gamma=1$.

\subsubsection{The $\gamma=1$ case}

To solve the HJB equation when $\gamma = 1$, we use the following ansatz:
\begin{eqnarray}
u\left(t,V,y\right) & = &
r(T-t) + \log(V) + \phi(t,y).
\label{eq:ansatz_CRRA_log}
\end{eqnarray}

\begin{prop}
\label{prop:u_phi_crra}Suppose there exists a function $\phi\in C^{1,2}\left(\left[0,T\right]\times\mathbb{R}^d\right)$
satisfying
\begin{eqnarray}
\partial_t \phi
+
        (\nabla_y \phi)'
	\!
	\left(r\vec{1}+\Sigma G - \frac{1}{2}\sigma \odot\sigma
	\!
	\right)
        \!+
\frac{1}{2}
	\mathrm{Tr}\left(
\Sigma
\nabla^2_{yy}\phi
	\right)
+\frac 12
G'\Sigma G
   &=& 0,\label{eq:HJB_crra_phi_log}
\end{eqnarray}
with terminal condition\textup{
\begin{eqnarray}
\forall y\in\mathbb{R}^d,\quad \phi\left(T,y\right) & = & 0.\label{eq:terminal_crra_phi_log}
\end{eqnarray}
}Then $u$ defined by (\ref{eq:ansatz_CRRA_log}) is solution of the HJB
equation (\ref{eq:HJB_crra_u}) with terminal condition (\ref{eq:terminal_CRRA_u}).\\

Moreover, the supremum in (\ref{eq:HJB_crra_u}) is achieved at
\begin{eqnarray}
\theta^{\star}(t,y) & = & G(t,y).\label{eq:optimizer_crra1_log}
\end{eqnarray}
\end{prop}

\begin{proof}
Let us consider $\phi\in C^{1,2}\left(\left[0,T\right]\times\mathbb{R}^d\right)$, solution of \eqref{eq:HJB_crra_phi_log} with terminal condition \eqref{eq:terminal_crra_phi_log}. For $u$ defined by \eqref{eq:ansatz_CRRA_log}, we have:
\begin{eqnarray*}
&&\partial_t u +
        (\nabla_y u)'
	\left(r\vec{1}+\Sigma G - \frac{1}{2}\sigma \odot\sigma
	\right)
        +\frac12
	\mathrm{Tr}\left(\Sigma\nabla^2_{yy}u
	\right)
        +r V\partial_V u
\\
&&+\sup\limits_{\theta\in \mathbb{R}^d} \left\{
V\partial_V u
\theta'
\Sigma G
+\frac{V^2}{2}
\theta'\Sigma\theta
\partial_{VV}^2u
+V\theta'\Sigma\partial_V\nabla_yu
        \right\}
\nonumber\\
&=&
-r + \partial_t \phi
+ (\nabla_y \phi)' \left(r\vec{1}+\Sigma G - \frac{1}{2}\sigma \odot\sigma
	\right)
+\frac 12
	\mathrm{Tr}\left(
\Sigma
\nabla^2_{yy}\phi
\right) + r+
\sup\limits_{\theta \in \mathbb{R}^d} \left\{
\theta'
\Sigma G
-\frac{1}{2}
\theta'\Sigma\theta
\right\}.
\end{eqnarray*}

The supremum in the above expression is reached at
\begin{align*}
\theta^\star(t,y)=
G(t,y).
\end{align*}

Therefore
\begin{eqnarray*}
&&\partial_t u +
        (\nabla_y u)'
	\left(r\vec{1}+\Sigma G - \frac{1}{2}\sigma \odot\sigma
	\right)
        +\frac12
	\mathrm{Tr}\left(\Sigma\nabla^2_{yy}u
	\right)
        +r V\partial_V u
\\
&&+\sup\limits_{\theta\in \mathbb{R}^d}
\left\{
V\partial_V u
\theta'
\Sigma G
+\frac{V^2}{2}
\theta'\Sigma\theta
\partial_{VV}^2u
+V\theta'\Sigma\partial_V\nabla_yu
        \right\}
\\
&=&
 \partial_t \phi
+ (\nabla_y \phi)' \left(r\vec{1}+\Sigma G - \frac{1}{2}\sigma \odot\sigma
	\right)
+\frac 12
	\mathrm{Tr}\left(
\Sigma
\nabla^2_{yy}\phi
\right)
+\frac{1}{2}
G'\Sigma G\\
&=&0.
\end{eqnarray*}

As it is straightforward to verify that $u$ satisfies the terminal condition
(\ref{eq:terminal_CRRA_u}), the result is proved.
\end{proof}

From the previous proposition, we see that solving the HJB equation \eqref{eq:HJB_crra_u} with terminal condition \eqref{eq:terminal_CRRA_u} boils down to solving \eqref{eq:HJB_crra_phi_log} with terminal condition \eqref{eq:terminal_crra_phi_log}. Because \eqref{eq:HJB_crra_phi_log} is a simple parabolic PDE, we can easily build a strong solution.

\begin{prop}
Let us define
\begin{equation}
\label{eq:FK-crra_log}
\phi : (t,y)\in[0,T]\times\mathbb{R}^d \mapsto \mathbb{E}\left[
    \int_t^T \frac{1}{2}G(s, Y^{t,y}_s)'\Sigma G(s, Y^{t,y}_s)ds\right],
 \end{equation}
where $\forall (t,y)\in[0,T]\times\mathbb{R}^d$, $\forall s \in [t,T]$,
\begin{align*}
Y_s^{t,y}&=y+\int_t^s
\left(
  r\vec{1}+\Sigma G(\tau,Y^{t,y}_{\tau}) - \frac12
  \sigma \odot \sigma
  \right) d\tau
+\sigma \odot
(\widehat W_s-\widehat W_t).
\end{align*}
Then $\phi$ is a $C^{1,2}([0,T]\times\mathbb{R}^d)$ function, solution of \eqref{eq:HJB_crra_phi_log} with terminal condition \eqref{eq:terminal_crra_phi_log}.\\
Furthermore,
\begin{equation}
\label{eq:gradphi_log}
\exists A_T > 0, \forall t \in [0,T], \forall y \in \mathbb{R}^d, \forall i \in \{1, \ldots, d\}, \norm{\nabla_y \phi(t,y)} \le A_T (1 + \norm{y}).
\end{equation}
\end{prop}

\begin{proof}
Because of the assumption \eqref{eq:lip} on $G$, the first part of the proposition is a consequence of classical results for parabolic PDEs and of the classical Feynman-Kac representation (see for instance \cite{af,ks}).\\

For the second part, we notice first that
$$\forall (t,y) \in [0,T]\times\mathbb{R}^d, \nabla_y \phi(t,y) = \mathbb{E}\left[\int_t^T \frac{1}{\gamma}D_y Y^{t,y}_sD_YG(s, Y^{t,y}_s)\Sigma G(s, Y^{t,y}_s)ds\right].$$
We have
$$d D_y Y^{t,y}_s = \Sigma D_Y G(t,Y_s^{t,y}) D_y Y^{t,y}_s ds, \quad D_y Y^{t,y}_t = I_d.$$
Because of the Lipschitz property of $G$ and Gr\"onwall inequality, $\sup_{s \in [t,T]} \norm{D_y Y^{t,y}_s}$ is uniformly bounded on $[0,T]\times\mathbb{R}^d$.
Therefore, by \eqref{eq:lip}, there exists a constant $C \ge 0$ such that
\begin{equation}
\label{eq:ineq}\forall (t,y) \in [0,T]\times\mathbb{R}^d, \forall s \in [t,T] \norm{D_y Y^{t,y}_sD_YG(s, Y^{t,y}_s)\Sigma G(s, Y^{t,y}_s)} \le C \norm{G(s, Y^{t,y}_s)}.
\end{equation}
By \eqref{eq:lip} there exists a constant $C' \ge 0$ such that
\begin{equation}
\norm{G(s,Y^{t,y}_s)} \le \norm{G(s,Y^{t,y}_s) - G(t,y)} + C' (1+ \norm{y}).
\label{eq:ineq2}
\end{equation}
But, by using Theorem \ref{propbayesdyn}
$$G(s,Y^{t,y}_s) - G(t,y) = \int_t^s D_YG(\tau, Y^{t,y}_\tau) (\sigma \odot d\widehat{W}_\tau).$$
Therefore, using the Lipschitz property of $G$ we see that $\mathbb{E}\left[\norm{G(s,Y^{t,y}_s) - G(t,y)}\right]$ is bounded by a constant that depends on $T$ only. Combining this result with Eqs. (\ref{eq:ineq}) and (\ref{eq:ineq2}), we obtain the property \eqref{eq:gradphi_log}.
\end{proof}

We now write a verification theorem and provide a result for solving the problem faced by the agent.

\begin{thm}
\label{verif_crra_gl_log}
Let us consider the function $\phi \in C^{1,2}([0,T]\times\mathbb R^d)$ defined by \eqref{eq:FK-crra_log}. Let us then consider the function $u$ defined by (\ref{eq:ansatz_CRRA_log}).\\

For all $\left(t,V,y\right)\in\left[0,T\right]\times\mathbb{R}_+^*\times\mathbb{R}^{d}$
and $\theta = (\theta_s)_{s \in [t,T]}\in\mathcal{A}^1_t$, we have
\begin{eqnarray}
\label{eq:CRRAverif_log}
\mathbb{E}
\left[
\log\left(V_T^{t,V,y,\theta}
\right)
\right]
&\leq&
u(t,V,y)
\end{eqnarray}
Moreover, equality in (\ref{eq:CRRAverif}) is obtained by taking the
optimal control $(\theta^{\star}_s)_{s \in [t,T]}\in\mathcal{A}^1_t$ given by \eqref{eq:optimizer_crra1_log}, i.e.,
\begin{equation}\label{eq:optcrraverif_log}\forall s \in [t,T], \theta^{\star}_s = G(s,Y^{t,y}_s).
\end{equation}
In particular $u=v$.
\end{thm}

\begin{proof}

The proof is similar to that of the $\gamma > 1$ case, therefore we do not detail all the computations.\\

From the Lipschitz property of $G$ stated in Eq. (\ref{eq:lip}), we see that $(\theta_s^\star)_{s \in [t,T]}$ is indeed admissible (i.e., $(\theta_s^\star)_{s \in [t,T]} \in \mathcal{A}^1_t$).\\

Let us then consider $\left(t,V,y\right)\in [0,T]\times\mathbb{R}_+^*\times\mathbb{R}^{d}$ and $\theta = (\theta_s)_{s \in [t,T]}\in\mathcal{A}^1_t$.\\

By It\=o's formula, we have for all $s\in[t,T]$
\begin{eqnarray*}
&&d\left(u\left(s,V_s^{t,V,y,\theta},Y_s^{t,y}\right)\right)
\\ & = &
\mathcal{L}^{\theta}u\left(s,V_s^{t,V,y,\theta},Y_s^{t,y}\right)ds\\
&&+
\left(
\partial_V u\left(s,V_s^{t,V,y,\theta},Y_s^{t,y}\right) \theta_s V_s^{t,V,y,\theta}
+
\nabla_y u\left(s,V_s^{t,V,y,\theta},Y_s^{t,y}\right)
\right)'
\left(
\sigma\odot d\widehat W_s
\right)
,\label{eq:verif_ito_crra_log}
\end{eqnarray*}
where
\begin{eqnarray*}
&&\mathcal{L}^{\theta} u
\left(s,V_s^{t,V,y,\theta},Y_s^{t,y}\right)\\
& = &
\partial_{t} u\left(s,V_s^{t,V,y,\theta},Y_s^{t,y}\right)
+\partial_V  u\left(s,V_s^{t,V,y,\theta},Y_s^{t,y}\right)
\left(
\theta_s'\Sigma G(s,Y_s^{t,y})+r
\right)V_s^{t,V,y,\theta}
\\&&+\nabla_y u\left(s,V_s^{t,V,y,\theta},Y_s^{t,y}\right) '
\left(
r\vec{1}+\Sigma G(s,Y_s^{t,y})-\dfrac{1}{2}\sigma\odot\sigma
\right)\\
&&+\dfrac{1}{2}V_s^{t,V,y,\theta}\partial_V\nabla_y u\left(s,V_s^{t,V,y,\theta},Y_s^{t,y}\right)' \Sigma \theta_s
+\dfrac{1}{2}\partial^2_{VV}u\left(s,V_s^{t,V,y,\theta},Y_s^{t,y}\right)
\left(V_s^{t,V,y,\theta}\right)^2\theta_s'\Sigma\theta_s
\\&&+\dfrac{1}{2}\text{Tr}\left(\Sigma\nabla^{2}_{yy}u\left(s,V_s^{t,V,y,\theta},Y_s^{t,y}\right)\right).
\end{eqnarray*}

Note that we have
\begin{eqnarray*}
&&\partial_V u\left(s,V_s^{t,V,y,\theta},Y_s^{t,y}\right)\theta_sV_s^{t,V,y,\theta}
+
\nabla_y u\left(s,V_s^{t,V,y,\theta},Y_s^{t,y}\right) \\
&=&\theta_s
+\nabla_y\phi\left(s,Y_s^{t,y}\right).
\end{eqnarray*}
Let us subsequently define, for all $s\in [t,T]$,
\begin{align*}
\kappa^\theta_s&=
\theta_s+
\nabla_y\phi\left(s,Y_s^{t,y}\right),
\end{align*}
and
\begin{align*}
\xi^\theta_{t,s}&=\exp
\left(
\int_t^s{\kappa^\theta_{\tau}}'
\left(\sigma \odot d\widehat W_\tau\right)
-\frac12
\int_t^s {\kappa^\theta_{\tau}}'\Sigma\kappa^\theta_{\tau}d\tau
\right).
\end{align*}

We have
\begin{eqnarray*}
d\left(
u\left(s,V_s^{t,V,y,\theta},Y_s^{t,y}\right)
\left(\xi_{t,s}^{\theta}\right)^{-1}\right) & = & \left(\xi_{t,s}^{\theta}\right)^{-1}\mathcal{L}^{\theta}u\left(s,V_s^{t,V,y,\theta},Y_s^{t,y}\right) ds.
\end{eqnarray*}

By definition of $u$, $\mathcal{L}^{\theta}u\left(s,V_s^{t,V,y,\theta},Y_s^{t,y}\right) \leq 0$ and $\mathcal{L}^\theta u\left(s,V_s^{t,V,y,\theta},Y_s^{t,y}\right)=0$ if $\theta_s=\theta_s^\star$. As a consequence, $\left(u\left(s,V_s^{t,V,y,\theta},Y_s^{t,y}\right)
\left(\xi_{t,s}^{\theta}\right)^{-1}
\right)_{s\in[t,T]}$ is nonincreasing, and therefore
\begin{eqnarray}
u\left(T,V_T^{t,V,y,\theta},Y_T^{t,y}\right) & \leq & u\left(t,V,y\right)\xi_{t,T}^{\theta},\label{eq:tp2_log}
\end{eqnarray}
with equality when $\left(\theta_s\right)_{s\in[t,T]}=\left(\theta_s^\star\right)_{s\in[t,T]}$.\\

Subsequently,
$$\mathbb{E}\left[\log\left(V_{T}^{t,V,y,\theta}\right)\right] = \mathbb{E}\left[u(T,V^{t,V,y,\theta}_T,Y^{t,y}_T)\right]\leq u(t,V,y)\mathbb{E}\left[\xi_{t,T}^\theta\right],
$$
with equality when $(\theta_s)_{s \in [t,T]} = (\theta_s^\star)_{s \in [t,T]}$.\\

$\left(\theta_s\right)_{s\in[t,T]}\in\mathcal{A}_{t}^{1}$ verifies the linear growth condition. Therefore, using Eq. (\ref{eq:gradphi_log}) and the same argument as in Theorem \ref{verif_cara_gl}, we see that $\left(\xi_{t,s}^{\theta}\right)_{s\in[t,T]}$ is a martingale
with $\mathbb{E}\left[\xi_{t,s}^{\theta}\right]=1$
for all $s\in[t,T]$.\\

We obtain
\begin{eqnarray*}
\mathbb{E}\left[\log\left(V_{T}^{t,V,y,\theta}\right)\right] & \leq & u\left(t,V,y\right),
\end{eqnarray*}
with equality when $\left(\theta_s\right)_{s\in[t,T]}=\left(\theta_s^\star\right)_{s\in[t,T]}$.\\

We conclude that

$$u(t,V,y) = v(t,V,y) =\!\!\! \sup_{(\theta_s)_{s \in [t,T]} \in \mathcal{A}^1_t}\!\!\! \mathbb{E}\left[\log\left( V_{T}^{t,V,y,\theta}\right)\right] = \mathbb{E}\left[\log\left( V_{T}^{t,V,y,\theta^\star}\right)\right].$$
\end{proof}

\section{Optimal portfolio choice in the Gaussian case: a tale of two routes}

We showed in Section 3 that solving the optimal portfolio choice problem boils down to solving linear parabolic PDEs in the CARA and CRRA cases. One important case in which these PDEs can be solved in closed form is that of a Gaussian prior. Moreover, in the Gaussian prior case, there are two routes to solve the problem with PDEs because, as we shall see below, $\beta$ appears to be a far more natural state variable than $y$. In this section, we solve the optimal portfolio choice problem in the case of a Gaussian prior using these two different routes and we discuss two essential points: (i) the influence of online learning on the optimal investment strategy, and (ii) the occurrence of blowups in some CRRA cases.\\

\subsection{Bayesian learning in the Gaussian case}

Let us consider a non-degenerate multivariate Gaussian prior $m_\mu$, i.e.,

\begin{equation}
\label{eq:priorgauss} m_\mu(dz)=\frac{1}{(2\pi)^\frac d2|\Gamma_0|^\frac 12}\exp
\left(
-\frac12 (z-\beta_0)'\Gamma_0^{-1}(z-\beta_0)
\right)dz,
\end{equation}
where $\beta_0\in \mathbb{R}^d$ and $\Gamma_0\in S_d^{++}(\mathbb{R})$.

Our first goal is to obtain closed-form expressions for $F$ and $G$ in the Gaussian case. In order to obtain these expressions we shall use the following lemma:
\begin{lem}
\begin{align*}
\forall M \in S_d^{++}(\mathbb{R}), \forall N \in \mathbb{R}^d, \qquad \int_{\mathbb{R}^d} \exp
\left(
-x'Mx+x'N
\right)dx
=
\pi^{\frac{d}{2}}
|M|^{-\frac12}
\exp
\left(
\frac{1}{4}
N'
M^{-1}
N
\right).
\end{align*}
\end{lem}

\begin{proof} Using the canonical form of a polynomial of degree 2, we get
$$-x'Mx+x'N = -\left(x- \frac 12M^{-1}N\right)'M\left(x- \frac 12M^{-1}N\right) + \frac 14 N'M^{-1}N.$$
Therefore,
\begin{eqnarray*}
&&\int_{\mathbb{R}^d} \exp
\left(
-x'Mx+x'N
\right)dx\\
&=& \int_{\mathbb{R}^d} \exp
\left(
-\left(x- \frac 12M^{-1}N\right)'M\left(x- \frac 12M^{-1}N\right) + \frac 14 N'M^{-1}N
\right)dx\\
&=& (2\pi)^{\frac d2} |(2M)^{-1}|^{\frac 12} \exp
\left(
\frac{1}{4}
N'
M^{-1}
N
\right)\\
&=&\pi^{\frac{d}{2}}
|M|^{-\frac12}
\exp
\left(
\frac{1}{4}
N'
M^{-1}
N
\right).
\end{eqnarray*}
\end{proof}

We are now ready to derive the expressions of $F$ and $G$.

\begin{prop}
\label{prop:FandG}
For the multivariate Gaussian prior $m_\mu$ given by (\ref{eq:priorgauss}), $F$ and $G$ are given by:\\

$\forall t\in\mathbb{R}_+,\forall y\in\mathbb{R}^d,$
\begin{eqnarray}
F(t,y)\!\!&=&\!\!
\frac{\left|
\Gamma_0^{-1}+t\Sigma^{-1}
\right|
^{-\frac12}
}{|\Gamma_0|^{\frac 12}}
\exp
\left\{
-r\vec{1}'
\Sigma^{-1}
 \left[
        y-Y_0+
            \frac t2
            \sigma \odot \sigma
\right] + \frac t2 r^2 \vec{1}\Sigma^{-1}\vec{1}
\right.\nonumber
\\
&&
\left.
-\frac12
\beta_0'\Gamma_0^{-1}\beta_0
+\frac{1}{2}
\left[
\Sigma^{-1}
 \left(
        y-Y_0
            +\frac{t}{2}
            \sigma \odot \sigma
\right)
+
\Gamma_0^{-1}
\beta_0
\right]
'
\left(
\Gamma_0^{-1}+t\Sigma^{-1}
\right)
^{-1}
\right.\nonumber\\
&&\left.
\times \left[
\Sigma^{-1}
 \left(
        y-Y_0
            +\frac{t}{2}
            \sigma \odot \sigma
\right)
+
\Gamma_0^{-1}
\beta_0
\right]
\right\}
,\\
G(t,y)\!\!&=&\!\!
-r\Sigma^{-1}\vec{1}
+
\Sigma^{-1}
\left(
\Gamma_0^{-1}+t\Sigma^{-1}
\right)^{-1}
\left[
\Sigma^{-1}
\left(
y-Y_0+\frac{t}{2}\sigma\odot\sigma
\right)
+\Gamma_0^{-1}\beta_0
\right].\nonumber\\
\label{eq:G_Gaussian}
\end{eqnarray}
\end{prop}

\begin{proof}
$\forall t\in\mathbb{R}_+,\forall y\in\mathbb{R}^d$,
\begin{align*}
&F(t,y)=
\int_{\mathbb{R}^d}
\exp\left(
(z-r\vec{1})'
\Sigma^{-1}
    \left(
        y-Y_0+
        \left(
            -r\vec{1}
            +\frac12
            \sigma \odot \sigma
        \right)t
    \right)
-\frac12
(z-r\vec{1})' \Sigma^{-1} (z-r\vec{1})
t
\right)
    m_\mu(dz)
.
\end{align*}

Therefore,
\begin{eqnarray*}
F(t,y)&=&
\frac{1}{(2\pi)^\frac d2|\Gamma_0|^\frac 12}
\int_{\mathbb{R}^d}
\exp\left\{
(z-r\vec{1})'
\Sigma^{-1}
    \left[
        y-Y_0+
        \left(
            -r\vec{1}
            +\frac12
            \sigma \odot \sigma
        \right)t
    \right]\right.\\
& &\left.
-\frac12
(z-r\vec{1})'
\Sigma^{-1}
(z-r\vec{1})
t
-\frac12 (z-\beta_0)'\Gamma_0^{-1}(z-\beta_0)
\right\}
dz\\
&=&
\frac{\exp
\left(
D
\right)
}
{(2\pi)^\frac d2|\Gamma_0|^\frac 12}
\int_{\mathbb{R}^d}
\exp
\left(-
z'Mz+z'N
\right)dz,
\end{eqnarray*}
where
\begin{eqnarray*}
M&=&\frac12
\left(
\Gamma_0^{-1}+t\Sigma^{-1}
\right),\\
N&=&
\Sigma^{-1}
 \left[
        y-Y_0+
        \left(
            -r\vec{1}
            +\frac12
            \sigma \odot \sigma
        \right)t
    \right]
+r\Sigma^{-1}
\vec{1}
t
+
\Gamma_0^{-1}
\beta_0
\\
&=&\Sigma^{-1}
 \left[
        y-Y_0+
        \frac t2
            \sigma \odot \sigma
    \right]
+
\Gamma_0^{-1}
\beta_0,\\
\end{eqnarray*}
and
\begin{eqnarray*}
D&=&
-r\vec{1}'
\Sigma^{-1}
 \left[
        y-Y_0+
        \left(
            -r\vec{1}
            +\frac12
            \sigma \odot \sigma
        \right)t
\right]
-\frac t2 r^2
\vec{1}'\Sigma^{-1}\vec{1}
-\frac12
\beta_0'\Gamma_0^{-1}\beta_0\\
&=&
-r\vec{1}'
\Sigma^{-1}
 \left[
        y-Y_0
            +\frac t2
            \sigma \odot \sigma
\right]
+\frac t2 r^2
\vec{1}'\Sigma^{-1}\vec{1}
-\frac12
\beta_0'\Gamma_0^{-1}\beta_0
\end{eqnarray*}

Thanks to the above lemma, we have
\begin{eqnarray*}
 F(t,y)
&=&
\frac{\exp
\left(
D
\right)
}
{(2\pi)^\frac d2|\Gamma_0|^\frac 12}
\pi^{\frac{d}{2}}
|M|^{-\frac12}
\exp
\left(\frac{1}{4}
N'
M^{-1}
N
\right)\\
&=&
\frac{\left|
\Gamma_0^{-1}+t\Sigma^{-1}
\right|
^{-\frac12}
}{|\Gamma_0|^\frac 12}
\exp
\left\{
-r\vec{1}'
\Sigma^{-1}
 \left[
        y-Y_0+
            \frac t2
            \sigma \odot \sigma
\right]
+ \frac t2 r^2 \vec{1}\Sigma^{-1}\vec{1}
-\frac12
\beta_0'\Gamma_0^{-1}\beta_0
\right.
\\
&&
\left.
+\frac12
\left[
\Sigma^{-1}
 \left(
        y-Y_0
            +\frac{t}{2}
            \sigma \odot \sigma
\right)
+
\Gamma_0^{-1}
\beta_0
\right]
'
\left(
\Gamma_0^{-1}+t\Sigma^{-1}
\right)
^{-1}
\right.\\
&&\left.
\left[
\Sigma^{-1}
 \left(
        y-Y_0
            +\frac{t}{2}
            \sigma \odot \sigma
\right)
+
\Gamma_0^{-1}
\beta_0
\right]
\right\}.
\end{eqnarray*}

Differentiating $\log F$ brings
\begin{eqnarray*}
G(t,y)
&=&
-r\Sigma^{-1}\vec{1}
+
\Sigma^{-1}
\left(\Gamma_0^{-1}+t
\Sigma^{-1}
\right)^{-1}
\left[
\Sigma^{-1}
\left(
y-Y_0+\frac{t}{2}\sigma\odot\sigma
\right)
+\Gamma_0^{-1}\beta_0
\right].
\end{eqnarray*}
\end{proof}

Using Theorems \ref{propbayes} and \ref{propbayesdyn}, we now deduce straightforwardly the value of $\beta_t$ and its dynamics.

\begin{prop}
  \label{betagauss}
\begin{eqnarray}
\beta_{t} & = & \Gamma_{t}\left(\Sigma^{-1}\left(Y_t - Y_0+\dfrac{t}{2} \sigma \odot \sigma\right)+\Gamma_{0}^{-1}\beta_{0}\right), \label{eq:betagauss}\\
d\beta_t & = & \Gamma_{t}\Sigma^{-1}\left(\sigma\odot d\widehat{W}_{t}\right), \label{eq:betagaussdyn}
\end{eqnarray}
where $\Gamma_{t}  =  \left(\Gamma_{0}^{-1}+t\Sigma^{-1}\right)^{-1}.$
\end{prop}

\begin{rem}
Classical Bayesian analysis or application of classical filtering tools enables to prove that the posterior distribution of $\mu$ given $\mathcal{F}^S_t$ is in fact the Gaussian distribution $\mathcal{N}(\beta_t,\Gamma_t)$. In particular, it is noteworthy that the covariance matrix process $(\Gamma_t)_{t \in \mathbb R_+}$ is deterministic.\\
\end{rem}

The above analysis shows that, in the Gaussian prior case, the problem can be written with two different sets of state variables: $(y,V)$ or $(\beta,V)$. We can consider indeed that the problem is described, as in Section 3,  by the stochastic differential equations
\begin{equation}
\label{system1}
\left \{
\begin{array}{lcl}
    dY_t & = &
  \left(
  r\vec{1}+\Sigma G\left(t,Y_t\right) - \frac12
  \sigma \odot \sigma
  \right) dt
+\sigma \odot d\widehat W_t\\
dV_t &=& \left(M_{t}'\Sigma G(t,Y_t) +rV_{t}\right)dt+M_{t}'\left( \sigma \odot d\widehat{W}_{t}\right),\\
\end{array}
\right.
\end{equation}
or alternatively by the following stochastic differential equations
\begin{equation}
\label{system2}
\left \{
\begin{array}{lcl}
d\beta_t &=& \Gamma_{t}\Sigma^{-1}\left(\sigma\odot d\widehat{W}_{t}\right)\\
dV_t &=& \left(M_{t}'(\beta_t - r\vec{1}) +rV_{t}\right)dt+M_{t}'\left( \sigma \odot d\widehat{W}_{t}\right).\\
\end{array}
\right.
\end{equation}

In what follows, we are going to solve the optimal portfolio choice problem in the Gaussian prior case by using alternatively the two different routes associated with these two ways of describing the dynamics of the system.\\

\begin{rem}
It is noteworthy that the dynamics of $(\beta_t)_{t \in \mathbb R_+}$ in the Gaussian case, as written in Eq.~(\ref{system2}), does not involve any term in $Y$. From Theorem \ref{propbayesdyn}, we see that this is related to the fact that the matrix $D_y G(t,\cdot)$ is independent of $y$ in the Gaussian case. A natural question is whether or not this property is specific to a Gaussian prior distribution. In fact, the answer is positive. If indeed $D_y G(t,\cdot)$ is independent of $y$, then $\log F(t,\cdot)$ is a polynomial of (maximum) degree 2, i.e.,
\begin{eqnarray*}
F(t,y)=\exp\left(
A(t)+B(t)'y+y'C(t)y
\right),
\end{eqnarray*}
where $A(t)\in\mathbb{R},B(t)\in\mathbb{R}^d,$ and $C(t)\in S_d(\mathbb{R})$. Since
\begin{flalign*}
    F(0,y)&=\int_{\mathbb{R}^d} \exp\left(
(z-r\vec{1})'\Sigma^{-1}
(y-Y_0)
    \right)m_\mu(dz)=\exp(A(0)+B(0)'y+y'C(0)y),
\end{flalign*}
the Laplace transform of $m_\mu$ is the exponential of a polynomial of (maximum) degree 2, and $m_\mu$ is therefore Gaussian (possibly degenerate, even in the form of a Dirac mass).
\end{rem}

Before we solve the PDEs in the CARA and CRRA cases in the next subsections, let us state some additional properties that will be useful to simplify future computations.\\

\begin{prop}
\label{diffprop}
The dynamics of the conditional covariance matrix process $(\Gamma_t)_{t \in \mathbb{R}_+}$ is given by:\\
\begin{equation}
\label{eq:dGamma}
d\Gamma_{t}  =  -\Gamma_{t}\Sigma^{-1}\Gamma_{t}dt.
\end{equation}
The first order partial derivatives of $G$ are given by:\\

$\forall t \in \mathbb{R}_+, \forall y \in \mathbb{R}^d,$
\begin{eqnarray}
\label{eq:dyG} D_y G(t,y) & = & \Sigma^{-1} \Gamma_t \Sigma^{-1},\\
\label{eq:dtG} \partial_t G(t,y) & = & - \Sigma^{-1} \Gamma_t G(t,y) - D_y G(t,y) \left(r \vec{1} - \frac 12 \sigma\odot \sigma \right).
\end{eqnarray}
\end{prop}

\begin{proof}
Eq. (\ref{eq:dGamma}) is a simple consequence of the definition of $\Gamma_{t}$.\\

Eq. (\ref{eq:dyG}) derives from the differentiation of Eq. (\ref{eq:G_Gaussian}) with respect to $y$.\\

For Eq. (\ref{eq:dtG}), we use Eqs. (\ref{eq:G_Gaussian}) and (\ref{eq:dGamma}) to obtain
\begin{eqnarray*}
&&\partial_t G(t,y)\\
&=&
-\Sigma^{-1} \Gamma_t \Sigma^{-1} \Gamma_t\left[ \Sigma^{-1}
\left(
y-Y_0+\frac{t}{2}\sigma\odot\sigma
\right)
+\Gamma_0^{-1}\beta_0
\right] + \frac 12 \Sigma^{-1} \Gamma_t \Sigma^{-1} \sigma \odot \sigma \\
&=& -\Sigma^{-1} \Gamma_t\left( G(t,y) + r \Sigma^{-1} \vec{1} \right)  + \frac 12 \Sigma^{-1} \Gamma_t \Sigma^{-1} \sigma \odot \sigma\\
&=& -\Sigma^{-1} \Gamma_t G(t,y) - \Sigma^{-1} \Gamma_t \Sigma^{-1}\left(r \vec{1} - \frac 12 \sigma\odot \sigma \right)\\
&=& -\Sigma^{-1} \Gamma_t G(t,y) - D_y G(t,y) \left(r \vec{1} - \frac 12 \sigma\odot \sigma \right).
\end{eqnarray*}
\end{proof}

\begin{rem}
From Eq. (\ref{eq:dyG}), we see that $$\sup_{(t,y) \in [0,T]\times \mathbb{R}^d} \norm{D_y G(t,y)} \le  \sup_{t \in [0,T]} \norm{\Sigma^{-1} \Gamma_t \Sigma^{-1}} < +\infty.$$ Therefore Gaussian priors satisfy \eqref{eq:lip} as announced in Section 2.\\
\end{rem}

We are now ready to solve the PDEs and derive the optimal portfolios in the CARA and CRRA cases.

\subsection{Portfolio choice in the CARA case}

\subsubsection{The general method with $y$}

Following the results of Section 3, solving the optimal portfolio choice of the agent in the CARA case boils down to solving the linear parabolic PDE (\ref{eq:HJB_cara_phi}) with terminal condition (\ref{eq:terminal_CARA_phi}).\\

Because $G(t,\cdot)$ is affine in $y$ for all $t\in [0,T]$ in the Gaussian case, we easily see from the Feynman-Kac representation (\ref{eq:FK-cara}) that for all $t\in [0,T], \phi(t,\cdot)$ is a polynomial of degree $2$ (in $y$). However looking for that polynomial of degree $2$ in $y$ by using the PDE (\ref{eq:HJB_cara_phi}) or Eq. (\ref{eq:FK-cara}) is cumbersome. As we shall see, the main reason for this is that $\beta$ is in fact a more natural variable to solve the problem than $y$. In fact, a better ansatz than a general polynomial of degree $2$ in $y$ is the following:
\begin{equation}\label{eq:ansatz-phi-cara}
  \phi(t,y) = a(t) + \frac 12 G(t,y)' B(t) G(t,y),
\end{equation}
where $a(t) \in \mathbb{R}$ and $B(t) \in S_d(\mathbb{R})$.\\

We indeed have the following proposition:
\begin{prop}
\label{prop:phi_a_b_CARA1}Assume there exists $a\in C^{1}\left(\left[0,T\right]\right)$
and $B\in C^{1}\left(\left[0,T\right],S_{d}\left(\mathbb{R}\right)\right)$
satisfying the following system of linear ODEs (for $t \in [0,T]$):
\begin{equation}
\left\{ \begin{array}{rcl}
\dot{a}\left(t\right)+\dfrac{1}{2}\mathrm{Tr}\left(\Gamma_{t}\Sigma^{-1}B\left(t\right)\Sigma^{-1}\Gamma_{t}\Sigma^{-1}\right) & = & 0\\
\dot{B}\left(t\right)-\Gamma_{t}\Sigma^{-1}B\left(t\right)-B\left(t\right)\Sigma^{-1}\Gamma_{t} + \dfrac{1}{\gamma}\Sigma & = & 0,
\end{array}\right.\label{eq:ode_cara1}
\end{equation}
with terminal condition
\begin{equation}
\left\{ \begin{array}{rcl}
a\left(T\right) & = & 0\\
B\left(T\right) & = & 0.
\end{array}\right.\label{eq:terminal_ab_cara1}
\end{equation}
Then, the function $\phi$ defined by (\ref{eq:ansatz-phi-cara}) satisfies (\ref{eq:HJB_cara_phi}) with terminal condition (\ref{eq:terminal_CARA_phi}).
\end{prop}

\begin{proof}
Let us consider $(a,B)$ solution of (\ref{eq:ode_cara1}) with terminal condition \eqref{eq:terminal_ab_cara1}. For $\phi$ defined by (\ref{eq:ansatz-phi-cara}), we have, by using the formulas of Proposition \ref{diffprop} that
\begin{eqnarray*}
&&\partial_t\phi
+\left(\nabla_y\phi\right)'
	\left(
	r\vec{1}
	- \frac{1}{2}
	\sigma\odot \sigma
	\right)
+\frac12
	\mathrm{Tr}
	\left(\Sigma
	\nabla^2_ {yy} \phi
	\right)
+
\frac{1}{2\gamma}
G'\Sigma G\\
&=& \dot{a} + \frac 12 G'\dot{B} G + \frac 12\partial_t G'BG + \frac 12 G'B\partial_tG + (D_yGBG)'\left(
	r\vec{1}
	- \frac{1}{2}
	\sigma\odot \sigma
	\right)\\
&& + \frac12
	\mathrm{Tr}
	\left(\Sigma
	D_yGBD_yG
	\right) + \frac{1}{2\gamma}
G'\Sigma G\\
&=& \dot{a} + \frac 12 G'\dot{B} G - \frac 12 G'\Gamma_t\Sigma^{-1}B G - \frac 12 G'B \Sigma^{-1} \Gamma_t G + \frac12
	\mathrm{Tr}
	\left(\Gamma_t \Sigma^{-1}
	B\Sigma^{-1} \Gamma_t \Sigma^{-1}
	\right) + \frac{1}{2\gamma}
G'\Sigma G\\
&=& \left(\dot{a} + \frac12
	\mathrm{Tr}
	\left(\Gamma_t \Sigma^{-1}
	B\Sigma^{-1} \Gamma_t \Sigma^{-1}
	\right)\right) + \frac 12 G'\left(\dot{B} - \Gamma_t\Sigma^{-1}B - B \Sigma^{-1} \Gamma_t + \frac 1\gamma \Sigma  \right)G\\
&=&0.
\end{eqnarray*}
Therefore $\phi$ is solution of the PDE (\ref{eq:HJB_cara_phi}) and it satisfies obviously the terminal condition (\ref{eq:terminal_CARA_phi}).
\end{proof}

The system of linear ODEs \eqref{eq:ode_cara1} with terminal condition \eqref{eq:terminal_ab_cara1} can be solved in closed form. This is the purpose of the following proposition.\\

\begin{prop}
\label{prop:ab_cara}The functions $a$ and $B$ defined (for $t \in [0,T]$) by
\begin{eqnarray*}
a\left(t\right) & = & \dfrac{1}{2\gamma}\int_{t}^{T}\mathrm{Tr}\left(\Sigma^{-1}\left(\Gamma_{s}-\Gamma_{T}\right)\right)ds\\
B\left(t\right) & = & \dfrac{1}{\gamma}\Sigma\left(\Gamma_{t}^{-1}-\Gamma_{t}^{-1}\Gamma_{T}\Gamma_{t}^{-1}\right)\Sigma
\end{eqnarray*}
satisfy the system (\ref{eq:ode_cara1}) with terminal condition
(\ref{eq:terminal_ab_cara1}).
\end{prop}

Wrapping up our results, we can state the optimal portfolio choice of an agent with a CARA utility function in the case of the Gaussian prior (\ref{eq:priorgauss}).

\begin{prop}
\label{opt1}
In the case of the Gaussian prior (\ref{eq:priorgauss}), the optimal strategy $(M^{\star}_t)_{t \in [0,T]}$ of an agent with a CARA utility function is given by
\begin{eqnarray}
M^{\star}_t & = & e^{-r\left(T-t\right)}\dfrac{1}{\gamma}\Sigma^{-1}\Gamma_{T}\Gamma_{t}^{-1}\Sigma G(t,Y_t).\label{eq:optcara1}
\end{eqnarray}
\end{prop}

\begin{proof} Let us consider $a$ and $B$ as defined in Proposition \ref{prop:ab_cara}. Then, let us define $\phi$ by (\ref{eq:ansatz-phi-cara}). We know from the verification theorem (Theorem \ref{verif_cara_gl}) and especially from Eq. (\ref{eq:optimizer_CARA1}) that
\begin{eqnarray*}
  M^{\star}_t &=& e^{-r(T-t)}\left(
    \frac{G(t,Y_t)}{\gamma}-\nabla_y\phi (t,Y_t)
    \right)\\
    &=& e^{-r(T-t)}\left(
    \frac{G(t,Y_t)}{\gamma}-D_y G(t,Y_t)B(t)G(t,Y_t)
    \right)\\
    &=& e^{-r(T-t)}\dfrac{1}{\gamma} \left(
    I_d-\Sigma^{-1} \Gamma_t \Sigma^{-1}\Sigma\left(\Gamma_{t}^{-1}-\Gamma_{t}^{-1}\Gamma_{T}\Gamma_{t}^{-1}\right)\Sigma
    \right)G(t,Y_t)\\
    &=& e^{-r(T-t)}\dfrac{1}{\gamma} \Sigma^{-1} \Gamma_{T}\Gamma_{t}^{-1} \Sigma G(t,Y_t).
\end{eqnarray*}
\end{proof}

We see from the form (\ref{eq:ansatz-phi-cara}) of the solution $\phi$ and from Eq. (\ref{eq:optcara1}) that $G(t,Y_t)$ rather than $Y_t$ itself is the driver of the optimal behavior of the agent at time $t$. Because of Eq. (\ref{eq:eq42}), this means that $\beta$ rather than $y$ is the natural variable for addressing the problem. In what follows, we solve the optimal portfolio choice problem in the case of a Gaussian prior by taking another route, on which the dynamics of the system is given by the stochastic differential equations (\ref{system2}) rather than (\ref{system1}).\\

\subsubsection{Solving the problem using $\beta$}

On our first route for solving the above optimal portfolio choice problem, the central equation was the HJB equation (\ref{eq:HJB_cara_u}) associated with the stochastic differential equations (\ref{system1}). Instead of using the stochastic differential equations (\ref{system1}), we now reconsider the problem in the Gaussian prior case by using the stochastic differential equations (\ref{system2}).\footnote{We omit the proofs in this subsection. They are similar to those presented in Section 3.}\\

The value function $\tilde{v}$ associated with the problem is now given by
\begin{eqnarray*}
\tilde{v}:\left(t,V,\beta\right)\in [0,T]\times\mathbb{R}\times\mathbb{R}^d & \mapsto & \sup_{M\in\tilde{\mathcal{A}}_{t}}\mathbb{E}\left[-\exp\left(-\gamma V_{T}^{t,V,\beta,M}\right)\right],
\end{eqnarray*}
where the set of admissible strategies is
\begin{eqnarray*}
\tilde{\mathcal{A}}_{t} & = & \left\{ \left(M_{s}\right)_{s\in\left[t,T\right]}, \mathbb{R}^{d}\textrm{-valued\;} \mathcal{F}^{S}\textrm{-adapted process}\right.\\
&& \left.\textrm{satisfying the linear growth condition with respect to } (\beta_s)_{s \in [t,T]}\right\},
\end{eqnarray*}
and where, $\forall t \in [0,T], \forall (V,\beta,M) \in \mathbb{R}\times\mathbb{R}^d \times \tilde{\mathcal{A}}_t, \forall s \in [t,T],$
\begin{flalign}
\beta_s^{t,\beta}&=\beta+\int_t^s \Gamma_\tau \Sigma^{-1} \left(\sigma \odot
d \widehat W_\tau\right),
&\\
V_{s}^{t,V,\beta,M} & =  V+\int_{t}^{s}\left(M_{\tau}'(\beta^{t,\beta}_\tau - r\vec{1})+rV_{\tau}^{t,V,\beta,M}\right)d\tau+\int_{t}^{s}M_{\tau}'(\sigma \odot d\widehat{W}_{\tau}).
\end{flalign}

It is noteworthy that for all $t \in [0,T]$, $\tilde{\mathcal{A}}_{t} = \mathcal{A}_t$. There is indeed no difference between the linear growth condition with respect to $\beta$ and the linear growth condition with respect to $Y$ in the Gaussian prior case. This is easy to see on Eq. (\ref{eq:betagauss}), recalling that $\beta_0$ and $Y_0$ are known constants.\\

The HJB equation associated with this optimization problem is
\vspace{1cm}
$$0= \partial_{t}\tilde{u}\left(t,V,\beta\right)+\dfrac{1}{2}\mathrm{Tr}\left(\Gamma_{t}\Sigma^{-1}\Gamma_{t}\nabla^2_{\beta\beta}\tilde{u}\left(t,V,\beta\right)\right)$$
\begin{equation}
+\sup_{M\in \mathbb{R}^d}\left\{ \left(M'\left(\beta-r\vec{1}\right)+rV\right)\partial_{V}\tilde{u}\left(t,V,\beta\right)+\dfrac{1}{2}M'\Sigma M\partial^2_{VV}\tilde{u}\left(t,V,\beta\right)+M'\Gamma_{t}\partial_{V} \nabla_\beta\tilde{u}\left(t,V,\beta\right)\right\},
\label{eq:HJB_cara_u_beta}
\end{equation}
with terminal condition
\begin{eqnarray}
\forall V\in\mathbb{R},\forall \beta\in\mathbb{R}^d,
\quad
\tilde{u}\left(T,V,\beta\right) & = & -\exp\left(-\gamma V\right).\label{eq:terminal_CARA_u_beta}
\end{eqnarray}

By using the ansatz \begin{eqnarray}
\label{eq:ansatz_cara_beta}
\tilde{u}\left(t,V,\beta\right) & = & -\exp\left[-\gamma\left(e^{r\left(T-t\right)}V+\tilde{\phi}\left(t,\beta\right)\right)\right],
\end{eqnarray} we obtain the following proposition:

\begin{prop}
\label{prop:u-phi_CARA_nd}Suppose there exists $\tilde{\phi}\in C^{1,2}\left(\left[0,T\right]\times\mathbb{R}^{d}\right)$
satisfying

$\forall (t,\beta) \in [0,T]\times\mathbb{R}^{d}$,
\begin{eqnarray}
\nonumber \partial_{t}\tilde{\phi}\left(t,\beta\right)+\dfrac{1}{2}\mathrm{Tr}\left(\Gamma_{t}\Sigma^{-1}\Gamma_{t}\nabla^2_{\beta\beta}\tilde{\phi}\left(t,\beta\right)\right)+\dfrac{1}{2\gamma}\left(\beta-r\vec{1}\right)'\Sigma^{-1}\left(\beta-r\vec{1}\right)&&\\
-\left(\Gamma_{t}\nabla_{\beta}\tilde{\phi}\left(t,\beta\right)\right)'\Sigma^{-1}\left(\beta-r\vec{1}\right) &=&  0,\label{eq:HJB_cara_phi_beta}
\end{eqnarray}
with terminal condition
\begin{eqnarray}
\forall \beta\in\mathbb{R}^d,
\quad
\tilde{\phi}\left(T,\beta\right) & = & 0.\label{eq:terminal_CARA_phi_beta}
\end{eqnarray}
Then $\tilde{u}$ defined by (\ref{eq:ansatz_cara_beta}) is solution of the HJB equation (\ref{eq:HJB_cara_u_beta}) with terminal
condition (\ref{eq:terminal_CARA_u_beta}).\\
Moreover, the supremum in (\ref{eq:HJB_cara_u_beta}) is achieved at
\begin{eqnarray}
M^{\star}(t,\beta) & = & e^{-r\left(T-t\right)}\Sigma^{-1}\left(\dfrac{\left(\beta-r\vec{1}\right)}{\gamma}-\Gamma_{t}\nabla_{\beta}\tilde{\phi}\left(t,\beta\right)\right).\label{optimizer_CARA1_beta}
\end{eqnarray}
\end{prop}
For solving Eq. \eqref{eq:HJB_cara_phi_beta} with terminal condition \eqref{eq:terminal_CARA_phi_beta}, it is natural\footnote{It is clear from the form of Eq. \eqref{eq:HJB_cara_phi_beta} that the solution is a polynomial of degree 2 in $\beta-r\vec{1}$.} to consider the following ansatz:
\begin{equation}\label{eq:ansatz-phi-cara_beta}
\tilde{\phi}\left(t,\beta\right)  =  \tilde{a}\left(t\right)+\dfrac{1}{2}\left(\beta-r\vec{1}\right)'\tilde{B}\left(t\right)\left(\beta-r\vec{1}\right),
\end{equation}
where $\tilde{a}(t) \in \mathbb{R}$ and $\tilde{B}(t) \in S_d(\mathbb{R})$.\\

The next proposition states the ODEs that $\tilde{a}$ and $\tilde{B}$ must satisfy.
\begin{prop}
\label{prop:phi_a_b_CARA_beta}Assume there exists $\tilde{a}\in{C}^{1}\left(\left[0,T\right]\right)$
and $\tilde{B}\in{C}^{1}\left(\left[0,T\right],S_{d}\left(\mathbb{R}\right)\right)$
satisfying the following system of linear ODEs (for $t \in [0,T]$):
\begin{equation}
\left\{ \begin{array}{rcl}
\dot{\tilde{a}}\left(t\right)+\dfrac{1}{2}\mathrm{Tr}\left(\Gamma_{t}\Sigma^{-1}\Gamma_{t}\tilde{B}\left(t\right)\right) & = & 0\\
\dot{\tilde{B}}\left(t\right)-\Sigma^{-1}\Gamma_{t}\tilde{B}\left(t\right)-\tilde{B}\left(t\right)\Gamma_{t}\Sigma^{-1} + \dfrac{1}{\gamma}\Sigma^{-1} & = & 0,
\end{array}\right.\label{eq:ode_cara_beta}
\end{equation}
with terminal condition
\begin{equation}
\left\{ \begin{array}{rcl}
\tilde{a}\left(T\right) & = & 0\\
\tilde{B}\left(T\right) & = & 0.
\end{array}\right.\label{eq:terminal_ab_cara_beta}
\end{equation}
Then, the function $\tilde{\phi}$ defined by (\ref{eq:ansatz-phi-cara_beta}) satisfies (\ref{eq:HJB_cara_phi_beta}) with terminal condition (\ref{eq:terminal_CARA_phi_beta}).
\end{prop}

The system of linear ODEs \eqref{eq:ode_cara_beta} with terminal condition \eqref{eq:terminal_ab_cara_beta} can be solved in closed form. This is the purpose of the following proposition.

\begin{prop}
\label{prop:ab_cara_beta}The functions $\tilde{a}$ and $\tilde{B}$ defined, for $t \in [0,T]$ by
\begin{eqnarray*}
\tilde{a}\left(t\right) & = & \dfrac{1}{2\gamma}\int_{t}^{T}\mathrm{Tr}\left(\Sigma^{-1}\left(\Gamma_{s}-\Gamma_{T}\right)\right)ds\\
\tilde{B}\left(t\right) & = & \dfrac{1}{\gamma}\left(\Gamma_{t}^{-1}-\Gamma_{t}^{-1}\Gamma_{T}\Gamma_{t}^{-1}\right)
\end{eqnarray*}
satisfy the system (\ref{eq:ode_cara_beta}) with terminal condition
(\ref{eq:terminal_ab_cara_beta}).
\end{prop}

We are now ready to state the main result of this subsection, whose proof is similar to that of Theorem \ref{verif_cara_gl}.

\begin{thm}
\label{verif_opt_beta}
Let us consider $\tilde{a}$ and $\tilde{B}$ as defined in Proposition \ref{prop:ab_cara_beta}. Let us then define $\tilde{\phi}$ by (\ref{eq:ansatz-phi-cara_beta}) and, subsequently, $\tilde{u}$ by (\ref{eq:ansatz_cara_beta}).\\

For all $\left(t,V,\beta\right)\in\left[0,T\right]\times\mathbb{R}\times\mathbb{R}^{d}$
and $M = (M_s)_{s \in [t,T]}\in\tilde{\mathcal{A}}_t$, we have
\begin{eqnarray}
\mathbb{E}\left[-\exp\left(-\gamma V_{T}^{t,V,\beta,M}\right)\right] & \leq & \tilde{u}\left(t,V,\beta\right).\label{eq:verif_ineq_cara_beta}
\end{eqnarray}
Moreover, equality in (\ref{eq:verif_ineq_cara_beta}) is obtained by taking the
optimal control $(M^{\star}_s)_{s \in [t,T]}\in\tilde{\mathcal{A}}_t$ given by
\begin{equation}
\forall s \in [t,T],\quad M^{\star}_s = e^{-r\left(T-s\right)}\dfrac{1}{\gamma}\Sigma^{-1}\Gamma_{T}\Gamma_{s}^{-1}(\beta_s - r \vec{1}).\label{eq:optcara1_beta}
\end{equation}
In particular $\tilde{u}=\tilde{v}$.
\end{thm}

\subsubsection{Comments on the results: understanding the learning-anticipation effect}

In the case of an agent maximizing an expected CARA utility objective function, the optimal portfolio allocation is given by
\begin{equation}
\label{M_cara_opt}
\forall t \in [0,T], \quad M_{t}^{\star}  =  e^{-r\left(T-t\right)}\dfrac{1}{\gamma}\Sigma^{-1}\Gamma_{T}\Gamma_{t}^{-1}\left(\beta_{t}-r\vec{1}\right).
\end{equation}

Of course, if $\mu$ was known, the optimal strategy would be
\begin{equation}
\label{M_cara_opt_mu}
\forall t \in [0,T], \quad M^{\star, \mu_\textrm{known}}_{t} = e^{-r\left(T-t\right)}\dfrac{1}{\gamma}\Sigma^{-1}\left(\mu_\textrm{known}-r\vec{1}\right).
\end{equation}

It is essential to notice that the optimal strategy does not boil down (except at time $t=T$) to the naive strategy
\begin{equation}
\label{M_cara_naive}
\forall t \in [0,T], \quad M_{t,\textrm{naive}} = e^{-r\left(T-t\right)}\dfrac{1}{\gamma} \Sigma^{-1}\left(\beta_{t}-r\vec{1}\right),
\end{equation} which consists in replacing, at time $t$, $\mu_\textrm{known}$ by the current estimator $\beta_t$ in Eq. (\ref{M_cara_opt_mu}).\\

The sub-optimality of the naive strategy comes from the fact that the agent does not only learn but knows that he will go on learning in the future, and uses that knowledge to design his investment strategy. We call this effect the learning-anticipation effect.\\

To better understand this learning-anticipation effect, it is interesting to study the case $d=1$. In that case, let us denote by $\sigma$ the volatility of the risky asset and let us assume that the prior distribution for $\mu$ is $\mathcal{N}(\beta_0, \nu_0^2)$, where $\nu_0 > 0$. The agent following the optimal strategy invests at time $t$ the amount
$$M_{t}^{\star} = e^{-r\left(T-t\right)}\frac{\sigma^2 + \nu_0^2 t}{\sigma^2 + \nu_0^2 T}\dfrac{\beta_{t}-r}{\gamma\sigma^{2}}$$
in the risky asset, whereas the naive strategy would consist instead in investing the amount
$$M_{t,\textrm{naive}} = e^{-r\left(T-t\right)}\dfrac{\beta_t-r}{\gamma\sigma^{2}}.$$

The magnitude of the learning-anticipation effect can be measured by the multiplier $\chi = \frac{\sigma^2 + \nu_0^2 t}{\sigma^2 + \nu_0^2 T}$. $\chi \in [0,1]$, and the further from $1$ the multiplier (i.e., the smaller in this case), the larger the learning-anticipation effect.\\

$\chi$ is an increasing function of $t$ with $\chi=1$ at time $t=T$. This means that the agent invests less (in absolute value) in the risky asset than he would if he opted for the naive strategy, except at time $T$ because there is nothing more to learn. In other words, he is prudent and waits for more precise estimates of the drift.\\

$\chi$ is also an increasing function of $\sigma$. The smaller $\sigma$, the more important the learning-anticipation effect. When volatility is low, it is really valuable to wait for a good estimate of $\mu$ before investing.\\

$\chi$ is a decreasing function of $\nu_0$ and $T$. The longer the investment horizon and the higher the uncertainty about the value of the drift, the stronger the incentive of the agent to start with a small exposure (in absolute value) in the risky asset and to observe the behavior of the risky asset before adjusting his exposure, \emph{ceteris paribus}.

\subsection{Portfolio choice in the CRRA case}

\subsubsection{The general method with $y$}

In Section 3, and more precisely in Theorem \ref{verif_crra_gl_log}, we have seen that an agent with a $\log$ utility function has an optimal investment strategy that depends on the prior only through $G$. There is therefore no need to solve PDEs.\\

In the CRRA case, when $\gamma \neq 1$, following the results of Section 3, we see that solving the optimal portfolio choice of the agent boils down to solving the linear parabolic PDE (\ref{eq:HJB_crra_phi}) with terminal condition (\ref{eq:terminal_crra_phi}).\\

In order to solve this equation, we consider the following ansatz:
\begin{equation}\label{eq:ansatz-phi-crra}
  \phi(t,y) = \exp\left(a(t) + \frac 12 G(t,y)' B(t) G(t,y)\right),
\end{equation}
where $a(t) \in \mathbb{R}$ and $B(t) \in S_d(\mathbb{R})$.\\

We have the following proposition:
\begin{prop}
\label{prop:phi_a_b_CRRA1}Assume there exists $a\in C^{1}\left(\left[0,T\right]\right)$
and $B\in C^{1}\left(\left[0,T\right],S_{d}\left(\mathbb{R}\right)\right)$
satisfying the following system of linear ODEs (for $t \in [0,T]$):
\begin{equation}
\left\{ \begin{array}{rcl}
\dot{a}\left(t\right)+\dfrac{1}{2}\mathrm{Tr}\left(\Gamma_{t}\Sigma^{-1}B\left(t\right)\Sigma^{-1}\Gamma_{t}\Sigma^{-1}\right) & = & 0\\
\dot{B}\left(t\right)+ \frac{1-\gamma}{\gamma}\Gamma_{t}\Sigma^{-1}B\left(t\right) + \frac{1-\gamma}{\gamma} B\left(t\right)\Sigma^{-1}\Gamma_{t} + B(t) \Sigma^{-1} \Gamma_{t}\Sigma^{-1} \Gamma_{t}\Sigma^{-1} B(t) + \dfrac{1-\gamma}{\gamma^2}\Sigma & = & 0,
\end{array}\right.\label{eq:ode_crra1}
\end{equation}
with terminal condition
\begin{equation}
\left\{ \begin{array}{rcl}
a\left(T\right) & = & 0\\
B\left(T\right) & = & 0.
\end{array}\right.\label{eq:terminal_ab_crra1}
\end{equation}
Then, the function $\phi$ defined by (\ref{eq:ansatz-phi-crra}) satisfies (\ref{eq:HJB_crra_phi}) with terminal condition (\ref{eq:terminal_crra_phi}).
\end{prop}

\begin{proof}
Let us consider $(a,B)$ solution of (\ref{eq:ode_crra1}) with terminal condition \eqref{eq:terminal_ab_crra1}. For $\phi$ defined by (\ref{eq:ansatz-phi-crra}), we have, by using the formulas of Proposition \ref{diffprop},
\begin{eqnarray*}
&&\partial_t\phi
+\left(\nabla_y\phi\right)'
	\left(
	r\vec{1} + \frac 1\gamma \Sigma G
	- \frac{1}{2}
	\sigma\odot \sigma
	\right)
+\frac12
	\mathrm{Tr}
	\left(\Sigma
	\nabla^2_ {yy} \phi
	\right)
+
\frac{1-\gamma}{2\gamma^2}
G'\Sigma G \phi\\
&=& \left(\dot{a} + \frac 12 G'\dot{B} G + \frac 12\partial_t G'BG + \frac 12 G'B\partial_tG\right)\phi + (D_yGBG)'\left(
	r\vec{1} + \frac 1\gamma \Sigma G
	- \frac{1}{2}
	\sigma\odot \sigma
	\right)\phi\\
&& + \frac12
	\mathrm{Tr}
	\left(\Sigma
	D_yGBD_yG
	\right)\phi + \frac12
	\mathrm{Tr}
	\left(\Sigma
	(D_yGBG)(D_yGBG)'
	\right)\phi + \frac{1-\gamma}{2\gamma^2}
G'\Sigma G \phi\\
&=& \left(\dot{a} + \frac 12 G'\dot{B} G - \frac 12 G'\Gamma_t\Sigma^{-1}B G - \frac 12 G'B \Sigma^{-1} \Gamma_t G\right) \phi + \frac 1\gamma G'B\Sigma^{-1} \Gamma_tG \phi\\
&&+ \frac12
	\mathrm{Tr}
	\left(\Gamma_t \Sigma^{-1}B\Sigma^{-1} \Gamma_t \Sigma^{-1}
	\right) \phi + \frac 12 G'B \Sigma^{-1} \Gamma_t \Sigma^{-1} \Gamma_t \Sigma^{-1} B G \phi + \frac{1-\gamma}{2\gamma^2}
G'\Sigma G \phi\\
&=& \left(\dot{a} + \frac12
	\mathrm{Tr}
	\left(\Gamma_t \Sigma^{-1}
	B\Sigma^{-1} \Gamma_t \Sigma^{-1}
	\right)\right)\phi\\
&&+ \frac 12 G'\left(\dot{B} - \Gamma_t\Sigma^{-1}B - B \Sigma^{-1} \Gamma_t + \frac 1\gamma B\Sigma^{-1} \Gamma_t + \frac 1\gamma \Gamma_t \Sigma^{-1} B\right.\\
&&\left. + B \Sigma^{-1} \Gamma_t \Sigma^{-1} \Gamma_t \Sigma^{-1} B + \frac{1-\gamma}{\gamma^2}
\Sigma \right)G \phi \\
&=&0.
\end{eqnarray*}
Therefore $\phi$ is solution of the PDE (\ref{eq:HJB_crra_phi}) and it satisfies obviously the terminal condition (\ref{eq:terminal_crra_phi}).
\end{proof}

The system of ODEs \eqref{eq:ode_crra1} is not a system of linear ODEs. The equation for $B$ is indeed a Riccati equation. Luckily, $t \mapsto - \frac 1\gamma \Sigma \Gamma_t^{-1} \Sigma$ is a trivial solution of the second differential equation of the system \eqref{eq:ode_crra1}. Therefore, using a classical trick of Riccati equations, we can look for a solution $B$ of the form
$$B(t) = - \frac 1\gamma \Sigma \Gamma_t^{-1} \Sigma + E(t)^{-1}, \quad \forall t \in [0,T]$$ where $E \in C^1([0,T],S_d(\mathbb{R}))$.\\

With this ansatz, looking for a solution $B$ to the above Riccati equation boils down to solving the linear ODE
\begin{equation}
\label{eq:E}\forall t \in [0,T], \quad \dot{E}(t) = \Sigma^{-1} \Gamma_t \Sigma^{-1} \Gamma_t \Sigma^{-1} - \left(\Sigma^{-1}\Gamma_t E(t) + E(t) \Gamma_t \Sigma^{-1} \right), \qquad E(T) = \gamma \Sigma^{-1}\Gamma_T \Sigma^{-1},
\end{equation}
and verifying that for all $t \in [0,T]$, $E(t)$ is invertible.\\

The unique solution to Eq. (\ref{eq:E}) is given in the following straightforward proposition:

\begin{prop}
\label{prop:E_crra}The function $E$ defined by
\begin{equation}
\label{eq:E_sol}
\forall t \in [0,T], \quad E(t) = \Sigma^{-1}\left(\Gamma_t + (\gamma-1) \Gamma_t \Gamma^{-1}_T \Gamma_t \right) \Sigma^{-1}
\end{equation}
is the unique solution of the Cauchy problem (\ref{eq:E}).
\end{prop}

Regarding the invertibility of $E(t)$ for all $t \in [0,T]$, we have the following result:

\begin{prop}
\label{prop:E_crra_inverse}
Let us consider $E$ as defined by Eq. (\ref{eq:E_sol}).\\
$E(t)$ is invertible for all $t \in [0,T]$ if and only if (i) $\gamma > 1$ or (ii) $\gamma < 1$ and $T < \frac{\gamma}{1-\gamma} \lambda_{\min}\left(\Sigma^{\frac12 }\Gamma_{0}^{-1}\Sigma^{\frac12 }\right)$, where the function $\lambda_{\min}(\cdot)$ maps a symmetric matrix to its lowest eigenvalue.\\
\end{prop}

\begin{proof}
Let us consider $t \in [0,T]$. $E(t)$ is invertible if and only if $\Gamma_{t}+ (\gamma - 1)\Gamma_{t}\Gamma_{T}^{-1}\Gamma_{t}$ is invertible.\\

If $\gamma>1$, then $\Gamma_{t}+ (\gamma - 1)\Gamma_{t}\Gamma_{T}^{-1}\Gamma_{t}$ is the sum of two positive definite symmetric matrices. It is therefore an invertible matrix.\\

If $\gamma<1$, then, using the definition of $(\Gamma_t)_{t \in [0,T]}$, we have
\begin{eqnarray*}
&&\Gamma_{t}+ (\gamma - 1)\Gamma_{t}\Gamma_{T}^{-1}\Gamma_{t}\\
&=& \Gamma_t \left(\Gamma_{t}^{-1}+ (\gamma - 1)\Gamma_{T}^{-1}\right)\Gamma_t\\
& = & \Gamma_t \left(\gamma \Gamma_{0}^{-1}+(t + (\gamma-1)T) \Sigma^{-1}\right)\Gamma_t\\
 & = & \gamma \Gamma_t \Sigma^{-\frac12}\left(\Sigma^{\frac12 }\Gamma_{0}^{-1}\Sigma^{\frac12 }+ \frac{t + (\gamma-1)T}{\gamma}I_d\right)\Sigma^{-\frac12}\Gamma_t.
\end{eqnarray*}
Therefore $E(t)$ is invertible if and only if $-\frac{t + (\gamma-1)T}{\gamma}$ is not eigenvalue of $\Sigma^{\frac12 }\Gamma_{0}^{-1}\Sigma^{\frac12}$, hence the result.
\end{proof}

The above result is very important. It means that when $\gamma < 1$ the solution of (\ref{eq:ode_crra1}) with terminal condition \eqref{eq:terminal_ab_crra1} blows up in finite time, in the sense that the solution can only be defined on an interval of the form $(\tau,T]$. If $T$ is small enough so that $\tau < 0$, then the solution exists on $[0,T]$. Otherwise, we are enable to solve (\ref{eq:ode_crra1}) with terminal condition \eqref{eq:terminal_ab_crra1} on $[0,T]$. When $d=1$, $B(t)$ is a scalar and it goes to $+\infty$ as $t \to \tau$. In particular, this means that the value function stops to be defined because it becomes infinite.\\

We are now ready to solve (\ref{eq:ode_crra1}) with terminal condition
(\ref{eq:terminal_ab_crra1}).

\begin{prop}
\label{prop:ab_crra}
Let us assume that either (i) $\gamma > 1$ or (ii) $\gamma < 1$ and $T < \frac{\gamma}{1-\gamma} \lambda_{\min}\left(\Sigma^{\frac12 }\Gamma_{0}^{-1}\Sigma^{\frac12 }\right)$.\\

Then, the functions $a$ and $B$ defined (for $t \in [0,T]$) by
\begin{eqnarray*}
a\left(t\right) & = & \dfrac{1}{2}\int_{t}^{T}\mathrm{Tr}\left(\Sigma^{-1}\left(-\frac 1\gamma\Gamma_{s}+\left(\Gamma_{s}^{-1}+(\gamma - 1)\Gamma_{T}^{-1}\right)^{-1}\right)\right)ds\\
B\left(t\right) & = & \Sigma\left(\left(\Gamma_{t}+(\gamma-1)\Gamma_{t}\Gamma^{-1}_{T}\Gamma_{t}\right)^{-1} - \frac 1\gamma \Gamma_t^{-1}\right)\Sigma
\end{eqnarray*}
satisfy the system (\ref{eq:ode_crra1}) with terminal condition
(\ref{eq:terminal_ab_crra1}).
\end{prop}

Wrapping up our results, we can state the optimal portfolio choice of an agent with a CRRA utility function with $\gamma \neq 1$ in the case of the Gaussian prior (\ref{eq:priorgauss}).

\begin{prop}
\label{opt1_crra}
Let us consider the Gaussian prior (\ref{eq:priorgauss}). Let us assume that either (i) $\gamma > 1$ or (ii) $\gamma < 1$ and $T < \frac{\gamma}{1-\gamma} \lambda_{\min}\left(\Sigma^{\frac12 }\Gamma_{0}^{-1}\Sigma^{\frac12 }\right)$.\\
Then, the optimal strategy $(\theta^{\star}_t)_{t \in [0,T]}$ of an agent with a CRRA utility function with $\gamma \neq 1$ is given by
\begin{eqnarray}
\theta^{\star}_t & = & \Sigma^{-1}\left(\Gamma_{t}^{-1}+\left(\gamma-1\right)\Gamma_{T}^{-1}\right)^{-1}\Gamma_{t}^{-1}\Sigma G(t,Y_t).\label{eq:optcrra1}
\end{eqnarray}
\end{prop}

\begin{proof} Let us consider $a$ and $B$ as defined in Proposition \ref{prop:ab_crra}. Then let us define $\phi$ by (\ref{eq:ansatz-phi-crra}). It is straightforward to see that $\phi$ satisfies \eqref{eq:gradlogphi}. Consequently, we know from the verification theorem (Theorem \ref{verif_crra_gl}) and especially from Eq. (\ref{eq:optcrraverif}) that
\begin{eqnarray*}
  \theta^{\star}_t &=& \frac{G(t,Y_t)}{\gamma}+\frac{\nabla_y\phi (t,Y_t)}{\phi (t,Y_t)}\\
    &=& \frac{G(t,Y_t)}{\gamma}+D_y G(t,Y_t)B(t)G(t,Y_t)\\
    &=& \left(
    \frac 1\gamma I_d+\Sigma^{-1} \Gamma_t \Sigma^{-1} \Sigma\left(\left(\Gamma_{t}+(\gamma-1)\Gamma_{t}\Gamma^{-1}_{T}\Gamma_{t}\right)^{-1} - \frac 1\gamma \Gamma_t^{-1}\right)\Sigma\right)G(t,Y_t)\\
    &=& \Sigma^{-1} \left(\Gamma^{-1}_{t}+(\gamma-1)\Gamma^{-1}_{T}\right)^{-1} \Gamma^{-1}_{t} \Sigma G(t,Y_t).
\end{eqnarray*}
\end{proof}

\begin{rem}
It is noteworthy that when $\gamma \to 1$, we recover the result of Theorem \ref{verif_crra_gl_log}, i.e. $\theta^{\star}_t = G(t,Y_t)$.
\end{rem}

As in the CARA case, we see from the form (\ref{eq:ansatz-phi-crra}) of the solution $\phi$ and from Eq. (\ref{eq:optcrra1}) that $G(t,Y_t)$ rather than $Y_t$ itself is the driver of the optimal behavior of the agent at time $t$. Because of Eq. (\ref{eq:eq42}), this means that $\beta$ rather than $y$ is the natural variable for addressing the problem. In what follows, we solve the optimal portfolio choice problem in the case of a Gaussian prior by taking another route, on which the dynamics of the system is given by the stochastic differential equations (\ref{system2}) rather than (\ref{system1}).\\

\subsubsection{Solving the problem using $\beta$}

On our first route for solving the above optimal portfolio choice problem, the central equation was the HJB equation (\ref{eq:HJB_crra_u}) associated with the stochastic differential equations (\ref{system1}). Instead of using the stochastic differential equations (\ref{system1}), we now reconsider the problem in the Gaussian prior case by using the stochastic differential equations (\ref{system2}).\footnote{We omit the proofs in this subsection. They are similar to those presented in Section 3.}\\

If $\gamma < 1$, we define for $t \in [0,T]$ the set
\begin{eqnarray*}
\mathcal{A}^\gamma_{t} & = & \left\{ \left(\theta_{s}\right)_{s\in\left[t,T\right]}, \mathbb{R}^{d}\textrm{-valued\;} \mathcal{F}^{S}\textrm{-adapted process}, \mathbb{E}\left[\int_t^T \theta^2_{s} ds\right] < +\infty \right\}.\\
\end{eqnarray*}

If $\gamma > 1$, we define for $t \in [0,T]$ the set
\begin{eqnarray*}
\tilde{\mathcal{A}}^\gamma_{t} & = & \left\{ \left(\theta_{s}\right)_{s\in\left[t,T\right]}, \mathbb{R}^{d}\textrm{-valued\;} \mathcal{F}^{S}\textrm{-adapted process}\right.\\
&& \left.\textrm{satisfying the linear growth condition with respect to } (\beta_s)_{s \in [t,T]}\right\}.
\end{eqnarray*}

As in the CARA case, we have in fact $\tilde{\mathcal{A}}^\gamma_{t} = \mathcal{A}^\gamma_{t}, \forall \gamma > 0$.\\

The value function $\tilde{v}$ associated with the problem is now given by
\begin{eqnarray*}
\tilde{v}:\left(t,V,\beta\right)\in [0,T]\times\mathbb{R}_+^*\times\mathbb{R}^d & \mapsto & \sup_{\theta\in\tilde{\mathcal{A}}^\gamma_{t}}\mathbb{E}\left[U^\gamma\left(V_{T}^{t,V,\beta,\theta}\right)\right],
\end{eqnarray*}
where, $\forall t \in [0,T], \forall (V,\beta,\theta) \in \mathbb{R}_+^*\times\mathbb{R}^d \times \tilde{\mathcal{A}}_t, \forall s \in [t,T],$
\begin{flalign}
\beta_s^{t,\beta}&=\beta+\int_t^s \Gamma_\tau \Sigma^{-1} \left(\sigma \odot
d \widehat W_\tau\right),
&\\
V_{s}^{t,V,\beta,\theta} & =  V+\int_{t}^{s}\left(\theta_{\tau}'(\beta^{t,\beta}_\tau - r\vec{1})+r\right)V_{\tau}^{t,V,\beta,\theta}d\tau+\int_{t}^{s}\theta_{\tau}' V_{\tau}^{t,V,\beta,\theta}(\sigma \odot d\widehat{W}_{\tau}).
\end{flalign}

The HJB equation associated with this optimization problem is
$$0= \partial_{t}\tilde{u}\left(t,V,\beta\right)+\dfrac{1}{2}\mathrm{Tr}\left(\Gamma_{t}\Sigma^{-1}\Gamma_{t}\nabla^2_{\beta\beta}\tilde{u}\left(t,V,\beta\right)\right)$$
\begin{equation}
+\sup_{\theta\in\mathbb{R}^d}\left\{ \left(\theta'\left(\beta-r\vec{1}\right)+r\right)V\partial_{V}\tilde{u}\left(t,V,\beta\right)+\dfrac{1}{2}\theta'\Sigma\theta V^{2}\partial^2_{VV}\tilde{u}\left(t,V,\beta\right)+\theta'\Gamma_{t}V\partial_{V} \nabla_\beta \tilde{u}\left(t,V,\beta\right)\right\} ,\label{eq:HJB_crra_u_beta}
\end{equation}
with terminal condition
\begin{eqnarray}
\forall V\in\mathbb{R}^*_+,\forall \beta\in\mathbb{R}^d,
\quad
\tilde{u}\left(T,V,\beta\right) & = & U^{\gamma}\left(V\right).\label{eq:terminal_crra_u_beta}
\end{eqnarray}

By using the ansatz \begin{eqnarray}
\label{eq:ansatz_crra_beta}
\tilde{u}\left(t,V,\beta\right) & = & U^\gamma\left(e^{r\left(T-t\right)}V\right)\tilde{\phi}\left(t,\beta\right)^\gamma,
\end{eqnarray} we obtain the following proposition:

\begin{prop}
\label{prop:u-phi_CRRA_beta}Suppose there exists $\tilde{\phi}\in C^{1,2}\left(\left[0,T\right]\times\mathbb{R}^{d}\right)$
satisfying

$\forall (t,\beta) \in [0,T]\times\mathbb{R}^{d}$,
\begin{eqnarray}
\partial_{t}\tilde{\phi}\left(t,\beta\right)+\dfrac{1}{2}\mathrm{Tr}\left(\Gamma_{t}\Sigma^{-1}\Gamma_{t}\nabla^2_{\beta\beta}\tilde{\phi}\left(t,\beta\right)\right)+\dfrac{1-\gamma}{2\gamma^2}\left(\beta-r\vec{1}\right)'\Sigma^{-1}\left(\beta-r\vec{1}\right)\nonumber \\
+\frac{1-\gamma}{\gamma}\left(\Gamma_{t}\nabla_{\beta}\tilde{\phi}\left(t,\beta\right)\right)'\Sigma^{-1}\left(\beta-r\vec{1}\right) & = & 0,\label{eq:HJB_crra_phi_beta}
\end{eqnarray}
with terminal condition
\begin{eqnarray}
\forall \beta\in\mathbb{R}^d,
\quad
\tilde{\phi}\left(T,\beta\right) & = & 1.\label{eq:terminal_CRRA_phi_beta}
\end{eqnarray}
Then $\tilde{u}$ defined by (\ref{eq:ansatz_crra_beta}) is solution of the HJB equation (\ref{eq:HJB_crra_u_beta}) with terminal
condition (\ref{eq:terminal_crra_u_beta}).\\
Moreover, the supremum in (\ref{eq:HJB_crra_u_beta}) is achieved at
\begin{eqnarray}
\theta^{\star}(t,\beta) & = & \Sigma^{-1}\left(\dfrac{\left(\beta-r\vec{1}\right)}{\gamma}+\Gamma_{t}\frac{\nabla_{\beta}\tilde{\phi}\left(t,\beta\right)}{\tilde{\phi}\left(t,\beta\right)}\right).\label{optimizer_CRRA1_beta}
\end{eqnarray}
\end{prop}
For solving Eq. \eqref{eq:HJB_crra_phi_beta} with terminal condition \eqref{eq:terminal_CRRA_phi_beta}, we consider the following ansatz:
\begin{equation}\label{eq:ansatz-phi-crra_beta}
\tilde{\phi}\left(t,\beta\right)  =  \exp\left(\tilde{a}\left(t\right)+\dfrac{1}{2}\left(\beta-r\vec{1}\right)'\tilde{B}\left(t\right)\left(\beta-r\vec{1}\right)\right),
\end{equation}
where $\tilde{a}(t) \in \mathbb{R}$ and $\tilde{B}(t) \in S_d(\mathbb{R})$.\\

The next proposition states the ODEs that $\tilde{a}$ and $\tilde{B}$ must satisfy.
\begin{prop}
\label{prop:phi_a_b_CRRA_beta}Assume there exists $\tilde{a}\in{C}^{1}\left(\left[0,T\right]\right)$
and $\tilde{B}\in{C}^{1}\left(\left[0,T\right],S_{d}\left(\mathbb{R}\right)\right)$
satisfying the following system of linear ODEs (for $t \in [0,T]$):
\begin{equation}
\left\{ \begin{array}{rcl}
\dot{\tilde{a}}\left(t\right)+\dfrac{1}{2}\mathrm{Tr}\left(\Gamma_{t}\Sigma^{-1}\Gamma_{t}\tilde{B}\left(t\right)\right) & = & 0\\
\dot{\tilde{B}}\left(t\right)+ \frac{1-\gamma}{\gamma}\Sigma^{-1}\Gamma_{t}B\left(t\right) + \frac{1-\gamma}{\gamma} B\left(t\right)\Gamma_{t}\Sigma^{-1} + B(t) \Gamma_{t}\Sigma^{-1} \Gamma_{t} B(t) + \dfrac{1-\gamma}{\gamma^2}\Sigma^{-1} & = & 0,
\end{array}\right.\label{eq:ode_crra_beta}
\end{equation}
with terminal condition
\begin{equation}
\left\{ \begin{array}{rcl}
\tilde{a}\left(T\right) & = & 0\\
\tilde{B}\left(T\right) & = & 0.
\end{array}\right.\label{eq:terminal_ab_crra_beta}
\end{equation}
Then, the function $\tilde{\phi}$ defined by (\ref{eq:ansatz-phi-crra_beta}) satisfies (\ref{eq:HJB_crra_phi_beta}) with terminal condition (\ref{eq:terminal_CRRA_phi_beta}).
\end{prop}

The system of linear ODEs \eqref{eq:ode_crra_beta} with terminal condition \eqref{eq:terminal_ab_crra_beta} can be solved in closed form on $[0,T]$ when there is no blowup. This is the purpose of the following proposition.

\begin{prop}
\label{prop:ab_crra_beta}
Let us assume that either (i) $\gamma > 1$ or (ii) $\gamma < 1$ and $T < \frac{\gamma}{1-\gamma} \lambda_{\min}\left(\Sigma^{\frac12 }\Gamma_{0}^{-1}\Sigma^{\frac12 }\right)$.\\

Then, the functions $\tilde{a}$ and $\tilde{B}$ defined, for $t \in [0,T]$ by\\
\begin{eqnarray*}
\tilde a\left(t\right) & = & \dfrac{1}{2}\int_{t}^{T}\mathrm{Tr}\left(\Sigma^{-1}\left(-\frac 1\gamma\Gamma_{s}+\left(\Gamma_{s}^{-1}+(\gamma - 1)\Gamma_{T}^{-1}\right)^{-1}\right)\right)ds\\
\tilde B\left(t\right) & = & \left(\left(\Gamma_{t}+(\gamma-1)\Gamma_{t}\Gamma^{-1}_{T}\Gamma_{t}\right)^{-1} - \frac 1\gamma \Gamma_t^{-1}\right)
\end{eqnarray*}
satisfy the system (\ref{eq:ode_crra_beta}) with terminal condition
(\ref{eq:terminal_ab_crra_beta}).
\end{prop}

We are now ready to state the main result of this subsection, whose proof is similar to that of Theorem \ref{verif_crra_gl}.

\begin{thm}
\label{verif_optcrra_beta}
Let us assume that either (i) $\gamma > 1$ or (ii) $\gamma < 1$ and $T < \frac{\gamma}{1-\gamma} \lambda_{\min}\left(\Sigma^{\frac12 }\Gamma_{0}^{-1}\Sigma^{\frac12 }\right)$.\\

Let us consider $\tilde{a}$ and $\tilde{B}$ as defined in Proposition \ref{prop:ab_crra_beta}. Let us then define $\tilde{\phi}$ by (\ref{eq:ansatz-phi-crra_beta}) and, subsequently, $\tilde{u}$ by (\ref{eq:ansatz_crra_beta}).\\

For all $\left(t,V,\beta\right)\in\left[0,T\right]\times\mathbb{R}_+^*\times\mathbb{R}^{d}$
and $\theta = (\theta_s)_{s \in [t,T]}\in\tilde{\mathcal{A}}_t$, we have
\begin{eqnarray}
\mathbb{E}\left[U^\gamma\left(V_{T}^{t,V,\beta,\theta}\right)\right] & \leq & \tilde{u}\left(t,V,\beta\right).\label{eq:verif_ineq_crra_beta}
\end{eqnarray}
Moreover, equality in (\ref{eq:verif_ineq_crra_beta}) is obtained by taking the
optimal control $(\theta^{\star}_s)_{s \in [t,T]}\in\tilde{\mathcal{A}}_t$ given by
\begin{equation}
\forall s \in [t,T],\quad \theta^{\star}_s = \Sigma^{-1}\left(\Gamma_{s}^{-1}+\left(\gamma-1\right)\Gamma_{T}^{-1}\right)^{-1}\Gamma_{s}^{-1}(\beta_s - r \vec{1}).\label{eq:optcrra1_beta}
\end{equation}
In particular $\tilde{u}=\tilde{v}$.
\end{thm}

\subsubsection{Comments on the results: beyond the learning-anticipation effect}

In the case of an agent maximizing an expected CRRA utility objective function, the optimal portfolio allocation is given by the formula
\begin{eqnarray*}
\theta_{t}^{\star} & = & \Sigma^{-1}\left(\Gamma_{t}^{-1}+\left(\gamma-1\right)\Gamma_{T}^{-1}\right)^{-1}\Gamma_{t}^{-1}\left(\beta_{t}-r\vec{1}\right),
\end{eqnarray*}
whenever either (i) $\gamma \ge 1$ or (ii) $\gamma < 1$ and $T < \frac{\gamma}{1-\gamma} \lambda_{\min}\left(\Sigma^{\frac12 }\Gamma_{0}^{-1}\Sigma^{\frac12 }\right)$.

When $\gamma \neq1$, the optimal strategy does not boil to the naive strategy $$\theta_{t,\textrm{naive}} = \dfrac{1}{\gamma} \Sigma^{-1}\left(\beta_{t}-r\vec{1}\right).$$ However, it does in the case of a logarithmic utility function (i.e., $\gamma = 1$). This means that there is no learning-anticipation effect in the case of an agent with a $\log$ utility.\\

As in the CARA case, it is interesting to analyze the formula in the one-asset case $d=1$. In that case, let us denote by $\sigma$ the volatility of the risky asset and let us assume that the prior distribution for $\mu$ is $\mathcal{N}(\beta_0, \nu_0^2)$, where $\nu_0 > 0$. The agent following the optimal strategy invests at time $t$ a proportion of his wealth
$$\theta_{t}^{\star} = \frac{\sigma^2 + \nu_0^2 t}{\sigma^2 + \nu_0^2 \frac{t+(\gamma-1)T}{\gamma}}\dfrac{\beta_{t}-r}{\gamma\sigma^{2}}$$
in the risky asset, whereas the naive strategy would consist instead in investing the proportion
$$\theta_{t,\textrm{naive}} =  \dfrac{\beta_t-r}{\gamma\sigma^{2}}.$$

When $\gamma > 1$, we observe a learning-anticipation effect similar to that of the CARA case. It is measured by the multiplier $\chi = \frac{\sigma^2 + \nu_0^2 t}{\sigma^2 + \nu_0^2 \frac{t+(\gamma-1)T}{\gamma}} \in  [0,1]$. $\chi$ is an increasing function of $t$. This means that the agent invests less (in absolute value) in the risky asset than he would if he opted for the naive strategy, except at time $T$ because there is nothing more to learn. $\chi$ is also an increasing function of $\sigma$. The smaller $\sigma$, the more important the learning-anticipation effect. When volatility is low, it is really valuable to wait for a good estimate of $\mu$ before investing. $\chi$ is eventually a decreasing function of $\nu_0$ and $T$. The longer the investment horizon and the higher the uncertainty about the value of the drift, the stronger the incentive of the agent to start with a small exposure (in absolute value) in the risky asset and to observe the behavior of the risky asset before adjusting his exposure, \emph{ceteris paribus}.\\

All the above effects are in line with the CARA case: the agent is prudent and waits to know more. However, the effects are reversed when $\gamma < 1$. In that case indeed, the multiplier $\chi$ ceases to be below $1$. Instead, it is above $1$. In fact, the very possibility that expected returns could be very high (or very low because we can short) creates an incentive for the agent to have a higher exposure in the risky asset. Then, as the uncertainty reduces through learning, the agent adjusts his position towards a milder one and ends up with the same position as in the naive strategy when $t=T$.\\

It is noteworthy that $\chi$ at time $0$ tends to $+\infty$ when $\gamma$ tends to $\frac{\nu_0^2}{\sigma^2 + \nu_0^2 T}$, and this corresponds to the threshold found in Proposition \ref{prop:E_crra_inverse} for the blowup occurring exactly at time $t=0$. This means, if $\beta_0 > r$, that the agent wants to borrow an infinite amount of money at time $0$ to invest in the risky asset.\\

This reversed phenomenon is linked to the qualitative difference between the power utility functions when $\gamma >1$, which are bounded from above, and the power utility functions when $\gamma <1$, which have no upper bound. This difference explains why, for $\gamma < 1$ and for a Gaussian prior distribution (which is unbounded), the multiplier $\chi$ and the value function can blowup to $+\infty$ and therefore stop to be defined if $T$ is too large (or equivalently if $\gamma$ is too small when $T$ is fixed).\\

\section{Optimal portfolio choice, portfolio liquidation, and portfolio transition with online learning and execution costs}

The results presented above have been obtained by using PDE methods only. It is noteworthy that one could have derived the same formulas by using the martingale method of Karatzas and Zhao \cite{karatzas1998bayesian}. However the martingale method requires a model in which there are martingales, and there are many problems in which martingales cannot be exhibited. The goal of this section is to show how PDEs can be used to address problems for which the martingale method cannot be applied.\\

The classical literature on portfolio choice and asset allocation mainly considers frictionless markets. In that case, both PDE methods and martingale methods can be used for solving the problem, because there exists an equivalent probability measure under which discounted prices, and therefore discounted portfolio values, are martingales. Martingale methods cannot be used however when one adds frictions in the model. In what follows, we consider frictions in the form of execution costs, as in optimal execution models \emph{à la} Almgren-Chriss (see \cite{almgren1999value,almgren2001optimal}). We show that the PDE method presented in the previous sections enables to address the optimal portfolio choice problem, but also optimal portfolio liquidation and optimal portfolio transition problems, when there are execution costs and when one learns the value of the drift over the course of the optimization problem.\\

We first present the modelling framework and a generic optimization problem encompassing the three types of problem we consider. We then derive the associated HJB equation and derive a simpler PDE using an ansatz. We then focus on the specific case in which (i) the prior distribution of the drift is Gaussian and (ii) the execution costs and penalty functions are quadratic, because in that case the PDE boils down to a system of ODEs that can be solved numerically. We then show some numerical examples for each of the problems.\\

\subsection{Notations and setup of the model}

\subsubsection{Price dynamics and Bayesian learning of the drift}

As above we consider a financial market with one risk-free asset and $d$ risky assets. In order to simplify the equations, we assume that the risk-free asset yields no interest. It is noteworthy that the model can easily be generalized to the case of a non-zero risk-free interest rate $r$.\\

We index by $i \in \left\{1,\hdots,d\right\}$ the $d$ risky assets. For $i\in\left\{1,\hdots,d\right\} $, the price of the $i^\text{th}$ risky asset $S^i$ has the following drifted Bachelier dynamics\footnote{Unlike in the previous sections where we used the classical Black-Scholes (log-normal) dynamics, we consider here the Bachelier dynamics. This dynamics is indeed standard in the optimal execution literature, although it raises the problem of negative prices.}
\begin{eqnarray}
\forall i\in\left\{1,\hdots,d\right\},\quad
dS^i_{t} & = & \mu^i dt+\sigma^i dW^i_{t},\label{eq:s_eds-1}
\end{eqnarray}
where the volatility vector $\sigma = (\sigma^{1}, \ldots, \sigma^d)'$ satisfies $\forall i \in \left\{1,\hdots,d\right\}, \sigma^{i} > 0$, and where the drift vector $\mu=(\mu^{1}, \ldots, \mu^d)'$ is unknown.\\

As above, we assume that the prior distribution of $\mu$, denoted by $m_\mu$, is sub-Gaussian.\\

Throughout, we shall respectively denote by $\rho = (\rho^{ij})_{1\le i,j \le d}$ and $\Sigma = (\rho^{ij}\sigma^{i}\sigma^{j})_{1\le i,j \le d}$ the correlation and covariance matrices associated with the dynamics of prices.\\

We introduce the process $(\beta_t)_{t\in\mathbb{R}_+}$ defined by
\begin{eqnarray}
    \forall t \in \mathbb{R}_+,\quad \beta_t&=&\mathbb{E}\left[\left.
\mu
\right|
 \mathcal{F}_t^S
\right].
\end{eqnarray}

We can state a result similar to that of Theorem \ref{propbayes}.
\begin{thm}
\label{prop:thmbeta}
Let us define
\begin{eqnarray}
 F:\quad(t,S)\in\mathbb{R}_+\times\mathbb{R}_+^d
\mapsto
\int_{\mathbb{R}^d}
\exp\left(
z'
\Sigma^{-1}
\left[
	  S-S_0
	-\frac{t}{2}
 	z
\right]
\right)
    m_\mu(dz).
\end{eqnarray}
$F$ is a well-defined finite-valued $C^\infty(\mathbb{R}_+\times\mathbb{R}^d)$ function.\\

We have
\begin{eqnarray}
\forall t\in\mathbb{R}_+,\quad \beta_t&=&
\Sigma
G
(t,S_t), \label{eq:eq52}\end{eqnarray}
where
\begin{eqnarray}
G&=&\frac{\nabla_S F}{F}.
\end{eqnarray}
\end{thm}

As in Section 2, we define the process $\left(\widehat W_t\right)_{t\in\mathbb{R}_+}$ by
\begin{eqnarray}
\forall i\in\left\{1,\hdots,d\right\},\forall t \in\mathbb{R}_+,\quad
    \widehat W_t^i& =& W_t^i+\int_0^t \frac{\mu^i-\beta^i_s}{\sigma^i} ds.
\end{eqnarray}

Using the same method as in Section 2, we can prove the following result on $\left(\widehat W_t\right)_{t\in\mathbb{R}_+}$:
\begin{prop}
$\left(\widehat{W}_{t}\right)_{t\in\mathbb{R}_+}$ is a Brownian motion
adapted to $\left(\mathcal{F}_{t}^{S}\right)_{t\in\mathbb{R}_+}$, with the same correlation structure as $\left({W}_{t}\right)_{t\in\mathbb{R}_+}$
\begin{eqnarray*}
\forall i,j\in \left\{1,\hdots,d\right\},
\quad
d\langle\widehat W^{i}, \widehat W^{j} \rangle_t
&=&
d\langle W^{i}, W^{j} \rangle_t.
\end{eqnarray*}
\end{prop}

The Brownian motion $\left(\widehat{W}_{t}\right)_{t\in\mathbb{R}_+}$ is used to re-write Eq. \eqref{eq:s_eds-1} as
\begin{eqnarray}
dS_t &=& \beta_t dt + \sigma \odot d\widehat{W}_{t}\\
     &=& \Sigma G(t,S_t) dt + \sigma \odot d\widehat{W}_{t}.
\end{eqnarray}

\subsection{Almgren-Chris modelling framework and optimization problems}

We consider the modelling framework introduced by Almgren and Chriss in \cite{almgren1999value,almgren2001optimal} (see also \cite{gueant2015ac, gueantlivre}). In this framework, we do not consider the Mark-to-Market (MtM) value of the portfolio as a state variable. Instead, we consider separately the position $q\in\mathbb{R}^d$ in the risky assets and the amount $X\in\mathbb{R}$ on the cash account.\\

Let us set a time horizon $T\in\mathbb{R}_+^*$. The strategy of the agent is described by the stochastic process $\left(v_{t}\right)_{t\in [0,T]} \in \mathcal{A}^{\textrm{AC}} = \mathcal{A}^{\textrm{AC}}_0$, where, for $t \in [0,T]$,
\begin{eqnarray*}
\mathcal{A}^{\textrm{AC}}_t &=& \left\lbrace (v_s)_{s\in [t,T]}, \mathbb{R}^d\textrm{-valued }\mathcal{F}^S\text{-adapted process},\right.\\
&& \left.\textrm{satisfying the linear growth condition with respect to } (S_s)_{s \in [t,T]}\right\}.
\end{eqnarray*}
This process represents the velocity at which the agent buys and sells the risky assets. In other words,
\begin{eqnarray}
q_{t} & = & q_{0}+\int_{0}^{t}v_{s}ds.
\end{eqnarray}
Now, for $v\in\mathcal{A}^{\textrm{AC}}$, the amount on the cash account evolves as
\begin{eqnarray}
dX_{t} & = & -v_{t}' S_{t} dt - \sum_{i=1}^dV^i_{t}L^i\left(\frac{v^i_{t}}{V^i_{t}}\right)dt,
\end{eqnarray}
where $\forall i\in \{1,\hdots,d\},(V^i_{t})_{t\in [0,T]}$ is a deterministic process, continuous\footnote{The results we obtain in this section can be generalized if the process is only piecewise continuous.} and bounded, modelling the market volume for the $i^\text{th}$ risky
asset,\footnote{This process can be set to very small values for modelling the night.} and where $(L^i)_{1\le i \le d}$ model execution costs. For each $i \in \{1, \ldots, d\}$, the execution cost function $L^i\in C(\mathbb{R},\mathbb{R}_{+})$ classically satisfies:
\begin{itemize}
\item $L^i(0)=0$,
\item $L^i$ is increasing on $\mathbb{R}_{+}$ and decreasing on $\mathbb{R}_{-}$,
\item $L^i$ is strictly convex,
\item $L^i$ is asymptotically superlinear, i.e.,
\begin{eqnarray*}
\lim_{|y|\to+\infty}\frac{L^i(y)}{|y|} & = & +\infty.
\end{eqnarray*}
\end{itemize}

\begin{rem}
In applications, $L^i$ is often a power function $L^i(y)=\eta^i\left|y\right|^{1+\phi^i}$
with $\phi^i>0$, or a function of the form $L^i(y)=\eta^i\left|y\right|^{1+\phi^i}+\psi^i|y|$
with $\phi^i,\psi^i>0$, where $\psi^i$ takes account of proportional costs
such as bid-ask spread or stamp duty. In the original Almgren-Chriss
 framework, the execution costs are quadratic. This corresponds
to $L^i(y)=\eta^iy^{2}$ ($\phi^i=1,\psi^i=0$).\
\end{rem}

Given $v\in \mathcal{A}^{\textrm{AC}}_t$, we define for $s \ge t$,
\begin{eqnarray}
X_{s}^{t,x,S,v} & = & x+\int_{t}^{s}
\left(
-v_{t}' S^{t,S}_{\tau}  - \sum\limits_{i=1}^d V^i_{\tau}L^i\left(\frac{v^i_{\tau}}{V^i_{\tau}}\right)
\right)
d\tau,\\
q_{s}^{t,q,v} & = & q+\int_{t}^{s}v_{\tau}d\tau,\\
S_{s}^{t,S} & = & S+\int_{t}^{s}
\Sigma G(\tau,S^{t,S}_\tau)
d\tau+\int_{t}^{s}\sigma\odot d\widehat{W}_{\tau}.
\end{eqnarray}

We assume that the agent has a constant absolute risk aversion denoted by $\gamma >0$. For an arbitrary initial state $(x_{0},q_0,S_0)$, the optimization problems we consider are of the following generic form:
\begin{equation}
\sup_{(v_t)_{t \in [0,T]}\in\mathcal{A}^{\textrm{AC}}}\mathbb{E}\left[-\exp\left(-\gamma\left(X_{T}^{0,x_0,S_0,v}+{q_{T}^{0,q_{0},v}}'
S_{T}^{0,S_{0}}-\ell\left(q_{T}^{0,q_{0},v}\right)\right)\right)\right],
\end{equation}
where the penalty function $\ell$ is assumed to be continuous and convex.\\

The choice of the penalty function $\ell$ depends on the problem faced by the agent:
\begin{itemize}
  \item In the case of a portfolio choice problem, we can assume that $\ell = 0$ or that $\ell$ penalizes illiquid assets (see for instance \cite{gueant2015ac, gueantlivre}).
  \item In the case of an optimal portfolio liquidation problem, we can assume that the penalty function is of the form $\ell(q) = \frac 12 q'Aq$ with $A\in S_d^{++}(\mathbb{R})$ such that the minimum eigenvalue of $A$ is large enough to force (almost complete) liquidation.\footnote{It is a relaxed form of the classical optimal liquidation problem.}
  \item In the case of an optimal portfolio transition problem, we can assume that the penalty function is of the form $\ell(q) = \frac 12 \left(q-q_{\mathrm{target}}\right)'A\left(q-q_{\mathrm{target}}\right)$ with $A\in S_d^{++}(\mathbb{R})$ such that the minimum eigenvalue of $A$ is large enough to force $q_T$ to be very close to the target $q_{\mathrm{target}}$.\footnote{It is a relaxed form of optimal transition problem.}
\end{itemize}

\subsection{The PDEs in the general case}

Let us introduce the value function $\mathcal{V}$ associated with the above generic problem.
\begin{eqnarray*}
\mathcal{V}&:&\left(t,x,q,S\right) \in[0,T]\times\mathbb{R}\times\mathbb{R}^d\times\mathbb{R}^d\\
&& \mapsto   \sup_{(v_s)_{s \in [t,T]}\in\mathcal{A}^{\textrm{AC}}_t}\mathbb{E}\left[-\exp\left(-\gamma\left(X_{T}^{t,x,S,v}+{q_{T}^{t,q,v}}'
S_{T}^{t,S}-\ell\left(q_{T}^{t,q,v}\right)\right)\right)\right].
\end{eqnarray*}

The HJB equation associated with the problem is
\begin{eqnarray}
\partial_{t}u
+G(t,S)'\Sigma
\nabla_Su
+\dfrac{1}{2}
\mathrm{Tr}
\left(
\Sigma\nabla^2_{SS}u
\right)&&\nonumber \\
+\sup_{v\in\mathbb{R}^d}\left\{ v'\nabla_qu-\left(v'S
+\sum\limits_{i=1}^d V^i_{t}L^i\left(\frac{v^i}{V^i_{t}}\right)\right)\partial_{x}u\right\} & = & 0,\label{eq:HJB_liq}
\end{eqnarray}
with terminal condition
\begin{eqnarray}
\forall (x,q,S)\in \mathbb{R}\times
\mathbb{R}^d\times
\mathbb{R}^d,\quad
u\left(T,x,q,S\right) &=& - \exp(-\gamma(x+q'S - \ell(q))).\label{eq:HJB_liq_fin}
\end{eqnarray}

For reducing the dimensionality of the problem, we consider the following ansatz
\begin{eqnarray}
u\left(t,x,q,S\right) &=& - \exp
\left(
-\gamma
\left(
x+q'S -\theta(t,q,S)
\right)
\label{eq:ansatz_liq}
\right).
\end{eqnarray}

We have the following result:\\

\begin{prop}
\label{prop:u-phi_liq}Suppose there exists $\theta\in{C}^{1,2,2}\left(\left[0,T\right]\times\mathbb{R}^d\times\mathbb{R}^d\right)$
satisfying
\begin{eqnarray}
\nonumber \partial_t \theta
+ G(t,S)'\Sigma
\left(
- q
+\nabla_S\theta
\right)
+\dfrac{1}{2}
\mathrm{Tr}
\left(
\Sigma
\nabla^2_{SS}\theta
\right)
&&\\
+\frac{\gamma}{2}
\left(
- q+\nabla_S\theta
\right)'
\Sigma
\left(
- q+\nabla_S\theta
\right) - \sum_{i=1}^d V_t^i H^i\left(
-\partial_{q^i} \theta
\right)&=&  0,\label{eq:HJB_liq_theta}
\end{eqnarray}
with terminal condition
\begin{eqnarray}
\forall (q,S)\in \mathbb{R}^d\times\mathbb{R}^d,\quad
\theta\left(T,q,S\right) & = &\ell(q),\label{eq:terminal_liq_theta}
\end{eqnarray}
where for all $i\in \{1,\ldots, d\}$, $H^i$ is the Legendre-Fenchel transform of $L^i$, i.e. $$H^i: p \in \mathbb{R} \mapsto \sup_{y \in \mathbb{R}} py - L^i(y).$$
Then $u$ defined by (\ref{eq:ansatz_liq}) is solution of the HJB
equation (\ref{eq:HJB_liq}) with terminal condition (\ref{eq:HJB_liq_fin}).\\

Moreover, the supremum in (\ref{eq:HJB_liq}) is achieved at $v^{\star}(t,q,S) = \left(v^{i\star}(t,q,S)\right)_{1 \le i \le d}$, where
\begin{eqnarray}
\forall i \in \{1, \ldots, d\}, v^{i\star} (t,q,S)& = & V_t^i {H^i}'\left(
-\partial_{q^i} \theta(t,q,S)
\right). \label{eq:optimizer_AC1}
\end{eqnarray}
\end{prop}

\begin{proof}
Let us consider $\theta \in {C}^{1,2,2}\left(\left[0,T\right]\times\mathbb{R}^d\times\mathbb{R}^d\right)$ solution of the PDE (\ref{eq:HJB_liq_theta}) with terminal condition~(\ref{eq:terminal_liq_theta}). For $u$ defined by (\ref{eq:ansatz_liq}), we have

\begin{eqnarray*}
&&\partial_{t}u
+ G(t,S)'\Sigma
\nabla_Su
+\dfrac{1}{2}
\mathrm{Tr}
\left(
\Sigma\nabla^2_{SS}u
\right)
+\sup_{v\in\mathbb{R}^d}\bigg\{ v'\nabla_qu-\left(v'S
+\sum\limits_{i=1}^d V^i_{t}L^i\left(\frac{v^i}{V^i_{t}}\right)\right)\partial_{x}u\bigg\}\\
=&&
\gamma\partial_t \theta u
+G(t,S)'\Sigma
\left(
-\gamma q
+\gamma\nabla_S\theta
\right)u
+\dfrac{1}{2}
\mathrm{Tr}
\left(
\Sigma
\left(
\gamma\nabla^2_{SS}\theta
+
\left(
-\gamma q+\gamma\nabla_S\theta
\right)
\left(
-\gamma q+\gamma\nabla_S\theta
\right)'
\right)
\right)u
\\
&&
-\gamma u \sup_{v\in\mathbb{R}^d}\left\{
 v'
\left(
S-\nabla_q\theta
\right)
-\left(v'S
+\sum\limits_{i=1}^d V^i_{t}L^i\left(\frac{v^i}{V^i_{t}}\right)\right)
\right\}
\\
=&&
\gamma u
\Bigg(
\partial_t \theta
+ G(t,S)'\Sigma
\left(
- q
+\nabla_S\theta
\right)
+\dfrac{1}{2}
\mathrm{Tr}
\left(
\Sigma
\nabla^2_{SS}\theta
\right)
+\frac{\gamma}{2}
\left(
- q+\nabla_S\theta
\right)'
\Sigma
\left(
- q+\nabla_S\theta
\right)
\\
&&
-\sup_{v\in\mathbb{R}^d}\left\{
- v'
\nabla_q\theta
-
\sum\limits_{i=1}^d V^i_{t}L^i\left(\frac{v^i}{V^i_{t}}\right)
\right\}
\Bigg)
\\
=&&
\gamma u
\Bigg(
\partial_t \theta
+G(t,S)'\Sigma
\left(
- q
+\nabla_S\theta
\right)
+\dfrac{1}{2}
\mathrm{Tr}
\left(
\Sigma
\nabla^2_{SS}\theta
\right)
+\frac{\gamma}{2}
\left(
- q+\nabla_S\theta
\right)'
\Sigma
\left(
- q+\nabla_S\theta
\right)\\
&&- \sum_{i=1}^d V_t^i H^i\left(
-\partial_{q^i} \theta
\right)
\Bigg)\\
=&&0.
\end{eqnarray*}
As it is straightforward to verify that $u$ satisfies the terminal condition
(\ref{eq:HJB_liq_fin}), the result is proved.
\end{proof}

The result of the above proposition means that for solving the HJB equation we can solve the simpler three-variable PDE (\ref{eq:HJB_liq_theta}) with terminal condition \eqref{eq:terminal_liq_theta}. However, Eq. (\ref{eq:HJB_liq_theta}) is not linear and corresponds to the equation of a zero-sum game between the agent and nature (see \cite{gueant2015option} for a similar equation in the case of option pricing with execution costs \textit{à la} Almgren-Chriss). Solving Eq. \eqref{eq:HJB_liq_theta} with terminal condition \eqref{eq:terminal_liq_theta} in the general case is out of the scope of this article. However, we can consider the special case where (i) the prior distribution of the drift is Gaussian and (ii) execution costs and penalty functions are quadratic as in the original Almgren-Chriss model, because solving the problem then boils down to solving a system of ODEs.\\

\subsection{The case of a Gaussian prior and quadratic costs}

Let us consider a non-degenerate multivariate Gaussian prior $m_\mu$, i.e.,
\begin{equation}
\label{eq:priorgauss_AC} m_\mu(dz)=\frac{1}{(2\pi)^\frac d2|\Gamma_0|^\frac 12}\exp
\left(
-\frac12 (z-\beta_0)'\Gamma_0^{-1}(z-\beta_0)
\right)dz,
\end{equation}
where $\beta_0\in \mathbb{R}^d$ and $\Gamma_0\in S_d^{++}(\mathbb{R})$.

By using Theorem \ref{prop:thmbeta}, we obtain a result similar to that of Proposition \ref{prop:FandG}.

\begin{prop}
\label{prop:FandG_AC}
For the multivariate Gaussian prior $m_\mu$ given by (\ref{eq:priorgauss_AC}), $G$ is given by
\begin{eqnarray}
\forall t\in\mathbb{R}_+,\forall S\in\mathbb{R}^d,\qquad  G(t,S)&=&
\Sigma^{-1}
\Gamma_t\left(
\Sigma^{-1}
\left(
S-S_0
\right)
+\Gamma_0^{-1}\beta_0
\right).\label{eq:G_Gaussian_AC}
\end{eqnarray}
\end{prop}




For carrying out computations, the following proposition will be useful.

\begin{prop}
\label{diffprop_AC}
The first order partial derivatives of $G$ are given by:\\

$\forall t \in \mathbb{R}_+, \forall S \in \mathbb{R}^d,$
\begin{eqnarray}
\label{eq:dyG_AC} D_S G(t,S) & = & \Sigma^{-1} \Gamma_t \Sigma^{-1},\\
\label{eq:dtG_AC} \partial_t G(t,S) & = & - \Sigma^{-1} \Gamma_t G(t,S).
\end{eqnarray}
\end{prop}

Let us assume, for each $i \in \{1, \ldots, d\}$, that $L^i\left(y\right)= \eta^i{y}^2$. Then, for each $i \in \{1, \ldots, d\}$,
$$H^i : p \in \mathbb{R} \mapsto \sup_{y \in \mathbb{R}} py - \eta^i{y}^2 = \frac{p^2}{4\eta^i}.$$

Let us also assume that $\ell(q) = \frac 12 \left(q-q_{\mathrm{target}}\right)'A\left(q-q_{\mathrm{target}}\right)$ with $A\in S_d^{++}(\mathbb{R})$, the choice of $A$ and $q_{\mathrm{target}}$ depending on the type of problem we consider:
\begin{itemize}
  \item $A = 0$ and $q_{\mathrm{target}}=0$ for an optimal portfolio choice problem.
  \item $A\in S_d^{++}(\mathbb{R})$ with a large minimum eigenvalue and $q_{\mathrm{target}}=0$ for an optimal portfolio liquidation problem.
  \item $A\in S_d^{++}(\mathbb{R})$ with a large minimum eigenvalue and $q_{\mathrm{target}}$ arbitrary for an optimal portfolio transition problem (towards the portfolio represented by $q_{\mathrm{target}}$).\\
\end{itemize}

In order to solve Eq. (\ref{eq:HJB_liq_theta}) with terminal condition \eqref{eq:terminal_liq_theta}, we consider the ansatz
\begin{equation}
\theta\left(t,q,S\right)
 = a\left(t\right) + \dfrac{1}{2} G(t,S)'b\left(t\right)G(t,S) + G(t,S)'c\left(t\right) q + \dfrac{1}{2}q'd\left(t\right)q +G(t,S)' e(t) +q' f(t),
\label{eq:ansatz_theta_good}
\end{equation}
where $a(t) \in \mathbb{R}$, $b(t) \in S_d(\mathbb{R})$, $c(t) \in M_d(\mathbb{R})$, $d(t) \in S_d(\mathbb{R})$, $e(t) \in \mathbb{R}^d$, and $f(t) \in \mathbb{R}^d$.\footnote{The function $d$ should not be confused with the number $d$ of risky assets.}

\begin{prop}
\label{prop:theta}Assume there exists $a\in C^{1}\left(\left[0,T\right]\right)$, $b \in C^{1}\left(\left[0,T\right],S_d(\mathbb{R})\right)$, $c \in C^{1}\left(\left[0,T\right],M_d(\mathbb{R})\right)$, $d \in C^{1}\left(\left[0,T\right],S_d(\mathbb{R})\right)$, $e \in C^{1}\left(\left[0,T\right],\mathbb{R}^d\right)$, and $f \in C^{1}\left(\left[0,T\right],\mathbb{R}^d\right)$
satisfying the following system of ODEs:
\begin{subnumcases}{}
\label{s:a}\dot{a}(t)\! +\! \dfrac{1}{2}\mathrm{Tr}\left( \Gamma_t \Sigma^{-1}b(t) \Sigma^{-1} \Gamma_t \Sigma^{-1}\right)\! +\! \frac \gamma 2 e(t)'\Sigma^{-1} \Gamma_t \Sigma^{-1} \Gamma_t \Sigma^{-1} e(t)\! -\! f(t)'N(t)f(t)\! =\! 0\\
\label{s:b}\dot{b}(t) + \gamma b(t) \Sigma^{-1} \Gamma_t \Sigma^{-1} \Gamma_t \Sigma^{-1} b(t)  - 2 c(t)'N(t)c(t) = 0\\
\label{s:c}\dot{c}(t) - \Sigma + \gamma b(t) \Sigma^{-1} \Gamma_t (-I_d + \Sigma^{-1} \Gamma_t \Sigma^{-1}c(t)) - 2c(t)'N(t)d(t) = 0\\
\label{s:d}\dot{d}(t) + \gamma \left(-I_d + c(t)'\Sigma^{-1} \Gamma_t\Sigma^{-1}\right)\Sigma \left(-I_d + \Sigma^{-1} \Gamma_t \Sigma^{-1}c(t)\right) - 2d(t)N(t)d(t)=0\\
\label{s:e}\dot{e}(t) + \gamma b(t) \Sigma^{-1} \Gamma_t  \Sigma^{-1} \Gamma_t \Sigma^{-1}e(t) -2c(t)'N(t)f(t)=0\\
\label{s:f}\dot{f}(t) + \gamma \left(-I_d + c(t)'\Sigma^{-1} \Gamma_t\Sigma^{-1}\right) \Gamma_t \Sigma^{-1}e(t) - 2d(t)N(t)f(t)=0,
\end{subnumcases}
with terminal condition
\begin{subnumcases}{}
\label{ct:a}a(T)=\frac 12 q_{\mathrm{target}}'A q_{\mathrm{target}}\\
\label{ct:b}b(T)=0\\
\label{ct:c}c(T)=0\\
\label{ct:d}d(T)=A\\
\label{ct:e}e(T)=0\\
\label{ct:f}f(T)=-A q_{\mathrm{target}},
\end{subnumcases}
where $N(t)$ is the diagonal matrix with diagonal $\left(\frac{ V_t^i}{4\eta_i}\right)_{1 \le i \le d}$.\\

Then, the function $\theta$ defined by (\ref{eq:ansatz_theta_good}) satisfies (\ref{eq:HJB_liq_theta}) with terminal condition \eqref{eq:terminal_liq_theta}.
\end{prop}

\begin{proof}
By using Eqs. (\ref{eq:dyG_AC}) and (\ref{eq:dtG_AC}), and noticing that $\partial_t G = -D_S \Sigma G$, we have
\begin{eqnarray*}
&&\partial_t \theta
+ G'\Sigma
\left(
- q
+\nabla_S\theta
\right)
+\dfrac{1}{2}
\mathrm{Tr}
\left(
\Sigma
\nabla^2_{SS}\theta
\right)
+\frac{\gamma}{2}
\left(
- q+\nabla_S\theta
\right)'
\Sigma
\left(
- q+\nabla_S\theta
\right) - \sum_{i=1}^d V_t^i H^i\left(
-\partial_{q^i} \theta
\right)\\
&=& \partial_t \theta
+ G'\Sigma
\left(
- q
+\nabla_S\theta
\right)
+\dfrac{1}{2}
\mathrm{Tr}
\left(
\Sigma
\nabla^2_{SS}\theta
\right)
+\frac{\gamma}{2}
\left(
- q+\nabla_S\theta
\right)'
\Sigma
\left(
- q+\nabla_S\theta
\right) - \nabla_q \theta' N \nabla_q \theta\\
&=& \dot{a} + \dfrac{1}{2} \partial_t G' bG + \dfrac{1}{2} G'\dot{b}G + \dfrac{1}{2} G'b\partial_t G +\partial_t G'cq + G'\dot{c}q + \dfrac{1}{2}q'\dot{d}q + \partial_t G'e + G' \dot{e} +q' \dot{f}\\
&&+ G'\Sigma \left(- q +  D_SGbG + D_SGc q + D_SGe\right) + \dfrac{1}{2}\mathrm{Tr}
\left( \Gamma_t \Sigma^{-1}b \Sigma^{-1} \Gamma_t \Sigma^{-1}
\right)\\
&& +  \frac{\gamma}{2} \left(- q +  D_SGbG + D_SGc q + D_SGe\right)'\Sigma\left(- q +  D_SGbG + D_SGc q + D_SGe\right)\\
&&- \left( cG + dq +f\right)' N \left( cG + dq +f\right)\\
&=& \dot{a} +  \dfrac{1}{2} G'\dot{b}G + G'\dot{c}q + \dfrac{1}{2}q'\dot{d}q + G' \dot{e} +q' \dot{f} - G'\Sigma q + \dfrac{1}{2}\mathrm{Tr}
\left( \Gamma_t \Sigma^{-1}b \Sigma^{-1} \Gamma_t \Sigma^{-1}
\right)\\
&& +  \frac{\gamma}{2} \left(- q +  D_SGbG + D_SGc q + D_SGe\right)'\Sigma\left(- q +  D_SGbG + D_SGc q + D_SGe\right)\\
&&- \left( cG + dq +f\right)' N \left( cG + dq +f\right)\\
&=& \left(\dot{a} + \dfrac{1}{2}\mathrm{Tr}
\left( \Gamma_t \Sigma^{-1}b \Sigma^{-1} \Gamma_t \Sigma^{-1}
\right) + \frac \gamma 2 e'D_SG \Sigma D_S G e - f'Nf\right)\\
&& + \dfrac{1}{2} G'\left(\dot{b} + \gamma b D_SG \Sigma D_SG b  - 2 c'Nc \right)G\\
&& + G'\left(\dot{c} - \Sigma + \gamma b D_SG \Sigma (-I_d + D_SGc) - 2c'Nd \right)q\\
&&+ \dfrac{1}{2}q'\left(\dot{d} + \gamma (-I_d + c'D_SG)\Sigma (-I_d + D_SGc) - 2dNd \right)q\\
&&+ G'\left(\dot{e} +\gamma b D_SG\Sigma D_SGe -2c'Nf\right) +q'\left( \dot{f} + \gamma (-I_d + c'D_SG)\Sigma D_SGe - 2dNf  \right)\\
&=&0.
\end{eqnarray*}
As it is straightforward to verify that $\theta$ satisfies the terminal condition
\eqref{eq:terminal_liq_theta}, the result is proved.
\end{proof}

The above system of ODEs deserves a few comments.\\

In fact, it can be decomposed into 3 sets of ODEs that can be solved one after the other: a first system of nonlinear ODEs \eqref{s:b}-\eqref{s:c}-\eqref{s:d} with the associated terminal conditions \eqref{ct:b}-\eqref{ct:c}-\eqref{ct:d} that defines $(b,c,d)$, a second system of linear ODEs \eqref{s:e}-\eqref{s:f} with the associated terminal conditions \eqref{ct:e}-\eqref{ct:f} that defines $(e,f)$ given $(b,c,d)$, and finally the simple ODE \eqref{s:a} with the associated terminal condition \eqref{ct:a} that defines $a$ given $(b,c,d,e,f)$.\\

The equation \eqref{s:a} for $a$ is trivial to solve. The second set of ODEs does not raise any difficulty because the ODEs are linear. In particular, if $q_{\textrm{target}}=0$, i.e., if we consider an optimal portfolio choice problem or an optimal portfolio liquidation problem, then the solution of the second system of linear ODEs is trivial: $(e,f) = (0,0)$.\\

Regarding the first set of equations, there exists a unique local solution $(b,c,d)$ by Cauchy-Lipschitz. In order to prove that $b$ and $d$ are symmetric matrices, we can proceed as follows: (i) replacing Eq. \eqref{s:c} by
\begin{equation}
\label{s:c2}
\dot{c}(t) - \Sigma + \frac \gamma 2 \left(b(t) + b(t)'\right) \Sigma^{-1} \Gamma_t (-I_d + \Sigma^{-1} \Gamma_t \Sigma^{-1}c(t)) - c(t)'N(t)\left(d(t)+ d(t)'\right) = 0,
\end{equation}
then (ii) considering the unique local solution $(\tilde{b},\tilde{c},\tilde{d})$ of \eqref{s:b}-\eqref{s:c2}-\eqref{s:d} with terminal conditions \eqref{ct:b}-\eqref{ct:c}-\eqref{ct:d}, then (iii) noticing that $(\tilde{b}',\tilde{c},\tilde{d}')$ is also a local solution of \eqref{s:b}-\eqref{s:c2}-\eqref{s:d} with terminal conditions \eqref{ct:b}-\eqref{ct:c}-\eqref{ct:d}, and therefore that $\tilde{b}=\tilde{b}'$ and $\tilde{d}=\tilde{d}'$ are symmetric, (iv) noticing that $(\tilde{b},\tilde{c},\tilde{d})$ is therefore a local solution of \eqref{s:b}-\eqref{s:c}-\eqref{s:d} with the associated terminal conditions \eqref{ct:b}-\eqref{ct:c}-\eqref{ct:d}, and (v) concluding therefore that $b=\tilde{b}$ and $d=\tilde{d}$ are symmetric.\\

Because of the local existence result, if $T$ is small enough, then there exist functions $a\in C^{1}\left(\left[0,T\right]\right)$, $b \in C^{1}\left(\left[0,T\right],S_d(\mathbb{R})\right)$, $c \in C^{1}\left(\left[0,T\right],M_d(\mathbb{R})\right)$, $d \in C^{1}\left(\left[0,T\right],S_d(\mathbb{R})\right)$, $e \in C^{1}\left(\left[0,T\right],\mathbb{R}^d\right)$, and $f \in C^{1}\left(\left[0,T\right],\mathbb{R}^d\right)$ satisfying the equations of Proposition \ref{prop:theta}. However, although we did not find any case of blowup numerically, a global existence result seems out of reach given the nature of system of ODEs.\\

Nevertheless, we can state a verification theorem that solves the problem when there exists a solution to the above system on $[0,T]$.

\begin{thm}
\label{verif_AC}
Assume there exist $a\in C^{1}\left(\left[0,T\right]\right)$, $b \in C^{1}\left(\left[0,T\right],S_d(\mathbb{R})\right)$, $c \in C^{1}\left(\left[0,T\right],M_d(\mathbb{R})\right)$, $d \in C^{1}\left(\left[0,T\right],S_d(\mathbb{R})\right)$, $e \in C^{1}\left(\left[0,T\right],\mathbb{R}^d\right)$, and $f \in C^{1}\left(\left[0,T\right],\mathbb{R}^d\right)$ satisfying the equations of Proposition \ref{prop:theta}. Let us then consider the function $\theta$ defined by \eqref{eq:ansatz_theta_good} and the associated function $u$ defined by \eqref{eq:ansatz_liq}.\\

For all $\left(t,x,q,S\right)\in\left[0,T\right]\times\mathbb{R}\times\mathbb{R}^{d}\times\mathbb{R}^{d}$
and $v = (v_s)_{s \in [t,T]}\in\mathcal{A}^{\textrm{AC}}_t$, we have
\begin{eqnarray}
\mathbb{E}\left[-\exp\left(-\gamma \left(X_{T}^{t,x,S,v}+{q_{T}^{t,q,v}}'
S_{T}^{t,S}-\ell\left(q_{T}^{t,q,v}\right)\right) \right)\right] & \leq & u\left(t,x,q,S\right).\label{eq:verif_ineq_liq}
\end{eqnarray}
Moreover, equality in (\ref{eq:verif_ineq_liq}) is obtained by taking the
optimal control $(v^{\star}_s)_{s \in [t,T]}\in\mathcal{A}^{\textrm{AC}}_t$ given by the closed-loop feedback formula
\begin{equation}
\label{closedloop}
\forall s \in [t,T], v^{\star}_s = \phi(s) q_s^{t,q,v^\star} + \psi(s,S^{t,S}_s),
\end{equation}
where $\phi: t \in \mathbb{R}_+ \mapsto - 2 N(t) d(t)$ and $\psi : (t,S) \in [0,T]\times\mathbb{R}^d \mapsto  - 2N(t)(c(t)G(t,S) + f(t))$.\\

In particular $u=\mathcal{V}$.
\end{thm}

\begin{proof}

Let us first prove that $(v_s^\star)_{s \in [t,T]}$ is well-defined and admissible (i.e., $(v_s^\star)_{s \in [t,T]} \in \mathcal{A}^{\textrm{AC}}_t$).\\

For that purpose, let us consider the Cauchy problem
$$\frac{d\tilde{q}_s}{ds} = \phi(s) \tilde{q}_s + \psi(s,S^{t,S}_s), \quad \tilde{q}_t = q.$$
Its unique solution is given by $$\forall s \in [t,T],\quad \tilde{q}_s = \exp\left({\int_t^s \phi(\tau) d\tau}\right) \left(q + \int_t^s \psi(\tau,S^{t,S}_\tau) \exp\left({-\int_t^\tau \phi(\zeta) d\zeta}\right) d\tau\right).$$
Then $v^{\star}$ is defined by $\dot{\tilde{q}}$ and can be written as
$$\forall s \in [t,T], \quad v^{\star}_s = \phi(s) \exp\left({\int_t^s \phi(\tau) d\tau}\right) \left(q + \int_t^s \psi(\tau,S^{t,S}_\tau) \exp\left({-\int_t^\tau \phi(\zeta) d\zeta}\right) d\tau\right) + \psi(s,S^{t,S}_s).$$
Given the definition of $\psi$ and the affine nature of $G$ with respect to $S$, $(v_s^\star)_{s \in [t,T]}$ satisfies the required linear growth condition to be in $\mathcal{A}^{\textrm{AC}}_t$.\\

Now, let us consider $\left(t,x,q,S\right)\in [0,T]\times\mathbb{R}\times\mathbb{R}^{d}\times\mathbb{R}^{d}$ and $v = (v_s)_{s \in [t,T]}\in\mathcal{A}^{\textrm{AC}}_t$.\\

By It\=o's formula, we have for all $s\in[t,T]$
\begin{eqnarray*}
&&du\left(s,X^{t,x,S,v}_s,q^{t,q,v}_s,S^{t,S}_s\right)\\
\!\!&=&\!\! \mathcal{L}^v u\left(s,X^{t,x,S,v}_s,q^{t,q,v}_s,S^{t,S}_s\right) ds
+ \nabla_Su\left(s,X^{t,x,S,v}_s,q^{t,q,v}_s,S^{t,S}_s\right)' \left(\sigma \odot d\widehat{W}_s\right).\\
\end{eqnarray*}
where
\begin{eqnarray*}
&&\mathcal{L}^v u\left(s,X^{t,x,S,v}_s,q^{t,q,v}_s,S^{t,S}_s\right)\\
&=& \partial_t u\left(s,X^{t,x,S,v}_s,q^{t,q,v}_s,S^{t,S}_s\right)+
G(s,S^{t,S}_s)'\Sigma\nabla_Su\left(s,X^{t,x,S,v}_s,q^{t,q,v}_s,S^{t,S}_s\right)\\
&&+ \dfrac{1}{2}
\mathrm{Tr}
\left(
\Sigma\nabla^2_{SS}u\left(s,X^{t,x,S,v}_s,q^{t,q,v}_s,S^{t,S}_s\right)\right) + v_s'\nabla_qu\left(s,X^{t,x,S,v}_s,q^{t,q,v}_s,S^{t,S}_s\right)\\ &&-\left(v_s'S^{t,S}_s
+\sum\limits_{i=1}^d V^i_{t}L^i\left(\frac{v_s^i}{V^i_{t}}\right)\right)\partial_{x}u\left(s,X^{t,x,S,v}_s,q^{t,q,v}_s,S^{t,S}_s\right).
\end{eqnarray*}
Note that we have
\begin{eqnarray*}
&&\nabla_Su\left(s,X^{t,x,S,v}_s,q^{t,q,v}_s,S^{t,S}_s\right)\\
&=& -\gamma u\left(s,X^{t,x,S,v}_s,q^{t,q,v}_s,S^{t,S}_s\right) \left( q_s^{t,q,v} - \nabla_S \theta\left(s,q^{t,q,v}_s,S^{t,S}_s\right)\right)\\
&=& -\gamma u\left(s,X^{t,x,S,v}_s,q^{t,q,v}_s,S^{t,S}_s\right) \left( q_s^{t,q,v} - \Sigma^{-1} \Gamma_t \Sigma^{-1}b(t)G(t,S_s^{t,S})\right.\\
& & \qquad\qquad\qquad\qquad\qquad\qquad \left.- \Sigma^{-1} \Gamma_t \Sigma^{-1}c(t) q^{t,q,v}_s - \Sigma^{-1} \Gamma_t \Sigma^{-1}e(t) \right).
\end{eqnarray*}
Let us subsequently define, for all $s\in [t,T]$,
\begin{align*}
\kappa^v_s&=
-\gamma \left( q_s^{t,q,v} - \Sigma^{-1} \Gamma_t \Sigma^{-1}b(t)G(t,S_s^{t,S}) - \Sigma^{-1} \Gamma_t \Sigma^{-1}c(t) q^{t,q,v}_s - \Sigma^{-1} \Gamma_t \Sigma^{-1}e(t) \right)
,
\end{align*}
and
\begin{align*}
\xi^v_{t,s}&=\exp
\left(
\int_t^s{\kappa^v_{\tau}}'
\left(\sigma \odot d\widehat W_\tau\right)
-\frac12
\int_t^s {\kappa^v_{\tau}}'\Sigma\kappa^v_{\tau}d\tau
\right).
\end{align*}
We have
\begin{eqnarray*}
d\left(
u\left(s,X^{t,x,S,v}_s,q^{t,q,v}_s,S^{t,S}_s\right)\left(\xi^v_{t,s}\right)^{-1}
\right) &=& \left(\xi^v_{t,s}\right)^{-1}
\mathcal{L}^vu\left(s,X^{t,x,S,v}_s,q^{t,q,v}_s,S^{t,S}_s\right)ds.
\end{eqnarray*}
By definition of $u$, $\mathcal{L}^vu\left(s,X^{t,x,S,v}_s,q^{t,q,v}_s,S^{t,S}_s\right)\leq 0$.\\

Moreover, by \eqref{eq:optimizer_AC1}, $\mathcal{L}^vu\left(s,X^{t,x,S,v}_s,q^{t,q,v}_s,S^{t,S}_s\right)=0$ if $v$ satisfies
\begin{eqnarray*} v_s = -2 N(s) \left(c(s)G(s,S^{t,S}_s) + d(s)q_s^{t,q,v} +f(s) \right) = \phi(s) q_s^{t,q,v} + \psi(s,S^{t,S}_s),
\end{eqnarray*}
which is the case when $(v_s)_{s \in [t,T]} = (v^{\star}_s)_{s \in [t,T]}$.\\

As a consequence,
$\left(
u\left(s,X^{t,x,S,v}_s,q^{t,q,v}_s,S^{t,S}_s\right)\left(\xi^v_{t,s}\right)^{-1}
\right)_{s\in[t,T]}$ is nonincreasing, and therefore
$$u\left(T,X^{t,x,S,v}_T,q^{t,q,v}_T,S^{t,S}_T\right) \le u(t,x,q,S)\xi^v_{t,T},$$
with equality when $(v_s)_{s \in [t,T]} = (v^{\star}_s)_{s \in [t,T]}$.\\

Subsequently,
\begin{eqnarray*}\mathbb{E}\left[-\exp\left(-\gamma \left(X_{T}^{t,x,S,v}+{q_{T}^{t,q,v}}'
S_{T}^{t,S}-\ell\left(q_{T}^{t,q,v}\right)\right)\right)\right] &=& \mathbb{E}\left[u\left(T,X^{t,x,S,v}_T,q^{t,q,v}_T,S^{t,S}_T\right)\right]\\
&\leq& u(t,x,q,S)\mathbb{E}\left[\xi^v_{t,T}\right],
\end{eqnarray*}
with equality when $(v_s)_{s \in [t,T]}= (v^{\star}_s)_{s \in [t,T]}$.\\

Because $v \in \mathcal{A}^{\textrm{AC}}_t$ satisfies the linear growth condition with respect to $(S^{t,S}_s)_{s \in [t,T]}$, so does $(q^{t,q,v}_s)_{s \in [t,T]}$. Therefore, using the same argument as in Theorem \ref{verif_cara_gl}, we see that $\left(\xi^v_{t,s}\right)_{s\in[t,T]}$ is a martingale
with $\mathbb{E}\left[\xi^v_{t,s}\right]=1$ for all $s\in[t,T]$.\\

We obtain
$$\mathbb{E}\left[-\exp\left(-\gamma \left(X_{T}^{t,x,S,v}+{q_{T}^{t,q,v}}'
S_{T}^{t,S}-\ell\left(q_{T}^{t,q,v}\right)\right)\right)\right] = \mathbb{E}\left[u\left(T,X^{t,x,S,v}_T,q^{t,q,v}_T,S^{t,S}_T\right)\right]\leq u(t,x,q,S),$$
with equality when $(v_s)_{s \in [t,T]}= (v^{\star}_s)_{s \in [t,T]}$.\\

We can conclude that
\begin{eqnarray*} \mathcal{V}(t,x,q,S) &=& \sup_{(v_s)_{s \in [t,T]} \in \mathcal{A}^{\textrm{AC}}_t}\!\!\! \mathbb{E}\left[-\exp\left(-\gamma \left(X_{T}^{t,x,S,v}+{q_{T}^{t,q,v}}'
S_{T}^{t,S}-\ell\left(q_{T}^{t,q,v}\right)\right) \right)\right]\\
&=& \mathbb{E}\left[-\exp\left(-\gamma \left(X_{T}^{t,x,S,v}+{q_{T}^{t,q,v^\star}}'
S_{T}^{t,S}-\ell\left(q_{T}^{t,q,v^\star}\right)\right) \right)\right]\\
&=&u(t,x,q,S).
\end{eqnarray*}
\end{proof}

\subsection{Numerical examples and comments}

We consider now three simple examples in order to illustrate the results obtained above. For these three examples, we consider one risky asset (stock) with the following characteristics:
\begin{itemize}
\item $S_0 = 50$ \euro,
\item $\mu=0.01$ \euro$\cdot\text{day}^{-1}$,
\item $\sigma=0.6$ \euro$\cdot\text{day}^{-1/2}$,
\item $V=4000000$ shares$\cdot$$\text{day}^{-1}$,
\item $L(y)=\eta|y|^{2}$ with $\eta=0.15$ \euro $\cdot\mbox{shares}^{-1}\cdot\text{day}^{-1}$.\\
\end{itemize}

The first problem we consider is an optimal portfolio choice problem (with $q_0 = 0$). The parameters are the following:
\paragraph{Objective function}
\begin{itemize}
\item $T=10$ days,
\item $\gamma=2\cdot10^{-7} \text{\euro}^{-1}$,
\item $\ell = 0$.
\end{itemize}

\paragraph{Bayesian prior $\mathcal{N}(\beta_0, \nu_0^2)$}
\begin{itemize}
\item $\beta_0 = 0.01$ \euro $\cdot\text{day}^{-1}$,
\item $\nu_0 = 0.03$ \euro $\cdot\text{day}^{-1}$.\\
\end{itemize}

Our methodology was first to approximate numerically the functions $a$, $b$, $c$, and $d$ (we know that $(e,f) = (0,0)$). Then, for different simulated paths of the stock price, we used Eq. (\ref{closedloop}) for finding -- in fact approximating numerically -- the optimal number of shares in the portfolio at each point in time (on a grid). The results are shown in Figure 1.\\

\begin{figure}[h!]
\centering
\includegraphics[width=14.5cm]{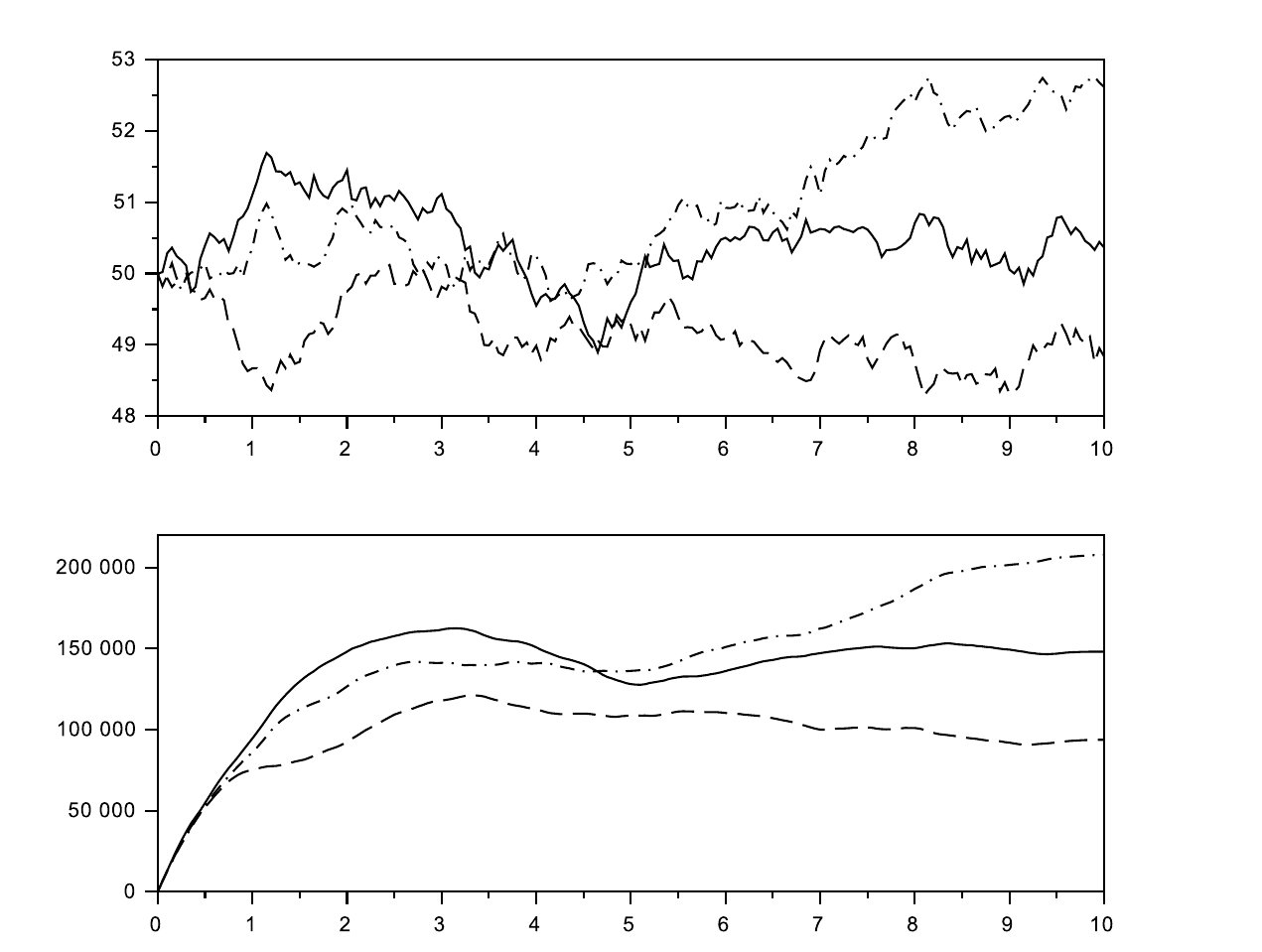}
\caption{Solution of the optimal portfolio choice problem for three trajectories of $S$. Top panel: price of the risky asset $S_t$. Bottom panel: Position $q_t$ in the risky asset.}
\end{figure}

Two things must be noticed in Figure 1. First, the agent builds a portfolio with a number of shares that lies around $q_{\text{opt}}$, where
$$q_{\text{opt}} = \frac{\mu}{\gamma \sigma^2} \simeq 138889$$
is the number of shares that would be optimal in the optimal portfolio choice model without uncertainty on $\mu$ and without execution costs. Second, the strategy followed by the agent looks like a trend-following strategy: the agent buys when the stock price increases and sells when the stock price decreases, though in a smooth manner. This is in fact quite natural given the dynamics of $(\beta_t)_{t\in\mathbb{R}_+}$.\\

The second problem we consider is an optimal portfolio liquidation problem (with $q_0 = 100000$ shares). The parameters are the following:
\paragraph{Objective function}
\begin{itemize}
\item $T=1$ day,
\item $\gamma=2\cdot10^{-6}\text{\euro}^{-1}$,
\item $A = 5 \cdot10^{-6}$ \euro $\cdot\text{share}^{-2}$.\footnote{The matrix $A$ is a scalar in the one-asset case.}
\end{itemize}

\paragraph{Bayesian prior $\mathcal{N}(\beta_0, \nu_0^2)$}
\begin{itemize}
\item $\beta_0 = 0.01$ \euro $\cdot\text{day}^{-1}$,
\item $\nu_0 = 0.1$ \euro $\cdot\text{day}^{-1}$.\\
\end{itemize}

We first approximated numerically the functions $a$, $b$, $c$, and $d$ (we know that $(e,f) = (0,0)$). Then, for different simulated paths of the stock price, we used Eq. (\ref{closedloop}) for approximating the optimal number of shares in the portfolio at each point in time (on a grid). The results are shown in Figure 2.\\

\begin{figure}[h!]
\centering
\includegraphics[width=14.5cm]{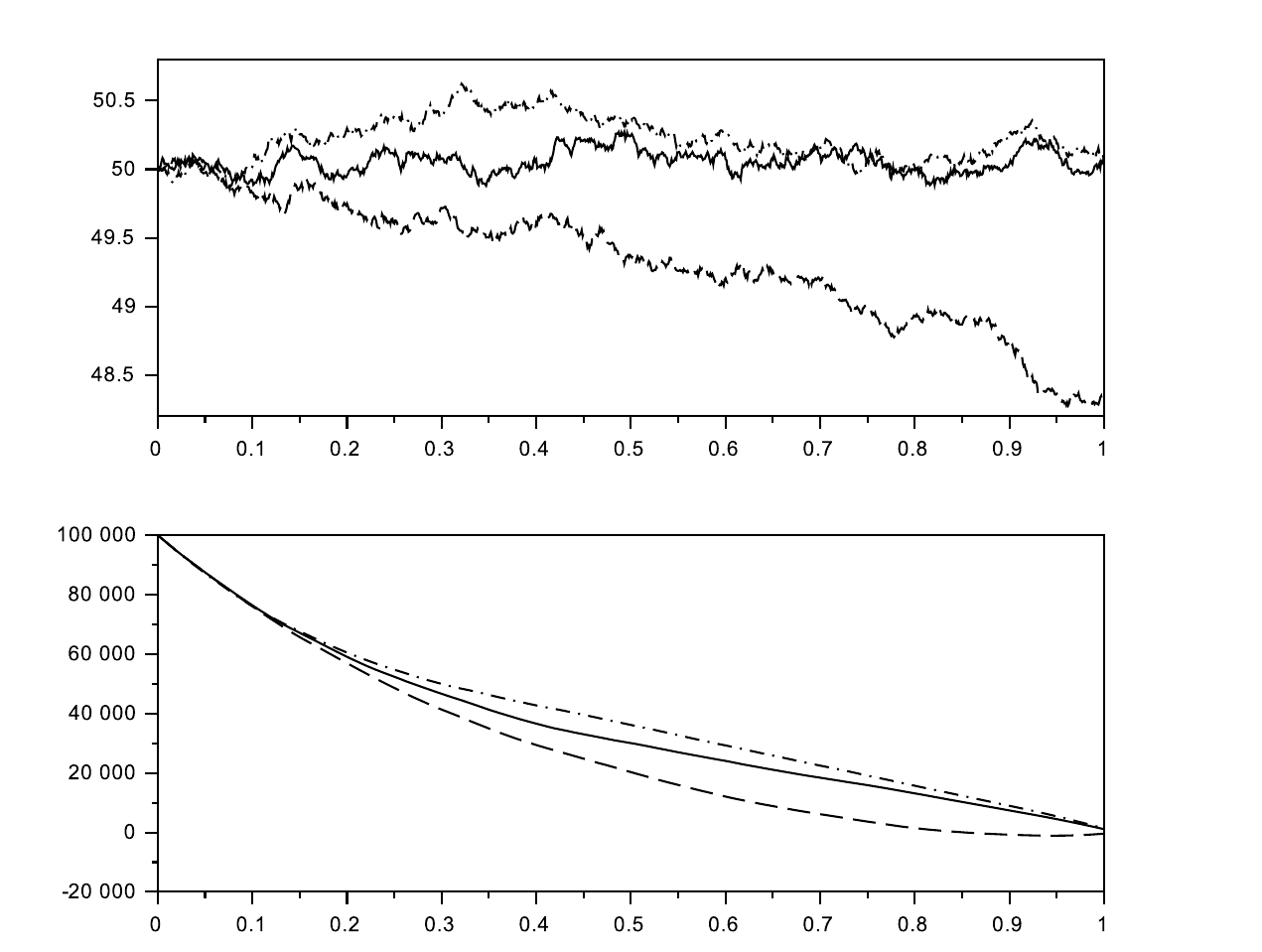}
\caption{Solution of the optimal portfolio liquidation problem for three trajectories of $S$. Top panel: price of the risky asset $S_t$. Bottom panel: Position $q_t$ in the risky asset.}
\end{figure}

We see in Figure 2 that the small value of $A$ we used is high enough to force complete liquidation in all of the three cases. We also see that the optimal (adaptive) strategy consists in liquidating at a faster pace for decreasing price trajectories than for increasing price trajectories. This is in line with the trend following effect exhibited in Figure 1.\\

The third problem we consider is an optimal portfolio transition problem (with $q_0 = 100000$ shares). The parameters are the following:
\paragraph{Objective function}
\begin{itemize}
\item $T=1$ day,
\item $\gamma=2\cdot10^{-6} \text{\euro}^{-1}$,
\item $q_\text{target} = 200000$ shares,
\item $A = 5 \cdot10^{-6}$ \euro $\cdot\text{share}^{-2}$.
\end{itemize}

\paragraph{Bayesian prior $\mathcal{N}(\beta_0, \nu_0^2)$}
\begin{itemize}
\item $\beta_0 = 0.01$ \euro $\cdot\text{day}^{-1}$,
\item $\nu_0 = 0.1$ \euro $\cdot\text{day}^{-1}$.\\
\end{itemize}

As above, we approximated numerically the functions $a$, $b$, $c$, $d$, $e$, and $f$, and then used Eq. (\ref{closedloop}) for approximating the optimal number of shares in the portfolio at each point in time (on a grid) for three different simulated paths of the stock price. The results are shown in Figure 3.\\

\begin{figure}[h!]
\centering
\includegraphics[width=14.5cm]{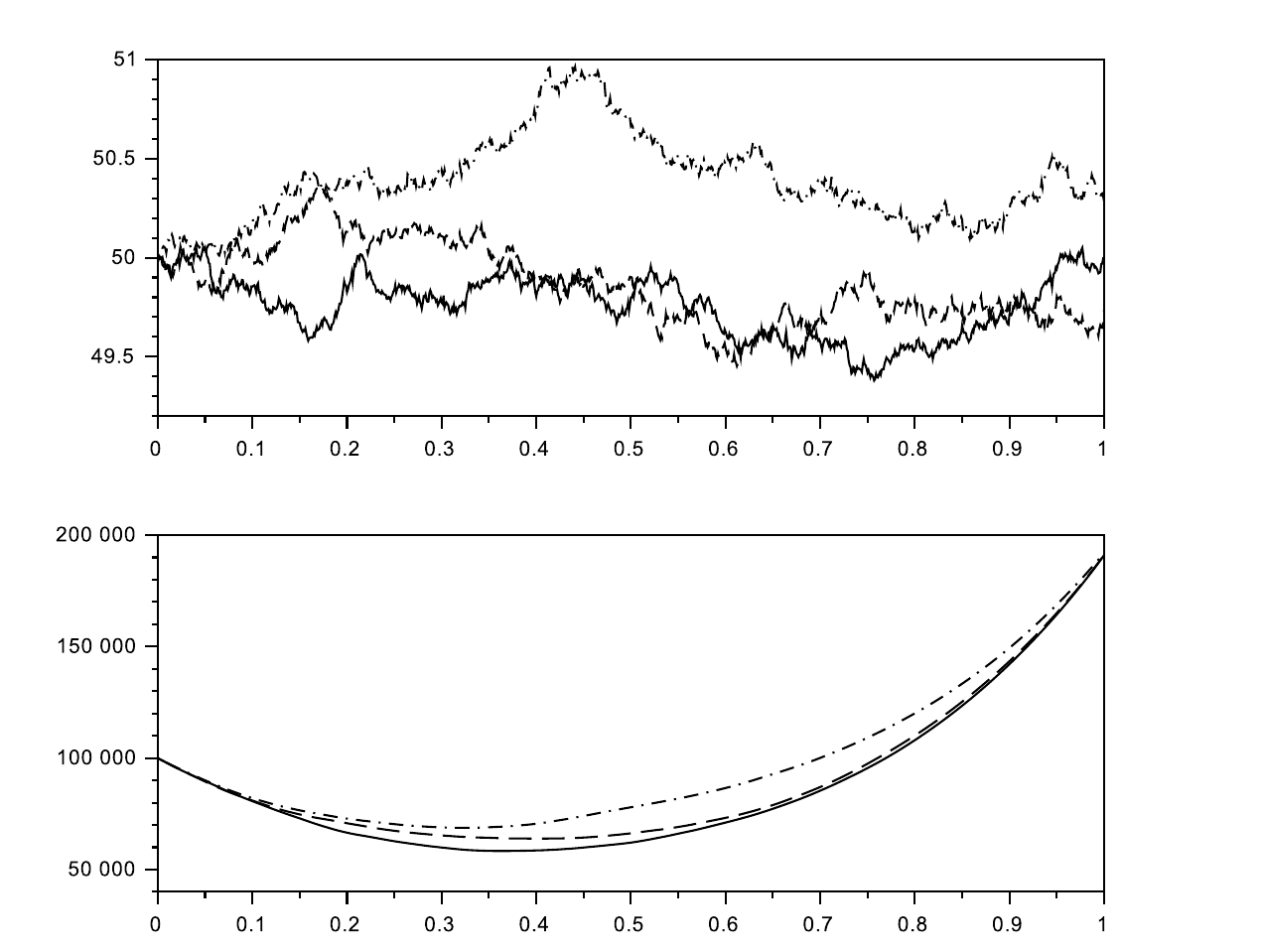}
\caption{Solution of the optimal portfolio transition problem for three trajectories of $S$. Top panel: price of the risky asset $S_t$. Bottom panel: Position $q_t$ in the risky asset.}
\end{figure}

We see in Figure 3 that the small value of $A$ we used is high enough to force complete transition from portfolio $q_0$ to portfolio $q_{\text{target}}$ in all of the three cases. In addition to the classical trend-following-like effect, we see in Figure 3 that the optimal strategy consists in selling shares before buying them back. In fact, the agent faces a trade-off because there are two opposite forces. When the final penalty is far away (i.e., at the beginning of the process), the agent faces a portfolio choice problem similar to the one tackled in the first example. Here, $$q_{\text{opt}} = \frac{\mu}{\gamma \sigma^2} \simeq  13889 < q_0.$$ Therefore, there is an incentive to sell shares at the beginning. After some time however, the final condition matters and the agent has to reach the target, hence the U-shaped trajectory.\\

These three examples illustrate the use of the PDE method for solving various problems under drift uncertainty.

\section*{Conclusion}

In this paper, we have presented a PDE method that can be used for addressing optimal portfolio choice, optimal portfolio liquidation, and optimal portfolio transition problems, when the expected returns of risky assets are unknown. The main idea is to use at the same time Bayesian (or more generally online) learning and dynamic programming techniques. Our approach goes beyond the martingale method of Karatzas and Zhao, because it can be used in more general models, for instance when a modelling framework \emph{à  la} Almgren-Chriss is considered. We believe that the use of Bayesian (or more generally online) learning in conjunction with stochastic optimal control enables to improve many models without increasing their dimensionality and we are looking forward to seeing other applications of the same method, especially in Finance.


\begin{thebibliography}{10}

\bibitem{almgren1999value}
Robert Almgren and Neil Chriss.
\newblock Value under liquidation.
\newblock {\em Risk}, 12(12):61--63, 1999.

\bibitem{almgren2001optimal}
Robert Almgren and Neil Chriss.
\newblock Optimal execution of portfolio transactions.
\newblock {\em Journal of Risk}, 3:5--40, 2001.

\bibitem{almgren2016option}
Robert Almgren and Tianhui~Michael Li.
\newblock Option hedging with smooth market impact.
\newblock {\em Market microstructure and liquidity}, 2(1), 2016.

\bibitem{almgren2006bayesian}
Robert Almgren and Julian Lorenz.
\newblock Bayesian adaptive trading with a daily cycle.
\newblock {\em The Journal of Trading}, 1(4):38--46, 2006.

\bibitem{bain}
Alan Bain and Dan Crisan.
\newblock Fundamentals of stochastic filtering.
\newblock {\em Springer}, 2009.

\bibitem{bdl}
Tomas Björk, Mark Davis, and Camilla Landén.
\newblock Optimal investment under partial information.
\newblock {\em Mathematical Methods of Operations Research}, 71(2):371--399, 2010

\bibitem{black1992global}
Fischer Black and Robert Litterman.
\newblock Global portfolio optimization.
\newblock {\em Financial analysts journal}, 48(5):28--43, 1992.

\bibitem{brendle2006portfolio}
Simon Brendle.
\newblock  Portfolio selection under incomplete information.
\newblock {\em Stochastic processes and their Applications}, 116(5):701--723, 2006.

\bibitem{cj}
Philippe Casgrain and Sebastian Jaimungal.
\newblock Trading Algorithms with Learning in Latent Alpha Models.
\newblock \emph{Working paper}, 2017

\bibitem{rogers2001}
L. Chris and G. Rogers.
\newblock The relaxed investor and parameter uncertainty.
\newblock {\em Finance and Stochastics}, 5(2):131--154, 2001.

\bibitem{cvitanic1992convex}
Jak{\v{s}}a Cvitani{\'c} and Ioannis Karatzas.
\newblock Convex duality in constrained portfolio optimization.
\newblock {\em The Annals of Applied Probability}, 767--818, 1992.


\bibitem{cvitanic2006dynamic}
  Jak{\v{s}}a Cvitani{\'c}, Ali Lazrak, Lionel Martellini, and Fernando Zapatero.
  \newblock Dynamic portfolio choice with parameter uncertainty and the economic value of analysts' recommendations
  \newblock {\em Review of Financial Studies}, 19(4):1113--1156, 2006.


\bibitem{danilova}
Albina Danilova, Michael Monoyios and Andrew Ng.
\newblock Optimal investment with inside information
and parameter uncertainty.
\newblock {\em Mathematics and Financial Economics}, 3(1):13--38, 2010.

\bibitem{davislleo}
Mark Davis and S\'ebastien Lleo.
\newblock Black–Litterman in continuous time: the case for filtering.
\newblock  {\em Quantitative Finance Letters}, 1(1), 30--35, 2013.

\bibitem{ekstrom}
Erik Ekstr\"om and Juozas Vaicenavicius.
\newblock Optimal Liquidation of an Asset under Drift Uncertainty.
\newblock {\em SIAM journal of financial mathematics}, 7(1):357--381, 2016.


\bibitem{joaquin}
Joaquin Fernandez-Tapia.
\newblock High-Frequency Trading with On-Line Learning.
\newblock\newblock {\em Working paper}, 2015.

\bibitem{fouque}
Jean-Pierre Fouque, Andrew Papanicolaou, and Ronnie Sircar.
\newblock Filtering and portfolio optimization with stochastic unobserved drift in asset returns.
\newblock {\em Communications in Mathematical Sciences}, 13(4):935--953, 2015.

\bibitem{fouque2}
Jean-Pierre Fouque, Andrew Papanicolaou, and Ronnie Sircar.
\newblock Perturbation analysis for investment portfolios under partial information with expert opinions.
\newblock {\em SIAM Journal on Control and Optimization}, 55(3), 1534-1566, 2017.

\bibitem{af}
Avner Friedman.
\newblock Partial differential equations of parabolic type.
\newblock {\em Courier Dover Publications}, 2008.

\bibitem{gueant2014optimal}
Olivier Gu{\'e}ant.
\newblock Optimal execution of asr contracts with fixed notional.
\newblock {\em Journal of Risk}, 19(3):77--99, 2017.

\bibitem{gueant2015ac}
Olivier Gu{\'e}ant.
\newblock Optimal execution and block trade pricing: a general framework.
\newblock {\em Applied Mathematical Finance}, 22(4), 2015.

\bibitem{gueant2015option}
Olivier Gu{\'e}ant and Jiang Pu.
\newblock Option pricing and hedging with execution costs and market impact.
\newblock {\em Mathematical Finance}, 27(3):803--831, 2017.

\bibitem{gueant2015accelerated}
Olivier Gu{\'e}ant, Jiang Pu, and Guillaume Royer.
\newblock Accelerated share repurchase: pricing and execution strategy.
\newblock {\em International Journal of Theoretical and Applied Finance},
  18(3), 2015.

\bibitem{gueantlivre}
Olivier Gu\'eant. The Financial Mathematics of Market Liquidity: From Optimal Execution to Market Making. CRC Press. 2016.

\bibitem{vroum}
Toshiki Honda.
\newblock Optimal portfolio choice for unobservable and regime-switching mean returns.
\newblock {\em Journal of Economic Dynamics and Control}, 28(1):45--78, 2003.

\bibitem{karatzas1987optimal}
Ioannis Karatzas, John~P. Lehoczky, and Steven~E. Shreve.
\newblock Optimal portfolio and consumption decisions for a ``small investor''
  on a finite horizon.
\newblock {\em SIAM journal on control and optimization}, 25(6):1557--1586,
  1987.

\bibitem{ks}
Ioannis Karatzas and Steven~E. Shreve.
\newblock Brownian motion and stochastic calculus (Vol. 113). Springer Science \& Business Media. 2012.

\bibitem{karatzas1998bayesian}
Ioannis Karatzas and Xiaoliang Zhao.
\newblock Bayesian adaptive portfolio optimization.
\newblock {\em Preprint, Columbia University}, 1998.

\bibitem{lak95}
Peter Lakner.
\newblock Utility maximization with partial information.
\newblock {\em Stochastic processes and their applications}, 56(2):247--273, 1995.

\bibitem{lak98}
Peter Lakner
\newblock Optimal trading strategy for an investor: the case of partial information.
\newblock {\em Stochastic processes and their applications}, 76(1):77--97, 1998.

\bibitem{laruelle}
Sophie Laruelle, Charles-Albert Lehalle, and Gilles Pagès.
\newblock Optimal posting price of limit orders: learning by trading.
\newblock {\em Mathematics and Financial Economics}, 7(3):359--403, 2013.

\bibitem{li}
Yongwu Li, Han Qiao, Shouyang Wang, and Ling Zhang.
\newblock Time-consistent investment
strategy under partial information.
\newblock {\em Insurance: Mathematics and Economics}, 65(C):187--197, 2015.

\bibitem{lipster}
Robert Liptser and Albert N. Shiryaev.
\newblock Statistics of Stochastics Processes: Vol 1\&2.
\newblock Springer. 2001


\bibitem{markowitz1952portfolio}
Harry M. Markowitz.
\newblock Portfolio selection.
\newblock {\em The journal of finance}, 7(1):77--91, 1952.

\bibitem{markowitz1999early}
Harry~M. Markowitz.
\newblock The early history of portfolio theory: 1600-1960.
\newblock {\em Financial Analysts Journal}, 55(4):5--16, 1999.

\bibitem{merton1969lifetime}
Robert~C. Merton.
\newblock Lifetime portfolio selection under uncertainty: The continuous-time
  case.
\newblock {\em The review of Economics and Statistics}, 247--257, 1969.

\bibitem{merton1971optimum}
Robert~C. Merton.
\newblock Optimum consumption and portfolio rules in a continuous-time model.
\newblock {\em Journal of economic theory}, 3(4):373--413, 1971.

\bibitem{monoyios2009}
Michael Monoyios.
\newblock Optimal investment and hedging under partial
and inside information.
\newblock {\em Radon Series on Computational and Applied Mathematics}, 8:371–-410, 2009.

\bibitem{ps2008}
Wolfgang Putschögl and J\"orn Sass.
\newblock Optimal consumption and investment under partial information.
\newblock {\em Decisions in Economics and Finance}, 31(2):137--170, 2008.

\bibitem{rieder2005portfolio}
Ulrich Rieder and Nicole B{\"a}uerle.
\newblock Portfolio optimization with unobservable Markov-modulated drift
  process.
\newblock {\em Journal of Applied Probability}, 362--378, 2005.

\bibitem{rishel1999optimal}
Raymond Rishel.
\newblock Optimal portfolio management with partial observations and power
  utility function.
\newblock In {\em Stochastic analysis, control, optimization and applications},
  605--619. Springer, 1999.

\bibitem{samuelson1969lifetime}
Paul~A. Samuelson.
\newblock Lifetime portfolio selection by dynamic stochastic programming.
\newblock {\em The review of economics and statistics}, 239--246, 1969.

\bibitem{sh2004}
J\"orn Sass and Ulrich Haussmann.
\newblock  Optimizing the terminal wealth under partial information: The drift process as a continuous time Markov chain.
\newblock {\em Finance and Stochastics}, 8(4):553--577, 2004.

\bibitem{sass2}
J\"orn Sass, Dorothee Westphal, and Ralf Wunderlich.
\newblock Expert Opinions and Logarithmic Utility Maximization for Multivariate Stock Returns with Gaussian Drift.
\newblock {\em International Journal of Theoretical and Applied Finance}, 20(04), 2017.

\bibitem{tobin1958liquidity}
James Tobin.
\newblock Liquidity preference as behavior towards risk.
\newblock {\em The review of economic studies}, 25(2):65--86, 1958.

\end{thebibliography}
\end{document}